\documentclass[12pt,report]{elsart}
\usepackage{epsfig,graphics,colordvi,color}

\usepackage{amsmath}\usepackage{amssymb}
\usepackage{hyperref}

\newcommand{\pslash}{\FMslash p}
\newcommand{\barqslash}{\FMslash {\bar q}}
\newcommand{\wslash}{\FMslash w}
\newcommand{\qslash}{\FMslash q}
\newcommand{\Qslash}{\FMslash Q}

\newcommand{\notsubset}{\;\FMslash \subset \;}
\def\rescale{\fontsize{10}{3}}
\begin{document}
\begin{frontmatter}
\Large{\title{Photon- and pion-nucleon interactions\\
in a  unitary and causal effective field theory \\
based on the chiral Lagrangian }}
\author[GSI,ITEP]{A. Gasparyan,}
\author[GSI]{and M.F.M. Lutz}
\address[GSI]{Gesellschaft f\"ur Schwerionenforschung (GSI),\\
Planck Str. 1, 64291 Darmstadt, Germany}
\address[ITEP]{Institute for Theoretical and Experimental Physics,\\
117259, B. Cheremushkinskaya 25, Moscow, Russia}
\begin{abstract}
We present and apply a novel scheme for studying photon- and pion-nucleon scattering
beyond the threshold region. Partial-wave amplitudes for the $\gamma\, N$ and $\pi \, N$
states are obtained by  an analytic extrapolation of subthreshold reaction amplitudes computed
in chiral perturbation theory, where the constraints set by  electromagnetic-gauge invariance, causality and
unitarity are used to stabilize the extrapolation. Based on the chiral Lagrangian we recover the empirical
s- and p-wave amplitudes up to energies $\sqrt{s}\simeq 1300$ MeV in terms of the
parameters relevant at order $Q^3$.
\end{abstract}
\end{frontmatter}

\tableofcontents
\section{Introduction}
\label{sec:1}

The study of photon- and pion-nucleon interactions has a long history in hadron physics.
In recent years such reactions have been
successfully used as a quantitative challenge of chiral perturbation theory ($\chi $PT), which is a systematic tool
to learn about low-energy QCD dynamics \cite{Bernard:1995dp,Bernard:2007zu,Pascalutsa:2006up}.
The application of $\chi $PT is
limited to the near threshold region. The pion-nucleon phase shifts have been analyzed
in great depth at subleading orders in the chiral expansion \cite{Bernard:1996gq,Fettes:1998ud,Fettes:2000xg}.
Pion photoproduction was studied in \cite{Bernard:1992nc,Bernard:1994gm,Bernard:1996ti,Fearing:2000uy}.
Compton scattering was considered in \cite{Bernard:1995dp,Beane:2004ra}.

Though there are many model computations (see e.g. \cite{Scholten:1996mw,Krehl:1999km,Penner:2002md,Gasparyan:2003fp,Shklyar:2004ba,Matsuyama:2006rp,Drechsel:2007if})
that address such reactions at energies significantly larger than threshold it is an open challenge to further
develop systematic effective field theories that have a larger applicability domain than $\chi $PT and are predictive
nevertheless. The inclusion of the isobar as an explicit degree of freedom in the chiral Lagrangian is investigated in
\cite{Datta:1996jb,Ellis:1997kc,Fettes:2000bb,Lutz:2001yb,Torikoshi:2002bt,Pascalutsa:2002pi,Pascalutsa:2005vq,Pascalutsa:2007yg,Long:2009wq}.
This leads to an extension of the applicability domain of the chiral Lagrangian which is based on power counting rules.
A successful description of scattering data in the isobar region requires the systematic summation of an infinite number of terms.
Such a summation may be motivated by generalized counting rules \cite{Pascalutsa:2002pi} to be applied directly to the S-matrix.
Alternatively, the required summation may be justified by the request that the process is described in accordance with the
unitarity constraint \cite{Lutz:2001yb,Meissner:1999vr,GomezNicola:1999pu,GomezNicola:2000wk,Nicola:2003zi}. The counting
rules are applied to irreducible diagrams only. An infinite number of reducible diagrams being summed by the unitarity request \cite{Lutz:2001yb}. This is analogous to the scheme proposed by Weinberg for the nucleon-nucleon scattering
problem \cite{Weinberg:1990rz,Weinberg:1991um}.

The purpose of this work is to develop a unified description of  photon and pion scattering off the nucleon based on
the chiral Lagrangian. We aim at a description from threshold up to and beyond the isobar region in terms of partial-wave
amplitudes that are consistent with the constraints set by causality and unitarity.  Our analysis is based on the chiral
Lagrangian with pion and nucleon fields truncated at order $Q^3$. We do not consider an explicit isobar field in the chiral
Lagrangian. The physics of the isobar resonance enters our scheme by an infinite summation of higher order counter terms
in the chiral Lagrangian. The particular summation is performed in accordance with unitarity and causality.

Our work is based on a scheme proposed in \cite{Lutz:Kolomeitsev:2010}. We develop a suitable extension to be
applied to pion-nucleon scattering, pion photoproduction and Compton scattering.
The scheme is based on an analytic extrapolation of subthreshold scattering amplitudes that is controlled
by constraints set by electromagnetic-gauge invariance, causality and unitarity.
Unitarized scattering amplitudes are obtained which have left-hand cut structures in accordance with causality.
The latter are solutions of non-linear integral equations that are solved
by $N/D$ techniques. The integral equations are imposed on partial-wave amplitudes that are free of kinematical zeros and
singularities. Such amplitudes are constructed in the Appendix. An essential ingredient of the scheme is the analytic
continuation of the generalized potentials that determine the partial-wave amplitudes via the non-linear integral equation.
We discuss the analytic structure of the generalized potentials in detail and construct suitable conformal mappings in terms of
which the analytic continuation is performed systematically. Contributions from far distant left-hand cut structures are 
represented by power series in the conformal variables. 

The relevant counter terms of the Lagrangian are adjusted to the empirical data available for
photon and pion scattering off the nucleon.
We focus on the s- and p-wave partial-wave amplitudes and
do not consider inelastic channels with two or more pions.
We recover the empirical s- and p-wave pion-nucleon
phase shifts up to about 1300 MeV quantitatively. The pion photoproduction process is analyzed in terms of its multipole
decomposition. Given the significant ambiguities in those multipoles we offer a more direct comparison of our results with
differential cross sections and polarization data. A quantitative reproduction of the data set up to energies of
about $\sqrt{s} \simeq 1300$ MeV is achieved.

\newpage

\section{Analytic extrapolation of subthreshold scattering amplitudes}
\label{sec:analitic_extrapolation}

In this section we perform a systematic analytic continuation of the reactions $\pi N \to \gamma N, \,\pi N $ and
$\gamma N \to \gamma N, \pi N $ based on the chiral Lagrangian \cite{Fettes:1998ud,Bernard:2007zu}.
After a review of the interaction terms  relevant at the accuracy level $Q^3$ we construct the tree-level amplitudes
for the three reactions. This specifies our notations and conventions. Explicit results for the one-loop diagrams that complement
the tree-level expressions to chiral order $Q^3$ are recalled in Appendix A. In section 2.1 a non-linear integral equation is presented
that combines the constraints of unitarity and causality in terms of a generalized potential. Solutions thereof will be constructed in
application of the N/D technique. Possible ambiguities of N/D technique and their relations to bare resonances or ghosts are discussed in
section 2.2. In section 2.3 we derive a faithful approximation of the generalized potential by
means of suitably constructed conformal mappings. Electromagnetic gauge invariance is kept rigourously.

The starting point of our studies is the chiral Lagrangian involving pion, nucleon and photon fields \cite{Fettes:1998ud,Bernard:2007zu}.
For the readers' convenience we collect all terms that contribute to order $Q^3$ for pion-nucleon scattering,
pion photoproduction and Compton scattering. From \cite{Fettes:1998ud,Beane:2004ra} we extract the terms
\allowdisplaybreaks[1]
\begin{eqnarray}
\mathcal{L}_{int}&=&
-\frac{1}{4\,f^2}\,\bar{N}\,\gamma^{\mu}\,\big( \vec{\tau} \cdot \big(\vec{\pi}\times
(\partial_\mu\vec{\pi})\big)\big) \,N +
\frac{g_A}{2\,f} \,\bar{N}\,\gamma_5\,\gamma^{\mu} \,\big( \vec{\tau}\cdot (\partial_{\mu}\vec{\pi} )\big) \,N
\nonumber \\
&-&e\,\Big\{ \big(\vec{\pi}\times(\partial_{\mu}\vec{\pi}) \big)_3
+ \bar{N}\,\gamma_\mu\, \frac{1+\tau_3}{2} \,N
- \frac{g_A}{2\,f} \,\bar{N}\,\gamma_5\,\gamma_{\mu}\,\big(\vec\tau\times\vec{\pi}\big)_3\,N \Big\} \,A^\mu
\nonumber\\
&-&\frac{e}{4\,m_N}\,\bar{N}\,\sigma_{\mu\nu}\,\frac{\kappa_s+\kappa_v\,\tau_3}{2}\,N\,F^{\mu\nu}+
\frac{e^2}{32\pi^2 f}\,\epsilon^{\mu\nu\alpha\beta}\,\pi_3\,F_{\mu\nu}\,F_{\alpha\beta}
\nonumber\\
&-&\frac{2\,c_1}{f^2}\,m_\pi^2\, \bar{N}\,( \vec{\pi}\cdot\vec{\pi})\,N -
\frac{c_2}{2\,f^2\,m_N^2}\,\Big\{\bar{N}\,(
\partial_{\mu}\,\vec{\pi})\cdot (\partial_{\nu}\vec{\pi})\,(\partial^\mu \partial^\nu N )+\rm{h.c.}\Big\}
\nonumber \\
&+& \frac{c_3}{f^2}\,\bar{N} \,(\partial_{\mu}\,\vec{\pi} )
\cdot (\partial^{\mu}\vec{\pi})\,N
-\frac{c_4}{2\,f^2}\,\bar{N}\,\sigma^{\mu\nu}\,\big(\vec{\tau} \cdot \big((\partial_{\mu}\vec{\pi})\times
(\partial_{\nu}\vec{\pi})\big)\big)\,N
\nonumber\\
&-&i\,\frac{d_1+d_2}{f^2\,m_N}\,
\bar{N}\,\big(\vec\tau\cdot \big((\partial_\mu \vec \pi )\times
(\partial_\nu\partial_\mu \vec\pi) \big)\big) \, (\partial^\nu N) + \rm{h.c.}
\nonumber \\
&+&\frac{i\,d_3}{f^2\,m_N^3}\,
\bar{N}\,\big(\vec \tau \cdot \big( (\partial_\mu\vec\pi )\times
(\partial_\nu \partial_\lambda\vec\pi )\big)\big)\,
(\partial^\nu\partial^\mu\partial^\lambda  N)
+\mbox{h.c.}
\nonumber\\
&-&2\,i\,\frac{m_\pi^2\,d_5}{f^2\,m_N}\,\bar{N}\,\big(\vec{\tau} \cdot \big(\vec{\pi}\times
(\partial_\mu\vec{\pi}) \big)\big)\,( \partial ^\mu N) +\rm{h.c.}
\nonumber\\
&-&\frac{i \,e }{f\,m_N }\, \epsilon^{\mu\nu\alpha\beta}\,\bar{N}\,\big( d_8\,
(\partial_\alpha \,\pi_3)  +d_9\,\big(\vec\tau \cdot (\partial_\alpha \vec \pi)\big) \big)\, (\partial_\beta\,N)\, F_{\mu\nu}+\mbox{h.c.}
\nonumber \\
&+&i\,\frac{d_{14}-d_{15}}{2\,f^2\,m_N}\,
\bar{N}\,\sigma^{\mu\nu}\,\big((\partial_\nu\vec\pi )\cdot
(\partial_\mu\partial_\lambda\vec\pi ) \big)\,(\partial^\lambda N)
+\mbox{h.c.}
\nonumber \\
&-&\frac{m_\pi^2\,d_{18}}{f} \,\bar{N}\,\gamma_5\,\gamma^{\mu} \, \big( \vec{\tau}\cdot (\partial_{\mu}\vec{\pi})\big) \, N
\nonumber \\
&+&\frac{e \,(d_{22}-2\,d_{21})}{2\,f}\,\bar{N}\,\gamma_5\,\gamma^\mu\,
\big(\vec\tau\times\partial^\nu \,\vec \pi\big)_3 \,N\, F_{\mu\nu}
\nonumber \\
&+&\frac{e\, d_{20}}{2  \,f\,m_N^2}\,\bar{N}\,\gamma_5\,\gamma^\mu\,
\big(\vec\tau \times (\partial_\lambda \,\vec \pi)\big)_3\,
(\partial^\nu\partial^\lambda N)\, F_{\mu\nu}+\mbox{h.c.} \,,
\label{Lagrangian}
\end{eqnarray}
with the field strength tensor $F^{\mu\nu}=\partial^{\mu} A^\nu-\partial^{\nu}A^\mu$. We use the convention $\epsilon^{0123}=+1$.
Most of the parameters introduced in (\ref{Lagrangian}) are reasonably known from $\chi $PT studies of  the pion-nucleon scattering
and pion-photo\-production processes close to threshold. We will connect to the different parameter sets suggested in the literature
in the result section when discussing our preferred parameter set.

\begin{table}[t]
\rescale
 \setlength{\tabcolsep}{2.4mm} \setlength{\arraycolsep}{2.4mm}
 \renewcommand{\arraystretch}{1.4}
\begin{center}
\begin{tabular}{|c||c|c|c|c|}
\hline
$I$ &$C^{(I)}_{-}$&$C^{(I)}_{+}$&
$C^{(I)}_{s,N}$&$C^{(I)}_{u,N}$ \\
\hline \hline
$\frac{1}{2}$&$2$&$1$&$3$&$-1$  \\
\hline
$\frac{3}{2}$&$-1$&$1$&$0$&$2$ \\
\hline
\end{tabular}
\end{center}
\caption{Isospin coefficients as introduced in (\ref{piN-tree-level}).}
\label{tab:isospin-piN}
\end{table}

In a strict chiral expansion the order $Q^3$ results are composed from a tree-level part and a one-loop part.
Both parts are invariant under electromagnetic gauge transformations separately.

The tree-level pion-nucleon scattering amplitude receives contributions from the Weinberg-Tomozawa term,
the s- and u-channel nucleon  exchange processes and  the $Q^2$ and $Q^3$ counter terms characterized by the parameters $c_1, ..., c_4$ and
$d_1+d_2,d_3,d_5,d_{14}-d_{15},d_{18}$.
We identify the relevant terms
\begin{eqnarray}
T^{(I)}_{\pi N \to \pi N}&=&
\parbox{2cm}{\includegraphics[width=2cm,clip=true]{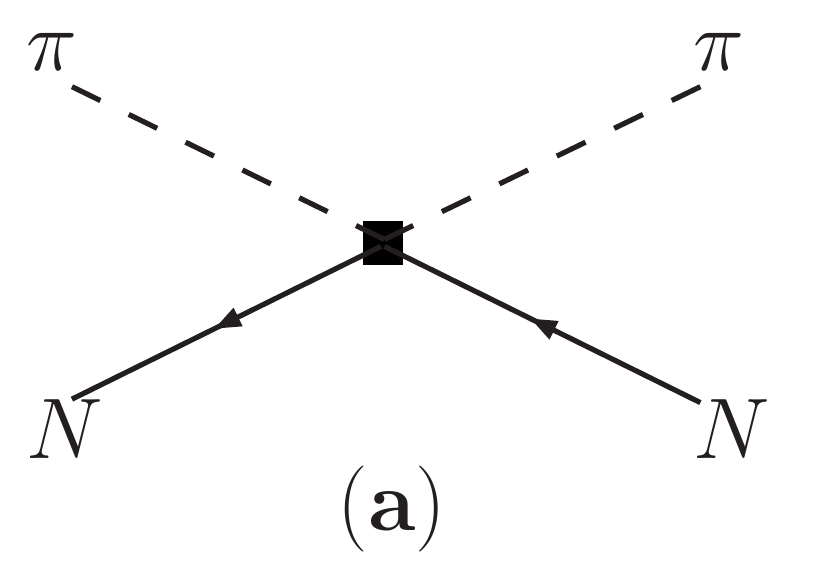}}
+\parbox{2cm}{\includegraphics[width=2cm,clip=true]{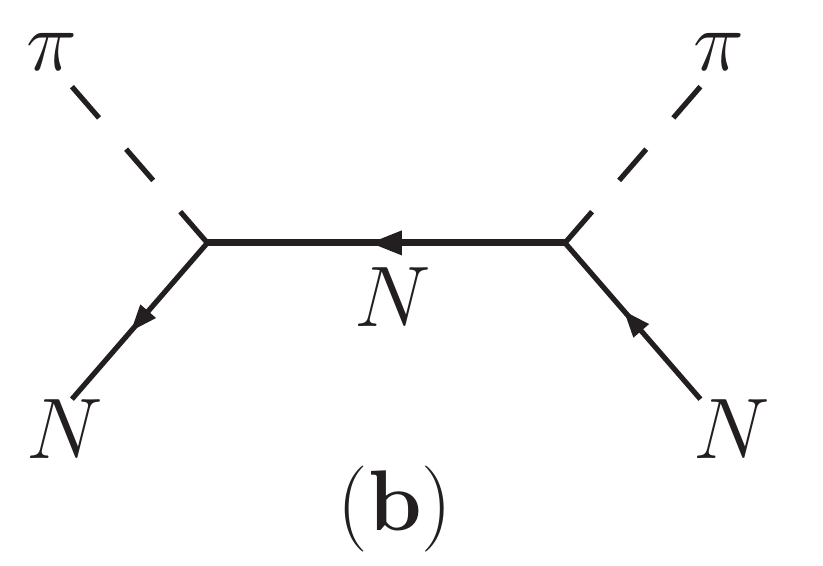}}
+\parbox{2cm}{\includegraphics[width=2cm,clip=true]{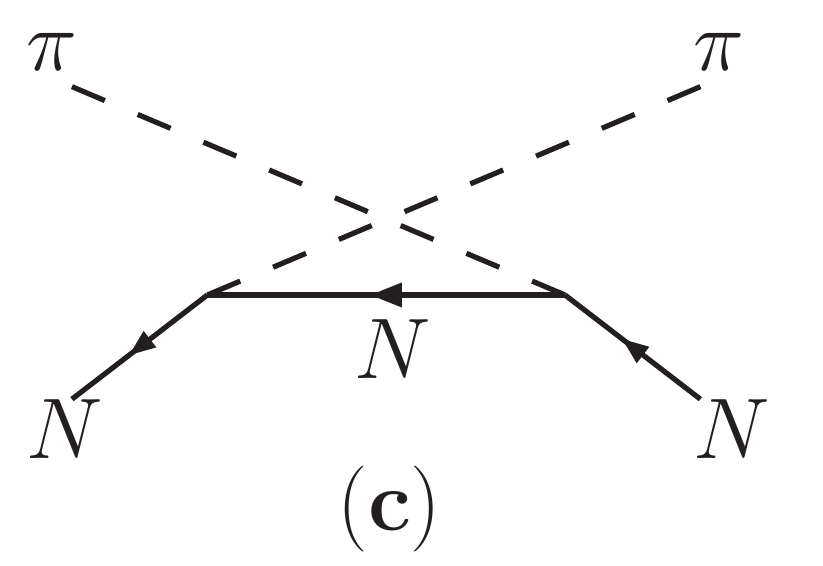}}
\nonumber\\
&=&\Big\{ \frac{1}{4\,f^2}\,(\qslash+\barqslash)+i\,\frac{c_4}{f^2}\,\bar{q}^\mu\,\sigma_{\mu\nu} \,q^\nu \,
+ 2 \,\frac{d_1+d_2}{f^2\,m_N}\,(\bar{q}\cdot q)\,( p\cdot q+ p\cdot \bar q)
\nonumber\\
&& \quad +\,2\, \frac{d_3}{f^2 \,m_N^3}\,(p \cdot q)\,(p\cdot\bar{q})\,(p\cdot q+p\cdot\bar{q})
\nonumber\\
&& \quad +\,4\,\frac{d_5}{f^2\,m_N}\, m_\pi^2 \,( p\cdot q+p\cdot\bar{q})
\Big\}\,C_{-}^{(I)}
\nonumber\\
& +& \Big\{ -4\,\frac{c_1}{f^2}\,m_\pi^2
+2\,\frac{c_2}{f^2\,m_N^2}\,(p\cdot q)(p\cdot\bar{q})
+2\,\frac{c_3}{f^2}\,(\bar{q}\cdot q )
\nonumber\\
&& \quad +\,i\,\frac{d_{14}-d_{15}}{f^2 m_N}\,(p\cdot q+p\cdot\bar{q})
\,\bar{q}^\mu\,\sigma_{\mu\nu} \,q^\nu \,\Big\} \,C_{+}^{(I)}
\nonumber\\
&-&\frac{(g_A-2 \,m_\pi^2 d_{18})^2}{4\,f^2}\,\gamma_5\,\barqslash\, S_N(p+q)\,\gamma_5\,\qslash \,C_{s,N}^{(I)}
\nonumber\\
&-& \frac{(g_A-2 \,m_\pi^2 d_{18})^2}{4\,f^2}\,\gamma_5\,\qslash \,S_N(p-\bar{q})\,\gamma_5\,\barqslash\, C_{u,N}^{(I)} \,,
\label{piN-tree-level}
\end{eqnarray}
with the  nucleon propagators $S_N(p)$, defined as follows
\begin{eqnarray}
S_N(p)=\frac{\pslash+m_N}{p^2-m_N^2+i\,\epsilon}\, .
\label{def-propagtors}
\end{eqnarray}
We use $p_\mu \,(\bar p_\mu)$ and $q_\mu\, (\bar q_\mu)$ for the initial (final) 4-momenta of the baryon and meson.
The pion-nucleon-scattering  amplitudes are projected onto good isospin $I$ where the various isospin
coefficients $C_{..}^{(I)}$ are collected in Tab. ~\ref{tab:isospin-piN}.
The contribution from the one-loop diagrams is recalled in Appendix A.

\begin{table}[t]
\rescale
\begin{center}
\begin{tabular}{|c||c|c|c|c|c|c|c|}
 \hline
$I,N$ &$C^{(I,N)}_{-}$&$C^{(I,N)}_{+}$ &$C^{(I,N)}_{0}$ &$C^{(I,N)}_{s,N}$&$\tilde C^{(I,N)}_{s,N}$&$C^{(I,N)}_{u,N}$&$\tilde C^{(I,N)}_{u,N}$ \\
\hline \hline
$\frac{1}{2}$, $p$&$\frac{2}{\sqrt{3}}$&$\frac{1}{\sqrt{3}}$ & $\sqrt{3}$&$\sqrt{3}$&$\frac{\sqrt{3}}{2}\,(\kappa_s+\kappa_v )$&$\frac{1}{\sqrt{3}}$& $\frac{1}{2\,\sqrt{3}}\,(3\,\kappa_s -\kappa_v )$ \\
\hline
$\frac{1}{2}$, $n$&$-\frac{2}{\sqrt{3}}$&$-\frac{1}{\sqrt{3}}$ & $\sqrt{3}$&$0$&$\frac{\sqrt{3}}{2}\,(\kappa_s-\kappa_v )$&$\frac{2}{\sqrt{3}}$& $\frac{1}{2\,\sqrt{3}}\,(3\,\kappa_s +\kappa_v )$ \\
\hline
$\frac{3}{2}$, $p$&$-\sqrt{\frac{2}{3}}$&$\sqrt{\frac{2}{3}}$ & $0$&$0$&$0$&$\sqrt{\frac{2}{3}}$&$\sqrt{\frac{2}{3}}\,\kappa_v$ \\
\hline
\end{tabular}
\end{center}
\caption{Isospin coefficients as introduced in (\ref{gN-piN-tree-level}).}
\label{tab:isospin-gN}
\end{table}
At tree-level the pion photoproduction amplitude is determined by the Kroll-Rudermann term, the
nucleon s- and u-channel exchange processes  and the t-channel pion exchange.
At chiral order $Q^3$ the counter terms $d_8,d_9,d_{18},2\,d_{21}-d_{22}$ contribute. It holds
\begin{eqnarray}
T_{\gamma p\to\pi N}^{(I,p),\;\mu}&=&\parbox{2cm}{\includegraphics[width=2cm,clip=true]{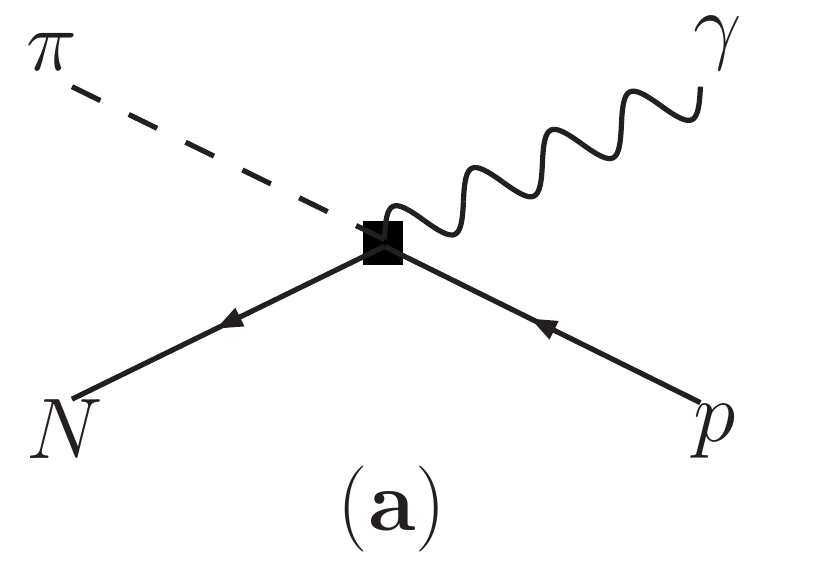}}
+\parbox{2cm}{\includegraphics[width=2cm,clip=true]{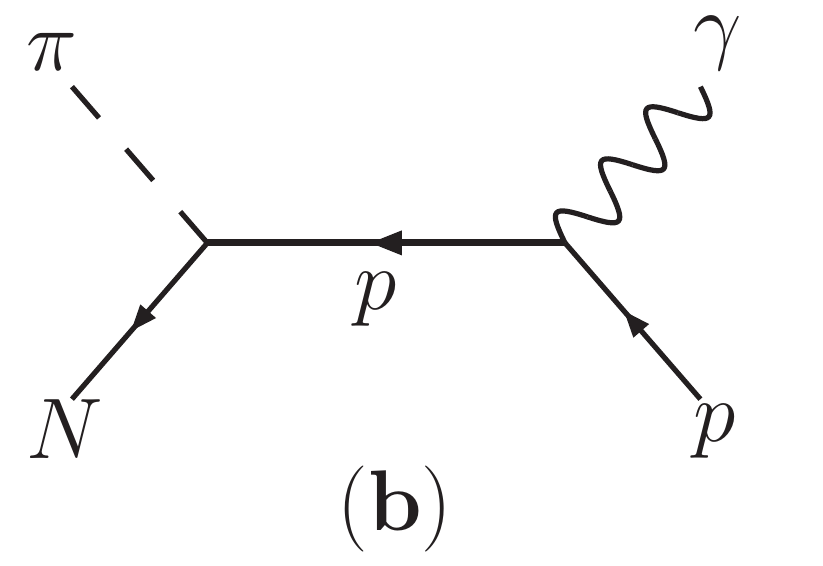}}
+\parbox{2cm}{\includegraphics[width=2cm,clip=true]{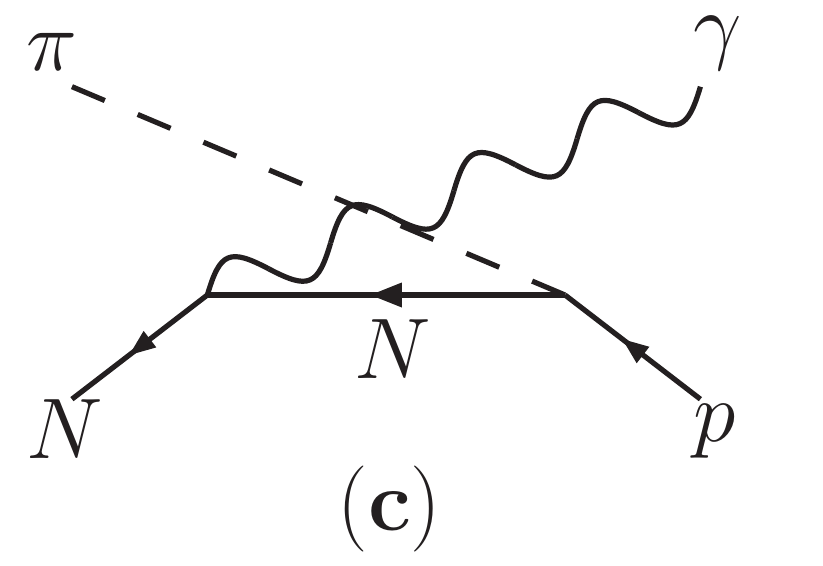}}
+ \parbox{2cm}{\includegraphics[width=2cm,clip=true]{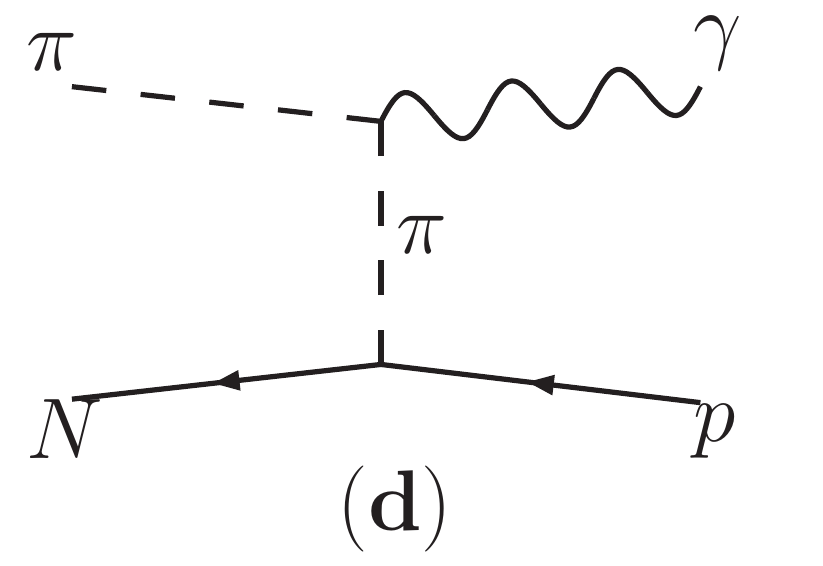}}
\nonumber\\
&=&-i\,\frac{e\,(g_A-2 \,m_\pi^2 d_{18})}{2\,f}\,\gamma_5\,\gamma^\mu\,C^{(I,p)}_{-}
\nonumber\\
&+&i\,\frac{e\,(g_A-2 \,m_\pi^2 d_{18})}{2\,f}\,\gamma_5\,\barqslash \,S_N(p+q)\left(
 \gamma^\mu \,C_{s,N}^{(I,p)}
+i\,\frac{\sigma^{\mu\nu}}{2\,m_N}\,q_\nu \,\tilde C^{(I,p)}_{s,N}\right)
\nonumber\\
&+&i\,\frac{e\,(g_A-2 \,m_\pi^2 d_{18})}{2\,f} \left( \gamma^\mu\,  C_{u,N}^{(I,p)}
+i\,\frac{\sigma^{\mu\nu}}{2\,m_N}\,q_\nu \, \tilde C^{(I,p)}_{u,N}\right)
S_N(p-\bar{q})\,\gamma_5\,\barqslash
\nonumber\\
&+&i\,\frac{e\,(g_A-2 \,m_\pi^2 d_{18})}{2\,f}\,
\frac{ \gamma_5\,\left( \barqslash-\qslash\right) \left(2\,\bar{q}-q\right)^\mu}{ (\bar{q}-q)^2-m_\pi^2+i\,\epsilon}\,C^{(I,p)}_{-}
 \nonumber\\
&+& \frac{4\, e }{f\,m_N }\,
\epsilon^{\mu\nu\alpha\beta}\,q_\nu \,\bar{q}_\alpha \,p_\beta\,\big[d_8\, C^{(I,p)}_{+}+d_9\,C^{(I,p)}_{0}\big]
\nonumber\\
&-&i\,\frac{e\,d_{20}}{2 \,f\,m_N^2}\,
\gamma_5\,\bar{q}_\alpha \,\big[\gamma^\mu\,(p^\alpha \,p^\beta+
\bar{p}^\alpha \,\bar{p}^\beta)-\gamma^\beta\,(p^\alpha \,p^\mu+\bar{p}^\alpha \,\bar{p}^\mu)\big]\, q_\beta\,C^{(I,p)}_{-}
\nonumber\\
&+&i\,\frac{e\,(d_{22}-2\,d_{21})}{2\, f}\,
\gamma_5\,\big[(\bar{q}\cdot q)\,\gamma^\mu-\qslash \,\bar{q}^\mu\big]\, C^{(I,p)}_{-}\,.
\label{gN-piN-tree-level}
\end{eqnarray}
The photoproduction amplitude  depends on the choice of the nucleon $(p,n)$ of the initial state: the amplitude
$T_{\gamma n\to\pi N}^{(I,n),\;\mu}$ is obtained from $T_{\gamma p\to\pi N}^{(I,p),\;\mu}$
by replacing the corresponding isospin coefficients of Tab. \ref{tab:isospin-gN}.

The proton Compton scattering is given
by nucleon s- and u-channel exchange contribution and
pion t-channel exchange.
\begin{eqnarray}
T_{\gamma p\to\gamma p}^{\mu\nu}&=&\parbox{2cm}{\includegraphics[width=2cm,clip=true]{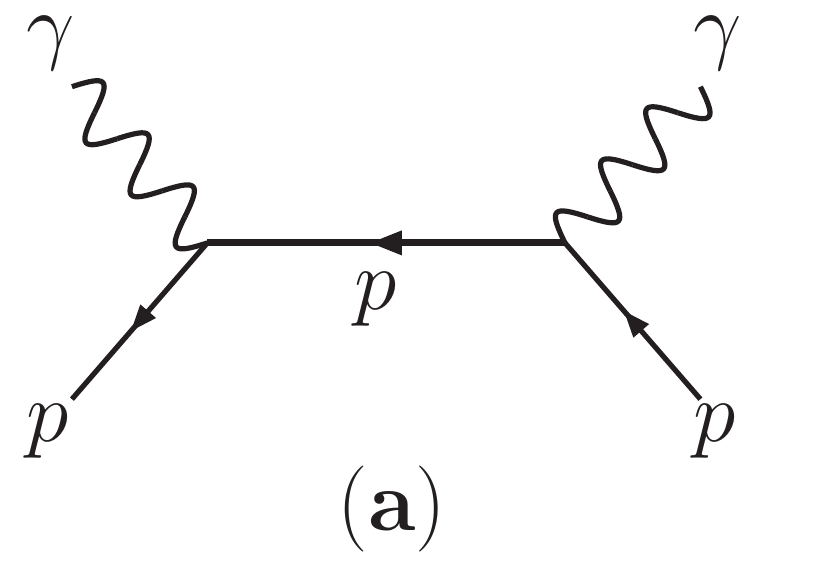}}
+\parbox{2cm}{\includegraphics[width=2cm,clip=true]{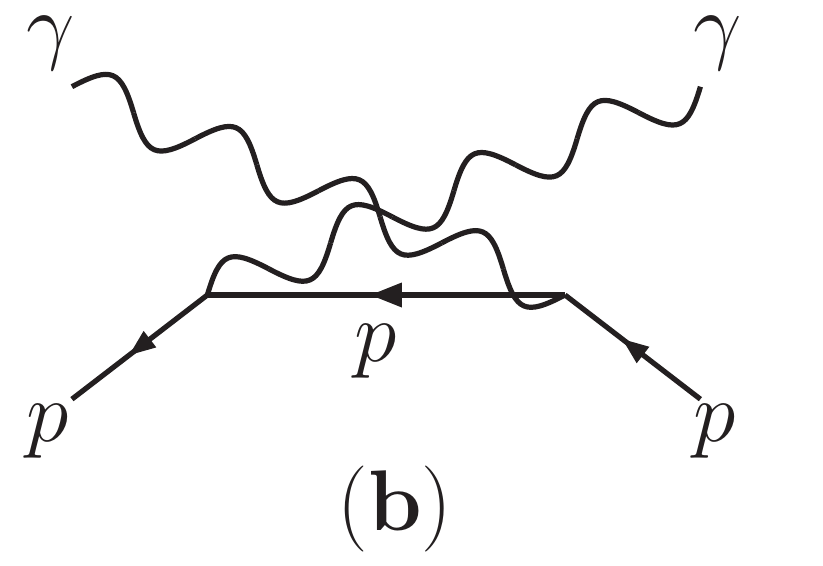}}
+\parbox{2cm}{\includegraphics[width=2cm,clip=true]{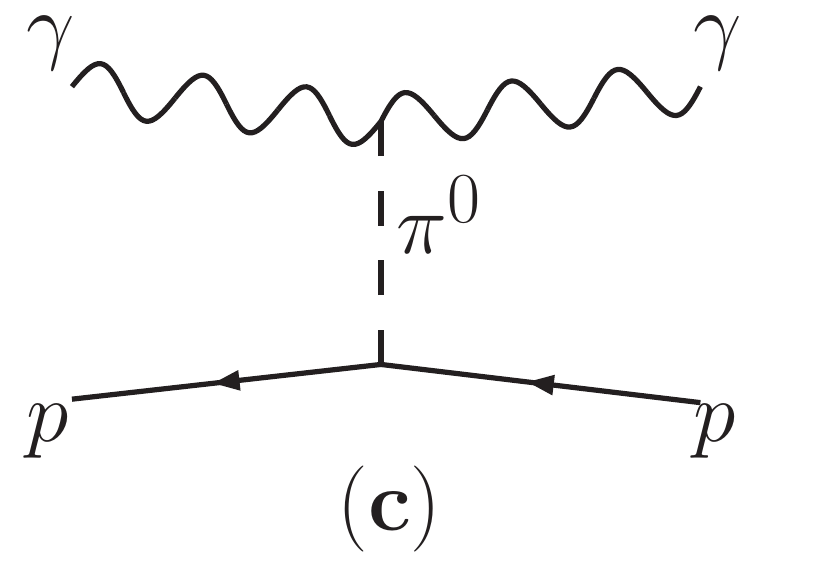}}
\nonumber\\
&=&-e^2\left(\gamma^{\mu}-i\,\frac{\kappa_p}{2\,m_N}\,\sigma^{\mu\alpha}\,\bar{q}_\alpha\right)
S_N(p+q)\left(\gamma^\nu-i\,\frac{\kappa_p}{2\,m_N}\,\sigma^{\beta\nu}\,q_\beta\right)
\nonumber\\
&-&e^2\left(\gamma^\nu-i\,\frac{\kappa_p}{2\,m_N}\,\sigma^{\beta\nu}\,q_\beta\right)
S_N(p-\bar{q})\left(\gamma^{\mu}-i\,\frac{\kappa_p}{2\,m_N}\,\sigma^{\mu\alpha}\,\bar{q}_\alpha\right)
\nonumber\\
&-&i\,\frac{e^2\,(g_A-2 \,m_\pi^2 d_{18})}{8\,\pi^2\, f^2}\,\gamma_5\,(\barqslash-\qslash)\,
\frac{\varepsilon^{\alpha\beta\mu\nu}\,q_\alpha \,\bar{q}_\beta}{(\bar q-q)^2-m_\pi^2+i\,\epsilon}\,,
\label{gp-gp-tree-level}
\end{eqnarray}
with the proton's anomalous magnetic moment
$\kappa_p = \frac{\kappa_s+\kappa_v}{2}$.

\subsection{Constraints from unitarity and causality}

Our approach is based on  partial-wave dispersion relations, for which the unitarity and causality constraints can be
combined in an efficient manner. Using a partial-wave decomposition simplifies calculations because of
angular momentum and parity conservation. For a suitably chosen partial-wave amplitude with angular
momentum $J$, parity $P$ and channel quantum numbers $a,b$ (in our case $\pi N$ and $\gamma N$ channels)
we separate the right-hand cuts from the left-hand cuts
\begin{eqnarray}
 T_{ab}^{(JP)}(\sqrt{s}\,)=U_{ab}^{(JP)}(\sqrt{s}\,)+\int_{\mu_{\rm thr}}^{\infty}\frac{dw}{\pi}\frac{\sqrt{s}-\mu_M}{w-\mu_M}
\frac{\Delta T_{ab}^{(JP)}(w)}{w-\sqrt{s}-i\epsilon}\,,
\label{disrel}
\end{eqnarray}
where the generalized potential, $U_{ab}^{(JP)}(\sqrt{s}\,)$, contains left-hand cuts only, by definition.
It is emphasized that the separation (\ref{disrel}) is gauge invariant. This follows since both contributions
in (\ref{disrel}) are strictly on-shell and characterized by distinct analytic properties \footnote{Gauge invariant approximations for the Bethe-Salpeter
equation are much more difficult to construct \cite{vanAntwerpen:1994vh}. This is because the scattering kernel
is required for off-shell kinematics and therefore is gauge dependent in this case.}.
The amplitude is considered as a function of $\sqrt{s}$ due to the MacDowell relations \cite{MacDowell:1959zza}.
A subtraction is made at $\sqrt{s}=\mu_M = m_N$ for the reasons to be discussed below.
The relation (\ref{disrel}) illustrates that the amplitude  possesses a unitarity cut along the positive real
axis starting from the lowest $s$-channel threshold. In our case the $\gamma N$ intermediate states induces a branch point at
$\sqrt{s} = \mu_{\rm thr} = m_N$, which defines the lowest s-channel unitarity threshold.
The structure of the left-hand cuts in $U_{ab}^{(JP)}(\sqrt{s}\,)$ can
be obtained by assuming the Mandelstam representation \cite{Mandelstam:1958xc,Ball:1961zza}
or examining the structure of Feynman diagrams in perturbation theory \cite{Nakanishi:1962}.
The form of the analyticity domain of  partial-wave amplitudes as implied
by basic principles of quantum field theory is discussed in \cite{Sommer:1970mr}.
Without specifying a particular structure of the left-hand singularities one can regard (\ref{disrel})
as a definition of $U_{ab}^{(JP)}(\sqrt{s}\,)$ \cite{Chen:1972rq,Johnson:1979jy}.
In what follows we will assume for definiteness the Mandelstam representation to hold.

The condition that the  scattering amplitude must be unitary
allows one to calculate the discontinuity along the right hand  cut
\begin{eqnarray}
 \nonumber \Delta T_{ab}^{(JP)}(\sqrt{s})&=&\frac{1}{2\,i}\left(T_{ab}^{(JP)}(\sqrt{s}+i\epsilon)-
T_{ab}^{(JP)}(\sqrt{s}-i\epsilon)\right)\\
&=&\sum_{c,d}\,T_{ac}^{(JP)}(\sqrt{s}+i\epsilon)\,\rho^{(JP)}_{cd}(\sqrt{s}\,)\,T_{db}^{(JP)}(\sqrt{s}-i\epsilon),
\label{discontinuity}
\end{eqnarray}
where, $\rho^{(JP)}_{cd}(\sqrt{s}\,)$, is the phase-space matrix. The sum in  (\ref{discontinuity}) runs over all possible
intermediate states. Combining (\ref{discontinuity}) and (\ref{disrel}) we arrive at the following non-linear integral equation for the
coupled-channel scattering amplitude
\begin{eqnarray}
 T_{ab}(\sqrt{s}\,)=U_{ab}(\sqrt{s}\,)+\sum_{c,d}\, \int_{\mu_{\rm thrs}}^{\infty}\frac{dw}{\pi}\,\frac{\sqrt{s}-\mu_M}{w-\mu_M}\,
\frac{T_{ac}(w)\,\rho_{cd}(w)\,T_{db}^*(w)}{w-\sqrt{s}-i\epsilon},
\label{def-non-linear}
\end{eqnarray}
where we suppressed reference to angular momentum and parity. Given any gauge-invariant approximation to the generalized potential
$U_{ab}(\sqrt{s}\,)$, the non-linear integral equation (\ref{def-non-linear}) generates a gauge-invariant coupled-channel scattering amplitude.

It is convenient to use partial-wave amplitudes free of kinematic constraints, such as zeros and singularities
at threshold and pseudothreshold. In Appendix~\ref{kinem_sing} partial-wave amplitudes are established, that
are kinematically unconstrained. The partial-wave amplitudes are scaled by powers of $\sqrt{s}$ such that the phase-space factors $\rho_{ab}(\sqrt{s}\,)$ approach
finite values for asymptotically large energies. This implies that the partial-wave scattering amplitudes are bounded asymptotically
in all partial waves. The generalized potentials are bounded modulo some logarithmic structures. All together we derive
\begin{eqnarray}
&&\rho_{\pm}^{\pi N,J}(\sqrt{s}\,)=  \frac{p_{\rm cm }\,(E\pm m_N)}{8\pi s}\left(\frac{p_{\rm cm}^2}{s}\right)^{J-1/2} \,,
\nonumber\\
&&  \rho^{\gamma N,J}_{\pm}(\sqrt{s}\,) =\frac{p_{\rm cm}}{8\pi\,\sqrt{s}}
\left(\frac{p_{\rm cm}^2}{s}\right)^{J-\frac{1}{2}}
\left(\begin{matrix}
\frac{p_{\rm cm}^2}{s}&\sqrt{\frac{J-\frac{1}{2}}{J+\frac{3}{2}}}
\frac{m_N p_{\rm cm}}{s}\\
\sqrt{\frac{J-\frac{1}{2}}{J+\frac{3}{2}}}
\frac{m_N p_{\rm cm}}{s}&
1+\frac{J-\frac{1}{2}}{J+\frac{3}{2}}\frac{m_N^2}{s}        \end{matrix}\right)\,,
\label{phase-space-text}
\end{eqnarray}
with the total angular momentum $J$ and parity $P= \pm $ for $J-\frac{1}{2}$ odd and $P= \mp $ for $J-\frac{1}{2}$ even.
In (\ref{phase-space-text}) we consider  $p_{\rm cm}$ and $E$, the relative momentum and the nucleon energy in the center-of-mass system,
as a function of Mandelstam's variable $s$.

Given an approximate generalized potential we will solve for the
partial-wave scattering amplitudes (\ref{def-non-linear}) with the phase-space distributions (\ref{phase-space-text}).
In the present rather exploratory study we take into account only the $\pi N$ intermediate states in (\ref{def-non-linear}).
The higher mass intermediate states are then effectively included in the potential $U_{ab}^{(JP)}(\sqrt{s})$. We also neglect
the $\gamma N$ intermediate state, since the effects thereof are suppressed by the square of the electromagnetic charge.

The non-linear integral equation (\ref{def-non-linear}) can be solved by means of $N/D$ techniques \cite{Chew:1960iv}.
This problem was studied extensively in the literature \cite{PhysRev.130.478,1971AnPhy..65..141R,Collins:1968,Stingl:1974im,Johnson:1979jy}.
The amplitude is represented as
\begin{eqnarray}
T_{ab}(\sqrt{s}\,)=\sum_c\,D^{-1}_{ac}(\sqrt{s}\,)\,N_{cb}(\sqrt{s}\,)\,,
\label{def-NoverD}
\end{eqnarray}
where $D_{ab}(\sqrt{s}\,)$ has no singularities  but the right-hand $s$-channel unitarity cuts.  The coupled-channel
unitarity condition is a consequence of the ansatz
\begin{eqnarray}
D_{ab}(\sqrt{s}\,)=\delta_{ab}-
\sum_{c}\int_{\mu_{ \rm thr}}^\infty \frac{dw}{\pi}\,\frac{\sqrt{s}-\mu_M}{w-\mu_M}\frac{N_{ac}(w)\,\rho_{cb}(w)}{w-\sqrt{s}}\,.
\label{def-D}
\end{eqnarray}
In contrast the branch points of $N_{ab}(\sqrt{s}\,)$ correspond to those of $U_{ab}(\sqrt{s}\,)$.
A perturbative evaluation of the functions $N_{ab}(\sqrt{s}\,)$ is futile since it is at odds with the causality constraint,
i.e. any finite truncation of the object $N_{ab}(\sqrt{s}\,)$ will violate the representation (\ref{disrel}). A summation of
an infinite set of terms is required to restore causality. This is readily achieved by considering the linear integral equation
\begin{eqnarray}
&&N_{ab}(\sqrt{s}\,)=U_{ab}(\sqrt{s}\,)
\nonumber\\
&& \qquad \qquad +\sum_{c,d} \int_{\mu_{\rm thr}}^\infty \frac{dw}{\pi}\,
\frac{\sqrt{s}-\mu_M}{w-\mu_M}\,\frac{N_{ac}(w)\,\rho_{cd}(w)\,[U_{db}(w)-U_{db}(\sqrt{s}\,)]}{w-\sqrt{s}}\,.
\label{Nequation0}
\end{eqnarray}
There is neither a guarantee that a solution of (\ref{def-non-linear}) exists, nor that a possible
solution is unique \cite{Johnson:1979jy,Castillejo:1955ed}. The question of existence and uniqueness is particularly cumbersome in
coupled-channel systems. The condition that the scattering amplitude may
be evaluated in powers of the generalized potential, at least close to the matching scale $\mu_M = m_N$ singles out the
unique solution we are interested in.  If the non-linear integral equation (\ref{def-non-linear}) admits such a solution
it is generated by (\ref{Nequation0}).

The linear equation (\ref{Nequation0}) leads to a solution always, which is unique for a reasonably behaved
generalized potential. However, the solution does not necessarily define a solution of the non-linear equation
(\ref{def-non-linear}).  The problem emerges when the function $\det D(\sqrt{s}\,)$ turns to zero at some energy point on
the first Rieman sheet. Then the ratio $N/D$ cannot be a solution to (\ref{def-non-linear}) since it contains a pole not present
in the original equation.
In order to find a solution to the non-linear equation the ansatz (\ref{def-D}) may be generalized to allow for Castillejo-Dalitz-Dyson (CDD) pole structures  \cite{Castillejo:1955ed,Johnson:1979jy}. Physically this points to the existence of a genuine bare resonance or to the effect of coupled channels not included explicitly.

Our strategy in the situation where the  function $\det D(\sqrt{s}\,)$ has an unphysical zero, is
to include one CDD pole in (\ref{identify-R}). It turns out that in all cases considered in this work, we are able to
construct such a solution at least in the energy region we are interested in. If the solution found does not satisfy
(\ref{def-non-linear}) at higher energies, one can regard it as a redefinition of the high-energy part of the potential
through equation (\ref{def-non-linear}).

It remains to specify the structure of the CDD pole terms. We consider the more general ansatz \cite{Johnson:1979jy}
\begin{eqnarray}
&& D_{ab}(\sqrt{s}\,)=\delta_{ab}-\frac{\sqrt{s}-\mu_M}{\sqrt{s}-M_{CDD}}\,R^{(D)}_{ab}
\nonumber\\
&& \qquad \qquad -\,\sum_c \int_{\mu_{\rm thr}}^\infty \frac{dw}{\pi}\,\frac{\sqrt{s}-\mu_M}{w-\mu_M}\frac{N_{ac}(w)\,\rho_{cb}(w)}{w-\sqrt{s}}\,,
\label{D-CDD-ansatz}
\end{eqnarray}
with a CDD pole mass parameter $M_{CDD}$ and some coupling matrix $R^{(D)}$. Though the combination (\ref{def-NoverD}) and
(\ref{D-CDD-ansatz}) satisfies the unitarity constraint (\ref{discontinuity}) it does not lead to a solution of the
non-linear equation (\ref{def-non-linear}) unless the linear equation (\ref{Nequation0}) is adjusted properly.

The problem arises how to link the CDD pole parameters with the parameters of the chiral Lagrangian. In order to establish such a link
we consider in a first step the case where the generalized potential has a pole below threshold. The assumption the pole
to sit below threshold is crucial here. This holds in the $P_{11}$ channel, where the generalized potential has a pole at the
physical nucleon mass.  Note that this is not the case in the $P_{33}$ channel. Assuming at first the absence of the Roper resonance
we demonstrate two equivalent ways of solving (\ref{def-non-linear}).  Either take the potential in the presence of the s-channel
pole and solve the original N/D equation (\ref{Nequation0}), or use an effective, pole-subtracted potential, and solve a
modified N/D equation. An appropriate generalization of (\ref{Nequation0}) is derived in the following. We consider the
decomposition
\begin{eqnarray}
&& U_{ab}(\sqrt{s}\,) = U^{\rm eff}_{ab}(\sqrt{s}\,) - \frac{g_a\,m\,g_b}{\sqrt{s}-m}\,\frac{\sqrt{s}-\mu_M}{m-\mu_M}\,,
\label{def-Ueff}
\end{eqnarray}
where the generalized potential has an s-channel pole term at $\sqrt{s}=m \leq \mu_{\rm thr}$. The strength of the pole term is
characterized by the coupling constants $g_a$.

Given the ansatz (\ref{def-NoverD}) together with (\ref{D-CDD-ansatz}) the original result based on the equations
(\ref{def-NoverD}, \ref{def-D}, \ref{Nequation0}) is recovered unambiguously by the following set of equations
\begin{eqnarray}
&& N_{ab}(\sqrt{s}\,)=U^{\rm eff}_{ab}(\sqrt{s}\,)
-\frac{\sqrt{s}-\mu_M}{\sqrt{s}-M_{CDD}}\,\Big[R^{(B)}_{ab}+\sum_c R^{(D)}_{ac}\,U^{\rm eff}_{cb} (\sqrt{s}\,)\Big]
\nonumber\\
&&\qquad  \;\,+\,\sum_{c,d}\int_{\mu_{\rm thr}}^\infty \frac{dw}{\pi}\,\frac{\sqrt{s}-\mu_M}{w-\mu_M}\,
\frac{N_{ac}(w)\,\rho_{cd}(w)\,[U^{\rm eff}_{db}(w)-U^{\rm eff}_{db}(\sqrt{s}\,)]}{w-\sqrt{s}}\,,
\label{CDD-ansatz}
\end{eqnarray}
with
\begin{eqnarray}
&& R^{(D)}_{ab}=\frac{m-M_{CDD}}{m-\mu_M}\left(\delta_{ab}-\sum_c\,\int_{\mu_{\rm thr}}^\infty \frac{dw}{\pi}\,\frac{m-\mu_M}{w-\mu_M}\,\frac{N_{ac}(w)\,\rho_{cb}(w)}{w-m}\right)\,,
\nonumber\\
&& R^{(B)}_{ab}=-\frac{\mu_M-M_{CDD}}{(\mu_M-m)^2}\,g_a\,m\,g_b
\nonumber\\
&&\qquad \,-\,\sum_{c,d}\,\int_{w_{thr}}^\infty \frac{dw}{\pi}\,\frac{w-M_{CDD}}{w-\mu_M}\,
\frac{N_{ac}(w)\,\rho_{cd}(w)}{(w-m)^2}\,g_d\,m\,g_b\,
\nonumber\\
&&\qquad \,+\,(m-M_{CDD})\,\sum_{c,d}\,\int_{\mu_{ \rm thr}}^\infty \frac{dw}{\pi}\,
\frac{N_{ac}(w)\,\rho_{cd}(w)\,U^{\rm eff}_{db}(w)}{(w-\mu_M)\,(w-m)}\,.
\label{identify-R}
\end{eqnarray}
The merit of the formal result (\ref{CDD-ansatz}, \ref{identify-R}) lies in its specification of the CDD pole parameters,
$R^{(D)}$ and $R^{(B)}$,  in terms of the parameters, $g_a, m$, characterizing the subthreshold pole term in the
generalized potential (\ref{def-Ueff}).  The CDD pole mass parameter, $M_{CDD}$, is irrelevant. By construction, the scattering
amplitude, which results from (\ref{def-NoverD}, \ref{D-CDD-ansatz}-\ref{identify-R}), does
not depend on the choice of $M_{CDD}$.

We conclude from the rewrite (\ref{CDD-ansatz},\ref{identify-R}) that introducing a CDD pole in the N/D
ansatz may be viewed as a modification of the generalized potential. Subtract from the original potential a pole term
according to (\ref{def-Ueff}) and then solve the system (\ref{D-CDD-ansatz},\ref{CDD-ansatz},\ref{identify-R}).

In a second step we consider the case of a resonance characterized by a pole on the second Rieman sheet. In this case the generalized
potential does not have a pole. If one assumed a pole at energies above threshold a contradiction with (\ref{def-non-linear}) arises.
Though the decomposition (\ref{def-Ueff}) does not exist we can use (\ref{identify-R}) to relate the CDD pole parameters with the
resonance parameters. The modified N/D equations (\ref{CDD-ansatz},\ref{identify-R}) are well defined for $m$ larger than threshold.
The resulting scattering  amplitude will still be independent on $M_{CDD}$ if the replacement $m \to m-i\,\epsilon$ is used
in (\ref{identify-R}) and if the real part of the integrals in (\ref{identify-R}) is taken. Still the problem remains how to link the
resonance parameters with the parameters of the chiral Lagrangian, in particular in our case where the Lagrangian does not involve any
resonance fields. To do so one may imagine incorporating a resonance field to the Lagrangian in an intermediate step.
If the resonance mass is chosen below threshold the identification of $U_{\rm eff}$ is defined unambiguously via (\ref{def-Ueff}). Next we integrate out the resonance field  by performing a chiral expansion of the resonance pole term as to absorb its effect into local counter terms. Finally the case of interest with the resonance mass chosen above threshold
follows by an analytic continuation in the resonance mass. That implies that the pole term in (\ref{def-Ueff}) must be expanded to the chiral order considered. This leads to an effective potential of the form
\begin{eqnarray}
U^{\rm eff}_{ab}(\sqrt{s}\,)=U_{ab}(\sqrt{s}\,)
- \frac{g_a\,m\,g_b}{m-\mu_M} \,\sum^n_{k=1}\left(\frac{\sqrt{s}-\mu_M}{m-\mu_M} \right)^k\,,
\label{matching-CDD}
\end{eqnarray}
as its proper generalization to the presence of a CDD pole. This
procedure guarantees that the presence of a CDD pole does not
renormalize the local counter terms to the given chiral order.
Though our chiral Lagrangian does not involve resonance fields the
parameters $g_a$ and $m$ have the interpretation of a resonance
coupling and mass parameter. In our scheme they arise as a
correlation of an infinite set of higher order counter terms, which
we sum in response of the causality request.

\subsection{Generalized potential, crossing symmetry and conformal mappings}
\label{sec:conformal_mapping}

At leading order one may try to identify the generalized potential, $U_{ab}(\sqrt{s}\,)$, with a partial-wave projected
tree-level amplitude.  After all the tree-level expressions do not show any right-hand unitarity
cuts. However, this is an ill-defined strategy since it would lead to an unbounded
generalized potential for which (\ref{def-non-linear}) does not allow any solution.
A remedy would be to restrict the integral in (\ref{def-non-linear})  to energies $\sqrt{s} < \Lambda_s$ with
a suitably chosen cutoff characterizing the size of the effective Hilbert space. However, even in this case
the generalized potential needs to be known reliably for energies $\sqrt{s}< \Lambda_s$. A priori this restricts the
choice of $\Lambda_s$ to energies where strict $\chi$PT converges, i.e. to energies where the summation
implied by (\ref{def-non-linear}) is irrelevant.

The key observation we exploit in this work is the fact that the solution of the nonlinear integral equation (\ref{def-non-linear})
requires the knowledge of the generalized potential for $\sqrt{s} > \mu_{\rm thr}$ only. Conformal mapping techniques may be used to
approximate the generalized potential in that domain efficiently based on the knowledge
of the generalized potential in a small subthreshold region around $\mu_E$ only, where it may be computed reliably in $\chi $PT.
To be specific we identify
\begin{eqnarray}
\mu_E = \left\{
\begin{array}{ll}
m_N+m_\pi \qquad &{\rm for } \qquad \pi N \to \pi N \\
m_N+ \frac{1}{2}\,m_\pi&{\rm for }\qquad \gamma N\to \pi N \\
m_N  &{\rm for }\qquad \gamma N\to \gamma N
\end{array}
\right. \,.
\label{def-muE}
\end{eqnarray}
A function $f(\sqrt{s}\,)$, analytic in a domain $\Omega$,
can be reconstructed unambiguously in terms of its derivatives at a point $\mu_E \in \Omega$.
The reconstruction is achieved by a suitable analytic and invertible function
$\xi(\sqrt{s}\,)$ with $\xi(\mu_E)=0$, which maps $\Omega$ onto the unit circle  in the complex
$\xi$-plane with $|\xi|<1$. Given such a conformal mapping the Taylor expanded function
\begin{eqnarray}
h(z) = f (\xi^{-1}(z))=  \sum \limits_{k=0}^\infty \,\frac{h_k}{k!}\,z^k\,,
\end{eqnarray}
converges inside the unit circle if expanded around $z=0$.
Thus the series
\begin{eqnarray}
f(\sqrt{s}\,)=\sum\limits_{k=0}^\infty \,\frac{h_k}{k!}\,\big[\xi(\sqrt{s}\,)\big]^k\,,
\label{def-conformal-mapping}
\end{eqnarray}
recovers  the original function $f(\sqrt{s}\,)$ in its analyticity domain. In contrast to a Taylor expansion of the function
$f(\sqrt{s}\,)$ around $\sqrt{s}=\mu_E$ the convergence is no longer limited by the circle touching the nearest branch point.
For an explicit example and illustration see e.g. \cite{Frazer:1961zz}.

Given a suitable conformal mapping, $\xi(\sqrt{s}\,)$,
we seek to establish a representation of the generalized potential of the form
\begin{eqnarray}
&& U(\sqrt{s}\,)=U_{\rm inside}(\sqrt{s}\,)+U_{\rm outside}(\sqrt{s}\,)\,,
\label{expansion} \\
&& U_{\rm outside}(\sqrt{s}\,)=\sum\limits_{k=0}^\infty {U_k}\,\big[\xi(\sqrt{s}\,)\big]^k\,, \qquad
U_k=\frac{d^k U_{\rm outside}(\xi^{-1}(\xi))}{k!\,d\xi^k}\Big|_{\xi=0}\,,
\nonumber
\end{eqnarray}
where we allow for an explicit treatment of cut structures that are inside the domain $\Omega$.
The splitting of the potential into the parts having only poles and cuts
inside and outside $\Omega$ is defined up to a polynomial in $\sqrt{s}$. Therefore $U_{\rm inside}(\sqrt{s}\,)$ can be chosen
to approach zero asymptotically. We insist that the series for $U_{\rm outside}(\sqrt{s}\,)$ is bounded asymptotically, when
truncated at any finite order. Then the expanded potential in (\ref{expansion}) is regular enough for the integral
equation (\ref{def-non-linear}) to be well defined and amenable to a solution via (\ref{Nequation0}).

\begin{figure}[t]
\begin{center}
\parbox[c]{9cm}{\includegraphics*[width=8cm,height=8cm]{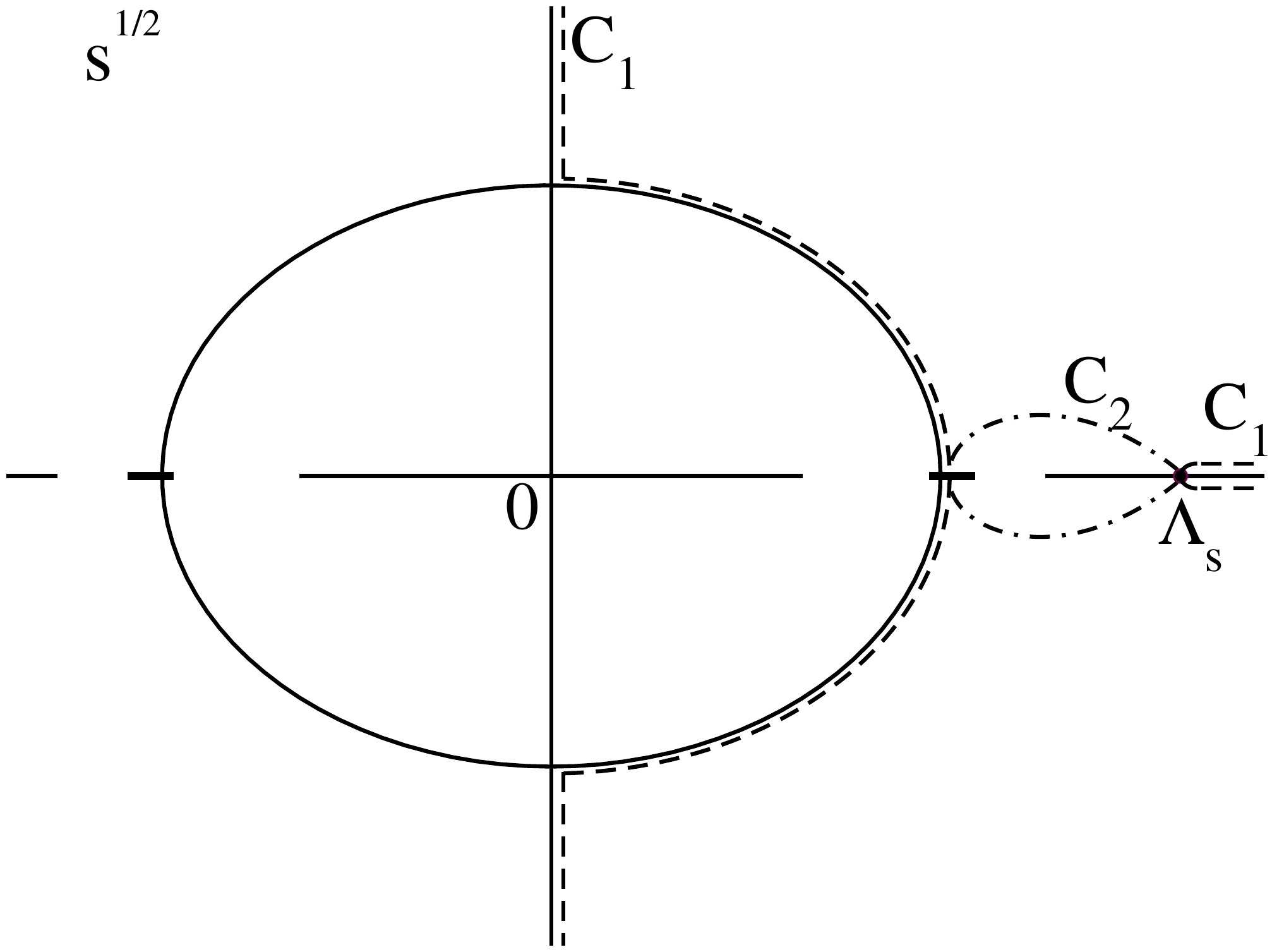}}
\parbox[c]{4cm}{ \caption{Singularity structure of the $\pi N \to \pi N$ amplitude (solid lines),
and domains of convergence (dashed lines) characterizing the expansion \eqref{expansion}.}\label{fig:piN}}
\end{center}
\end{figure}

In order to construct suitable conformal mappings it is necessary to
discuss in some detail the analytic structure of the considered amplitudes.
The singularities of the $\pi N$ scattering amplitude are given by
the solid lines shown in Fig.~\ref{fig:piN}  \cite{MacDowell:1959zza,PhysRev.126.1596}.
Apart from the s-channel unitarity cut extending from threshold, $\sqrt{s}=m_N+m_\pi$, to infinity,
there is the $u$-channel cut defined by the intervals
\begin{eqnarray}
&& 0 < \sqrt{s} < m_N-m_\pi \,,  \qquad
 m_N - \frac{m_\pi^2}{m_N} < \sqrt{s} < \sqrt{m_N^2+2\,m_\pi^2} \,,
\label{def-LambdapiN}
\end{eqnarray}
the whole imaginary axis  and the $t$-channel cut containing
the circle $|s|=m_N^2-m_\pi^2$, and also the nucleon
pole at $\sqrt{s}=\pm  m_N$. Note that each cut has its mirror partner ($\sqrt{s}\to-\sqrt{s}$)
because they originate from a Mandelstam representation written in terms of $s,t,u$.

In contrast to the $\pi N$ scattering amplitude its generalized potential is an analytic function of $\sqrt{s}$ for energies
larger than the s-channel threshold. Thus it may be Taylor expanded around the point $ \sqrt{s}=\mu_E $.
However, the convergence radius would be limited by the location of the nearest branch point. The radius of convergence may be increased 
if some cut structures can be evaluated explicitly, such that the nearest branch point of the residual potential is more distant. For 
instance one may consider the u-channel cut implied by the nucleon-exchange process explicitly. As a consequence the residual potential 
has its nearest u-channel branch point at $\sqrt{s} = m_N-m_\pi$. Similarly the t-channel cuts close to the real line may be evaluated 
in $\chi $PT approximatively, it would start to build up at order $Q^3$. However, there is a natural limit to this construction: the 
u-channel and t-channel cuts turn non-perturbative at some energy necessarily.

Since we do not know a priori where the u-channel cuts turn non-perturbative we insist on that the domain $\Omega$, which we 
seek to specify, excludes the u-channel unitarity branch cuts, i.e.
\begin{eqnarray}
(-m_N+m_\pi, m_N-m_\pi ) \notsubset \Omega \qquad \& \qquad (-i\,\infty, +i\,\infty) \notsubset \Omega \,.
\label{def-u-channel-Omega}
\end{eqnarray}
Incidentally such a construction lives in harmony with the crossing symmetry of the scattering amplitude. Given the assumption 
the generalized potential cannot be evaluated by means of the representation (\ref{expansion}) for $\sqrt{s} < m_N-m_\pi$.  The 
decomposition (\ref{expansion}) is faithful for energies $\sqrt{s} \in \Omega$ only. Nevertheless, the full scattering amplitude 
can be reconstructed  at $\sqrt{s}\leq m_N-m_\pi$ from the knowledge of the generalized potential at
$\sqrt{s}> m_N+m_\pi$ by a crossing transformation of the solution to (\ref{def-non-linear}). Such a construction is  
consistent with crossing symmetry if the solution to (\ref{def-non-linear}) and its crossing transformed form coincide in 
the region $m_N-m_\pi < \sqrt{s} < m_N+m_\pi$.
This is the case approximatively, if the matching scale $\mu_M$ in  (\ref{def-non-linear}) is constructed properly.
For $m_N-m_\pi < \mu_M < m_N+m_\pi$ the scattering amplitude remains perturbative in the matching interval, at least sufficiently 
close to $\sqrt{s} \sim \mu_M$. With our choice
\begin{eqnarray}
\mu_M = m_N \,,
\label{def-muM}
\end{eqnarray}
this is clearly the case (see also \cite{Lutz:2001yb}).

Before providing a first example for a suitable conformal mapping there is yet a further issue to be discussed. Since we neglect
intermediate states with mass larger than $m_N + 2\,m_\pi$ one may argue to incorporate a cutoff into (\ref{def-non-linear}) at
$\sqrt{s} \leq \Lambda_s \simeq m_N +2\,m_\pi $. This would induce a branch point of the generalized potential at
$\sqrt{s} = \Lambda_s$. If we insist on the condition
\begin{eqnarray}
(\Lambda_s, \infty ) \notsubset \Omega\,,
\label{def-Lambdas-Omega}
\end{eqnarray}
the effect of higher mass states is part of $U_{\rm outside}(\sqrt{s}\,)$ in (\ref{expansion}). Thus their effect
renormalize the counter terms of the effective field theory, as they should.  Modulo subtle effects from the
chiral constraints it is impossible to discriminate the effect of a local counter term from the contribution of higher intermediate mass states
in the generalized potential.

Though it is possible to use a sharp cutoff $\Lambda_s$ in (\ref{def-non-linear}) and construct conformal mappings subject
to the conditions (\ref{def-u-channel-Omega}, \ref{def-Lambdas-Omega}), it is advantageous to use the limit $\Lambda_{s} \to \infty$
in  (\ref{def-non-linear}). This avoids an artificial behavior of the scattering amplitude close to the cutoff.
Rather than modifying the integral equation it suffices to adjust the generalized potential, $U(\sqrt{s}\,)$, such that
the influence from the region $\sqrt{s}> \Lambda_s$ is irrelevant. This is readily achieved by replacing $\xi(\sqrt{s}\,) \to \xi_\Lambda(\sqrt{s}\,)$
in (\ref{expansion}) with
\begin{eqnarray}
\xi_\Lambda(\sqrt{s}\,)=\left\{
\begin{array}{ll}
\xi(\sqrt{s}\,) \qquad & {\rm for} \qquad \sqrt{s}<\Lambda_s \\
\xi(\Lambda_s) \qquad & {\rm for} \qquad \sqrt{s}\geq\Lambda_s \ .
\end{array} \right. \,, \qquad \xi'(\Lambda_s) =0 \,.
\label{xiLambda}
\end{eqnarray}
where the condition $\xi'(\Lambda_s) =0$ guarantees a smooth behavior of $\xi_\Lambda(\sqrt{s}\,)$ at the cutoff scale.
Though it is not immediate from the nonlinear equation (\ref{def-non-linear}), our claim follows from (\ref{Nequation0}).
For a generalized potential that is constant for $\sqrt{s} > \Lambda_s$, the contribution of the integral $dw $ in (\ref{Nequation0})
from the region $w > \Lambda_s$ vanishes identically for external energies $\sqrt{s}> \Lambda_s$.

Consider the  dashed line $C_1$ in Fig. ~\ref{fig:piN} which encloses
a domain $\Omega = \Omega_1$ and meets our requirements (\ref{def-u-channel-Omega}, \ref{def-Lambdas-Omega}).
A conformal mapping of the domain $\Omega_1$ onto the unit circle  is readily constructed with
\begin{eqnarray}
&&\xi(\sqrt{s}\,)=\frac{\phi(\sqrt{s}\,)-\phi(\mu_E)}{\phi(\sqrt{s}\,)+\phi(\mu_E)}\,,
\qquad
\nonumber\\
&& \phi(\sqrt{s}\,)=\frac{\sqrt{\left(s\,\Lambda_s^2-(m_N^2-m_\pi^2)^2\right)
\left(\Lambda_s^2-s\right)}}{s-m_N^2+m_\pi^2}.
\label{mapping1}
\end{eqnarray}
Note that the construction of (\ref{expansion}) does not require an explicit representation of the inverse function
$\xi^{-1}(z)$. The inverse of (\ref{mapping1}) is not particularly illuminating and therefore not given here. The
coefficients $U_k$ may be derived in application of the chain rule with
\begin{eqnarray}
\frac{d}{d z}\,\xi^{-1}(z ) =   \left(\frac{d }{ d\sqrt{s}}\,\xi(\sqrt{s}\,) \right)^{-1} \,.
\label{chain-rule}
\end{eqnarray}
It is not always convenient to map the unit circle onto the largest
possible analyticity domain of the generalized potential, as we succeeded to do by means of (\ref{mapping1})
in the case of  elastic $\pi N$ scattering. For other reactions finding such a transformation
can be quite complicated.

\begin{figure}[t]
\begin{center}
\parbox[c]{9.5cm}{\includegraphics*[width=8cm,height=8cm]{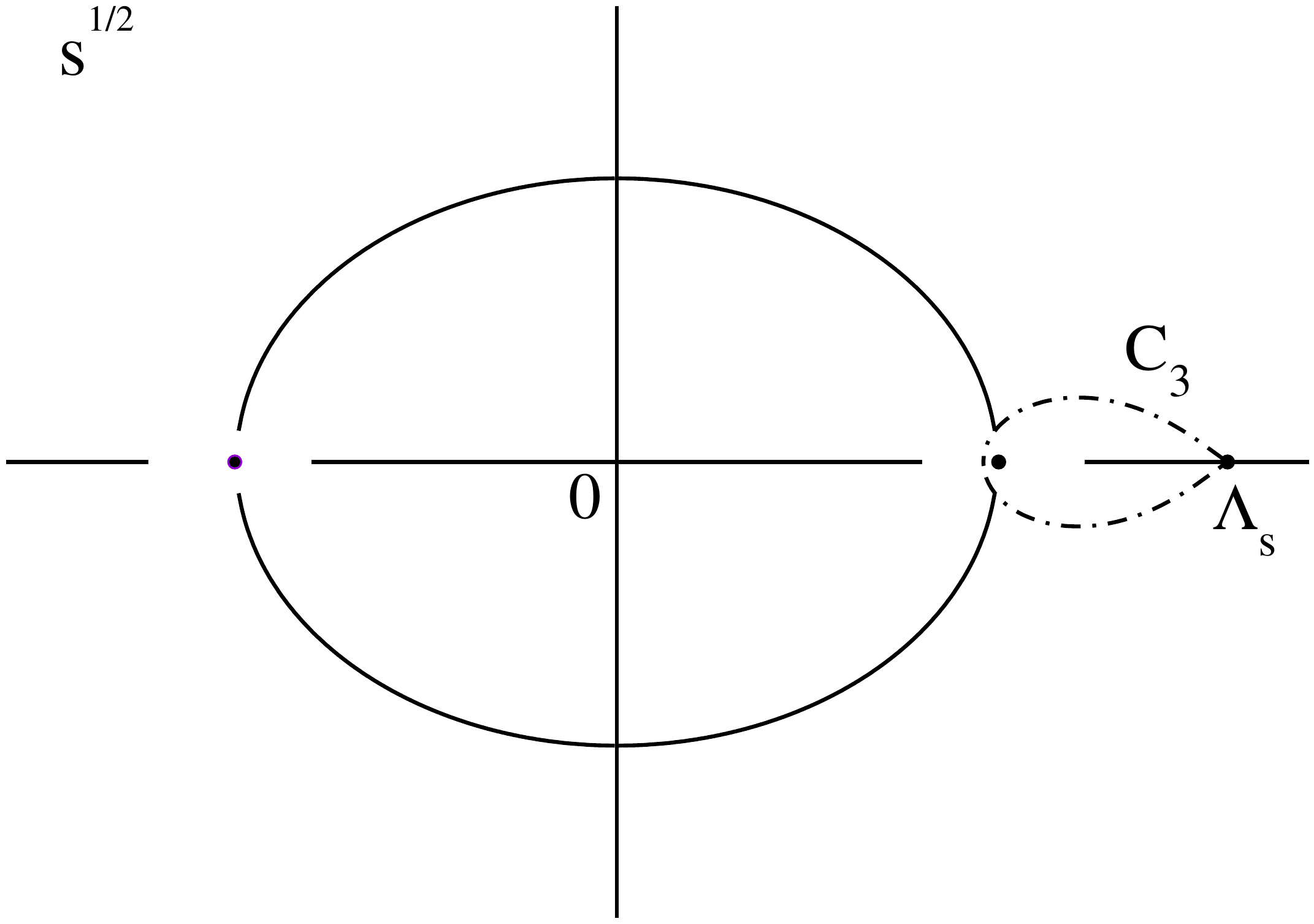}}
\parbox[c]{4cm}{ \caption{Singularity structure of the $\gamma N\to\pi N$ amplitude (solid lines and dots),
and domains of convergence (dashed lines) characterizing the expansion \eqref{expansion}.}\label{fig:gammaN}}
\end{center}
\end{figure}

The representation (\ref{expansion}) need not converge in parts of the complex plane  very distant from the physical region.
Therefore it suffices to use an universal mapping for $\pi N\to\pi N$, $\gamma N\to\pi N$ and $\gamma N\to\gamma N$ reactions.

The transformation function $\xi(\sqrt{s}\,)$ is chosen to be a superposition of the function $(\Lambda_s-\sqrt{s}\,)^2$ and a
M\"obius transformation,
\begin{eqnarray}
&& \xi(\sqrt{s}\,)=\frac{a_1\left(\Lambda_s-\sqrt{s}\,\right)^2-1}
{(a_1-2\,a_2)\left(\Lambda_s-\sqrt{s}\,\right)^2+1}\,,
\nonumber\\
&& a_1=\frac{1}{\left(\Lambda_s-\mu_E\right)^2}\,, \qquad
a_2=\frac{1}{\left(\Lambda_s-\Lambda_0\right)^2}\,.
\label{mapping2}
\end{eqnarray}
For $\Lambda^2_0=m_N^2-m_\pi^2$ it maps the interior of the contour $C_2$ in Fig.~\ref{fig:piN} onto
the unit circle. Thus it meets our conditions (\ref{def-u-channel-Omega}-\ref{def-Lambdas-Omega}), in particular it holds
$\xi'(\Lambda_s) =0$. Though not necessary, it is convenient to have available the inverse mapping with
\begin{eqnarray}
&& \xi^{-1} (x) = \Lambda_s -\frac{1+x}{\sqrt{(1+x)\,[(1-x)\,a_1 +2\,a_2\,x]}}\,.
\label{def-inverse}
\end{eqnarray}

The particular form of the conformal mapping used in this work is irrelevant. We constructed various mappings
that conform with the physical requirement as discussed above. Our final results show a very minor dependence on
the choice thereof, that can be compensated by a slight change of free parameters.

We turn to the analytic structure of the $\gamma N\to\pi N$ amplitude \cite{PhysRev.126.1596}, which
is illustrated in Fig.~\ref{fig:gammaN}. The unitarity $s$- and $u$-channel
cuts are the same as in the case of $\pi N$ amplitude, except that
the part of the $u$-channel cut lying on the real axis is given by
\begin{eqnarray}
&& 0 < s < \frac{m_N}{m_N+m_\pi}\,\Big( m_N^2-m_N \,m_\pi-m_\pi^2\Big)\,.
\label{def-LambdagammaN}
\end{eqnarray}
The $t$-channel cut from multiple pion exchanges consists  of the whole imaginary axis
and two arcs defined by
\begin{eqnarray}
&& s=\frac{2\,m_N^2+m_\pi^2-m_t^2}{2}
\pm i\,\frac{\left(m_t^2-m_\pi^2\right)\sqrt{4\,m_N^2-m_t^2}}{2\,m_t}\,, \quad
\nonumber\\
&& m_t \in (2\,m_\pi, 2\,m_N )\,.
\label{def-arcs}
 \end{eqnarray}
In addition there are the $t$-channel singularities from the one-pion exchange, that includes
the imaginary axis, the line $0<s< (m_N-m_\pi)^2 $ and a pole at $s= m_N^2$.
We use the conformal mapping (\ref{mapping2}) with $\Lambda_0 \simeq 901$ MeV chosen such that the
contour $C_3$ in Fig. ~\ref{fig:gammaN} touches the edge of the $t$-channel cut.

\begin{figure}[t]
\begin{center}
\parbox[c]{9cm}{\includegraphics*[width=8cm,height=8cm]{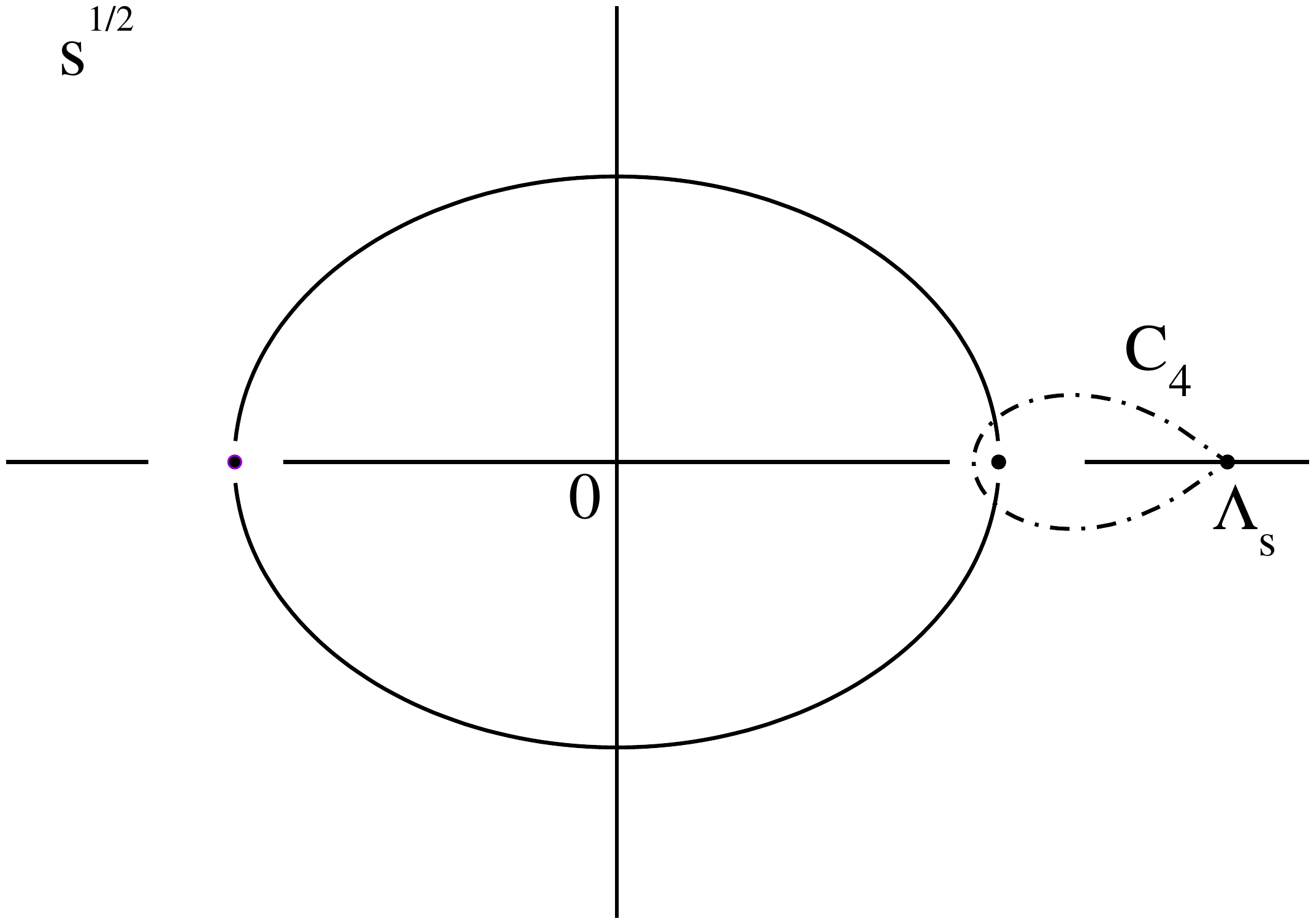}}
\parbox[c]{4cm}{ \caption{Singularity structure of the $\gamma N\to\gamma N$ amplitude (solid lines and dots),
and domains of convergence (dashed lines) characterizing the expansion \eqref{expansion}.}\label{fig:compton}}
\end{center}
\end{figure}

The singularity structure of the $\gamma N\to\gamma N$ amplitude
is presented on Fig.~\ref{fig:compton}.
The unitarity $s$- and $u$-channel
cuts are again the same as in the case of $\pi N$ amplitude, except that
the part of the $u$-channel cut lying on the real axis is given by
\begin{eqnarray}
0 < s < \frac{m_N^4}{(m_N+m_\pi)^2} \,.
\label{def-LambdagammaNb}
\end{eqnarray}
The $t$-channel cut from the one-pion exchange consists of the whole imaginary axis and parts of
the circle defined by
\begin{eqnarray}
s=m_N^2-\frac{m_t^2}{2}\pm i\,\frac{m_t}{2}\sqrt{4\,m_N^2-m_t^2}\,,\qquad
m_t \in (m_\pi , 2\,m_N)\,.
\label{t-channel-cut-Compton}
\end{eqnarray}
The cut lines of a multiple $n$-pion exchange include the whole imaginary axis and a part of the arcs
(\ref{t-channel-cut-Compton}) with $m_t \geq n \,m_\pi$. We use the conformal
mapping (\ref{mapping2}) with $\Lambda_0\simeq 877$ MeV, that maps
the interior of the contour $C_4$ on Fig.~\ref{fig:compton} onto the unit circle.
The particular value for $\Lambda_0$  makes the contour $C_4$ not to cross
the part of the $t$-channel cut, corresponding to two and more-pion exchanges.

In the generalized potential based on the tree-level scattering amplitudes
(\ref{piN-tree-level}, \ref{gN-piN-tree-level}, \ref{gp-gp-tree-level}) only the
$s$- and $u$-channel nucleon exchange and the t-channel pion-exchange
contribute to $U_{\rm inside}(\sqrt{s}\,)$ in (\ref{expansion}). The corresponding
singularities must be separated in 'inside' and 'outside' parts as shown in Appendix~\ref{singularities}.
The 'outside' part of the potential is expanded as $U_{\rm outside}(\sqrt{s}\,)$ in \eqref{expansion}.
Given the particular choices of the conformal mappings the one-loop expressions for the three reactions
contribute to $U_{\rm outside}(\sqrt{s}\,)$ only. The order to which  $U_{\rm outside}(\sqrt{s}\,)$  must be 
expanded is discussed in the result section \ref{results}. The goal is to correlate the order of the 
expansion (\ref{expansion}) with the chiral order as to obtain a generalized potential accurate 
in the region $\mu_M \sqrt{s} < \Lambda_s$ uniformly. Then the typical error may be identified in the chiral 
domain around $\sqrt{s} \simeq \mu_M$.

There is one peculiarity appearing when we consider the one-loop contributions to the generalized potential.
We use a heavy baryon prescription for loop diagrams because it is well established for all reaction channels considered.
Since the potential must possess only left-hand singularities it is necessary to subtract cut structures associated
with the s-channel unitarity cut. The latter is responsible for the imaginary part of the one-loop expression at
energies $\sqrt{s}> m_\pi+m_N$.  In accordance with (\ref{def-non-linear}) we subtract from the one-loop expressions
the integral
\begin{eqnarray}
\sum_{c,d}\, \int_{\mu_{\rm thrs}}^{\infty}\frac{dw}{\pi}\,\frac{\sqrt{s}-\mu_M}{w-\mu_M}\,
\frac{U^{(1)}_{ac}(w)\,\rho_{cd}(w)\,U^{(1)}_{db}(w)}{w-\sqrt{s}-i\epsilon}\,,
\label{subtraction}
\end{eqnarray}
in a given partial wave, where $U^{(1)}(\sqrt{s})$ is the leading order tree-level potential. In (\ref{subtraction}) the reference to 
angular momentum and parity is omitted. The integral (\ref{subtraction}) is finite since the leading-order potential is asymptotically 
bounded. This is a consequence of the representation (\ref{expansion}). The one-loop expression after subtraction of (\ref{subtraction}) 
still has a residual imaginary part at $\sqrt{s}> m_\pi+m_N$. It diminishes systematically if the order of the truncation 
in (\ref{expansion}) and the order of chiral expansion is increased. We neglect this  imaginary part as higher order and decompose 
the residual loop-expression according to (\ref{expansion}). The truncation of the sum in (\ref{expansion}) is uniform for all chiral 
moments of the potential, in particular for the leading-order term specifying the subtraction integral (\ref{subtraction}).

\clearpage

\newpage

\section{Numerical results }
\label{results}

In this section we present our numerical results for the three reactions $\pi N \to \pi N$, $\gamma N \to \pi N$ and
$\gamma N \to \gamma N$, where we assume the $\pi N$ channel in an s- or p-wave state.
We consider the energy region from threshold up to $1.3$ GeV, which is motivated by the
expected reliability of the two-channel approximation (the inelasticities in the $\pi N$ scattering are small at these energies \cite{Koch:1985bn,Arndt:2006bf}). A detailed comparison with available experimental data is provided.
We aim for a description of the experimental data set by solving the integral
equation (\ref{def-non-linear}) with the generalized potential
calculated in perturbation theory based on the chiral Lagrangian (\ref{Lagrangian}).
The generalized potential is analytically continued beyond the threshold region
and expanded systematically in application of suitably constructed conformal mappings as discussed
in section~\ref{sec:conformal_mapping}.

\subsection{$\pi N$ elastic scattering}

The $\pi N$ sector is developed independently from the $\gamma N$ channel as it is treated to first order
in the electric charge, $e \simeq 0.303$. The chiral Lagrangian (\ref{Lagrangian})
provides several free parameters. To order $Q$ the two parameters, $f$ and $g_A$, are relevant, where
$f \simeq 92.4$ MeV  is identified with the pion decay constant and $g_A \simeq 1.26$ with
the axial-vector coupling constant of the nucleon. Owing to the Goldberger-Treiman
relation $f\,g_{\pi NN} = g_A \,m_N$ one may use alternatively the empirical pion-nucleon coupling constant
\begin{eqnarray}
\frac{g^2_{\pi N N}}{4\pi} \simeq 13.54\,,
\label{def-gpinn}
\end{eqnarray}
as input \cite{Timmermans:1990tz}. We do not count the masses of the pion and nucleon, $m_\pi$ and $m_N$, as free
parameters. They are assumed to take their empirical values properly isospin averaged. At chiral order $Q^2$ four additional
parameters $c_1,c_2,c_3,c_4$ arise. There remain the four counter terms proportional to
$d_1+d_2, d_3, d_5, d_{14}-d_{15}$ that turn relevant at chiral order $Q^3$. The parameter $d_{18}$ parameterizes
the degree of violation of the  Goldberger-Treiman relation. A study of pion-nucleon scattering only does not
determine $g_A$ so that we use (\ref{def-gpinn}) as input. Given the empirical value for $g_{\pi NN}$ the parameter $d_{18}$
enters no longer as a free parameter in our study.

The generalized potential in (\ref{def-non-linear}) is decomposed into an inside and outside part based on the conformal
mapping (\ref{mapping2}). While the inside part of the generalized potential is determined by the
pion-nucleon coupling constant $g_{\pi NN}$, the outside part depends on the various counter terms and the order at which
the expansion in (\ref{expansion}) is truncated. As discussed in the Appendix B we perform a summation of the outside
part as is implied by the contribution of the u-channel nucleon-exchange cut for $\sqrt{s} > m_N-m_\pi^2/m_N$. This controls
large cancelations amongst the inside and outside parts of the potential that arise for large angular momentum.
We will determine the truncation order of the residual contributions in (\ref{expansion}) by the number of free counter
terms contributing to the outside part of the potential. The original potential contains polynomial terms in $\sqrt{s}$ that
are unphysical at large energies. In contrast the extrapolated potential is bounded at each order in the expansion of
(\ref{expansion}). Thus, there is an
issue how many terms in the expansion should be considered. If too many terms are included the resulting potential would
be unphysically large, even though the potential would be bounded asymptotically. Since the outside part of the potential
is governed by left-hand cuts that are far distant, the expansion (\ref{expansion}) should
converge quickly if applied to the full potential. This is nicely confirmed by the following  phenomenological study.

\begin{figure}
\includegraphics*[width=13cm]{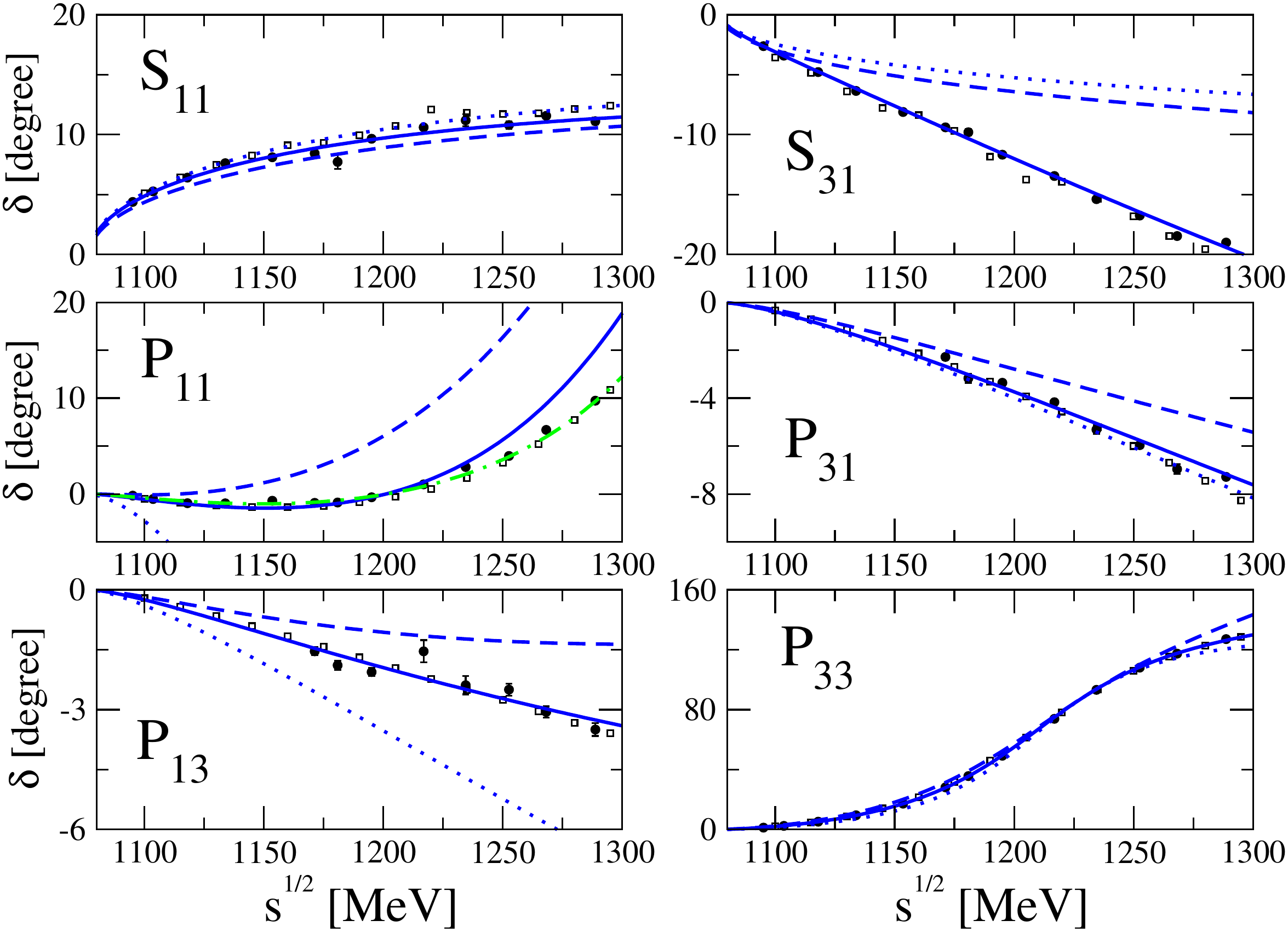}
\caption{Results of the fit for $\pi N$ $S$ and $P$-wave phase shifts.
The solid curves correspond to the full $Q^3$ results, the dashed curves
to $Q^2$ results, and the dotted curves to $Q^1$ calculation.
The dash-dotted line in the $P_{11}$ phase denote a phenomenological fit as explained it text.
The data are from \cite{Koch:1985bn}(circles) and \cite{Arndt:2006bf}(squares).}
\label{fig:piN1}
\end{figure}

In a first step we consider the inside part of the potential as determined by the pion-nucleon coupling constant (\ref{def-gpinn})
but keep the values $U_k$ in (\ref{expansion}) unrelated to the parameters of the chiral Lagrangian.
We use $\Lambda_s = 1500$ MeV, but checked that moderate variations with 1300 MeV $<\Lambda_s<$ 1600 MeV
lead to almost identical results. Within this phenomenological framework we study the relevance of higher order
terms in the expansion of the residual outside part of the potential.  For the fits we take the energy independent phase-shift
analysis \cite{Arndt:2006bf}. The analysis from \cite{Koch:1985bn} is used for comparison.

The empirical partial waves $S_{11}, P_{31}$ and  $P_{13}$ can
be well reproduced from threshold to $\sqrt{s} \leq  1300$ MeV
with only the leading term in the expansion. The  $S_{31}$ partial wave requires two terms.
The results for these four partial waves are shown
by solid curves in Fig.~\ref{fig:piN1} and compared with the empirical phase shifts \cite{Koch:1985bn,Arndt:2006bf}.

The phase-shifts are obtained as solutions of the linear equation (\ref{Nequation0}). The original non-linear
equation (\ref{def-non-linear}) is satisfied for the $S_{11}$ and $P_{13}$ partial waves  for energies from threshold
to $\sqrt{s} \leq 1300$ MeV with a violation smaller than 1$\%$. The deviation is somewhat larger for the
$P_{31}$ and $S_{31}$ waves, for which we obtain about 10$\%$ at the upper end. It is possible to introduce
a number of CDD poles to get solutions that satisfy the causality constraint more precisely, however,
this would introduce additional free parameters without clear physical meaning. The small deviations obtained
we take as a signal of higher energy effects not yet controlled in the present work.

One cannot obtain a satisfactory fit for $P_{11}$ and $P_{33}$ partial waves based on the linear equation (\ref{Nequation0})
without severe violation of the causality relation (\ref{def-non-linear}). This reflects the presence of
low lying resonances in these partial waves, the $\Delta$(1232) and the N(1440) resonances.
In the $P_{11}$ partial wave a numerically large contribution of the nucleon pole graph to the potential makes it
impossible for the solution (\ref{def-non-linear}) to exist without a CDD pole. Similar findings were discussed previously
in models \cite{Zachariasen:1961,Bart:1969ue} also based on a single-channel approximation.
After inclusion of one CDD pole, we can reproduce the $P_{33}$ partial wave with only the zeroth order coefficient
in the expansion (\ref{expansion}). The $P_{11}$ partial wave is well described in the presence
of one CDD pole if the corresponding potential is expanded to first order. The results are shown in
Fig.~\ref{fig:piN1} by the solid line ($P_{33}$) and dash-dotted line ($P_{11}$). In both cases the
set of  equations (\ref{D-CDD-ansatz}, \ref{CDD-ansatz}, \ref{identify-R}) provides solutions of
(\ref{def-non-linear}) accurate at the 1$\%$ level.

We return to our effective field theory which relates the expansion coefficients in (\ref{expansion}) to the parameters
of the chiral Lagrangian (\ref{Lagrangian}). At chiral order $Q^2$ there are four counter terms, $c_1,..,c_4$
which contribute to the two s- and four p-wave potentials. There is no contribution to the potentials of higher partial waves.
The coefficients $U_k$ in (\ref{expansion}) posses a chiral decomposition into powers of the pion mass modulo logarithm terms.
For the p-waves we have to truncate the expansion (\ref{expansion}) at leading order, $k=0$, since any additional term may be
altered arbitrarily by higher order counter terms. For the s-wave potentials the second term with, $k=1$, may be considered
in addition, since the $Q^2$ counter terms determine the slope of the s-wave potential to leading order in a chiral expansion.
However, it is unclear a priori whether one should go to the maximum order introduced above. Since the terms with $k\geq 1$ affect
the high energy part of the potential we find it advantageous to consider such terms only at an order where they are determined
quite precisely. To be on the safe side, we restrict ourselves to the zeroth order in the expansion (\ref{expansion})
at chiral order $Q^2$. At the next order $Q^3$ there are additional counter terms available that lead to a precise determination of the
slope of the s-wave potentials. Therefore it is justified to consider the terms with $k=1$ in (\ref{expansion}). For the p-waves
we keep the  $k=0$ terms only, since the slopes of the p-wave potentials receive contributions from unknown $Q^4$
counter terms.

\begin{table}[t]
\begin{center}
\begin{tabular}{|c|c|c|c|c|}
\hline
$c_1$ [$\rm{GeV}^{-1}$]&$c_2$ [$\rm{GeV}^{-1}$]&$c_3$ [$\rm{GeV}^{-1}$]&$c_4$ [$\rm{GeV}^{-1}$]\\
\hline
$-1.31$&$2.03$&$-5.11$&$3.59$ \\
\hline
$\bar d_1+\bar d_2$ [$\rm{GeV}^{-2}$]&$\bar d_3$ [$\rm{GeV}^{-2}$]&$\bar d_5$ [$\rm{GeV}^{-2}$]&$\bar d_{14}-\bar d_{15}$ [$\rm{GeV}^{-2}$]\\
\hline
$0.13$&$0.79$&$0.68$&$-0.49$\\
\hline
$m_{P_{33}}$ [MeV]&$g^{\pi N}_{P_{33}}$&$m_{P_{11}}$ [MeV]&$g^{\pi N}_{P_{11}}$\\
\hline
$1230.7$&$11.0$ &$1514$&$25$\\
\hline
\end{tabular}
\end{center}
\caption{Parameters obtained from a fit to the empirical s- and p-wave pion-nucleon phase shifts.}
\label{table:parameters}
\end{table}

We determine the chiral parameters by a fit to our phenomenological potentials, discussed above. At chiral order $Q^3$ it can be
reproduced identically with the exception of the $P_{11}$ potential. The numerical values of our parameters
are collected in Tab.~\ref{table:parameters}. The solid lines in Fig.~\ref{fig:piN1} show
the resulting phase shifts. All phase shifts coincide with the previously discussed phenomenological phases
with the exception of the $P_{11}$ phase, for which some further improvement is desirable.
An accurate reproduction of the phenomenological $P_{11}$ potential would require the inclusion of the term $U_k$ with
$k=1$ in (\ref{expansion}). Such a term is not generated in our approach convincingly. The $P_{11}$ needs further attention,
in particular the inclusion of the inelastic $\pi\pi N$ channel (see e.g. \cite{Krehl:1999km}, \cite{Gasparyan:2003fp}, \cite{Pearce:1989re}).

The convergence properties of our approach are illustrated in  Fig.~\ref{fig:piN1} by additional dashed and
dotted lines. The dashed lines correspond to the $Q^2$ calculation. It follows from the $Q^3$ result by switching
off the contribution of the one-loop diagrams together with the parameters,
$d_{1}+d_{2},d_3, d_5, d_{14}-d_{15}$. In addition the terms with $k=1$ in (\ref{expansion}) are dropped. The
dotted lines show the results with $c_i=0$, in addition, which  defines the order $Q$ result. We do not change the
CDD pole parameters  when going from $Q^3$ to $Q^2$ or $Q^1$  results. The solid, dashed and dotted
lines in Fig.~\ref{fig:piN1} are all obtained with the same parameters as given in Tab.~\ref{table:parameters}.

For all partial waves we obtain a convincing convergence pattern, although
a few comments must be made. In the $P_{11}$ partial wave there is a convergence,
but the $Q^1$ and $Q^2$ results are completely off the data. This is due to a significant cancelation between a large
contribution from the nucleon pole term and the counter terms. The $S_{31}$ potential  receives a sizeable contribution
from the first order expansion term, which is absent in the $Q^1$ and $Q^2$ lines.

\renewcommand{\arraystretch}{1.4}
\tabcolsep=1.3mm
\begin{table}[t]\begin{center}
\begin{tabular}{|c|ccc|}
\hline
 & current work &  KA86\protect\cite{Koch:1985bn} &
EM98\protect\cite{Matsinos:1997pb} \\
\hline
$a_{[S_-]}^{(\pi N)}$~[fm]& 0.116 & 0.130 &  0.109 $\pm$ 0.001  \\
$a_{[S_+]}^{(\pi N)}$~[fm]& 0.003 &-0.012 & 0.006 $\pm$ 0.001  \\
$b_{[S_-]}^{(\pi N)}$~[$m_\pi^{-3}]$& 0.010 & 0.008&   0.016 \\
$b_{[S_+]}^{(\pi N)}$~[$m_\pi^{-3}]$& -0.058 &-0.044&  -0.045 \\
\hline
$a_{[P_{11}]}^{(\pi N)}$~[$m_\pi^{-3}$]& -0.082 &
-0.078& -0.078 $\pm$ 0.003 \\
$a_{[P_{31}]}^{(\pi N)}$~[$m_\pi^{-3}$]& -0.048 &
-0.044& -0.043 $\pm$ 0.002 \\
$a_{[P_{13}]}^{(\pi N)}$~[$m_\pi^{-3}$]& -0.032 &
-0.030& -0.033 $\pm$ 0.003 \\
$a_{[P_{33}]}^{(\pi N)}$~[$m_\pi^{-3}$]& 0.193 &
0.214& 0.214 $\pm$ 0.002 \\
\hline
\end{tabular}
\end{center}
\caption{Pion-nucleon threshold parameters (see e.g. \cite{Lutz:2001yb}). }
\label{table:scattering_lengths}
\end{table}

In order to further scrutinize the consistency of our calculation
we provide the threshold values of the $s$- and $p$-wave amplitudes
as they come out of our calculation. In Tab.~\ref{table:scattering_lengths}
they are compared with the results of different partial-wave analyses \cite{Koch:1985bn,Matsinos:1997pb}.
Given the discrepancies among the analyses we find an acceptable overall pattern.
For the p-wave scattering volumes we agree with the analysis of \cite{Koch:1985bn} at the $10$\% level.
The s-wave parameters spread most widely in the isospin even channel. A direct extraction of
both scattering lengths from an analysis of pionic hydrogen and pionic deuterium data
based on $\chi$PT \cite{Meissner:2005ne} gives the following values
\begin{eqnarray}
a_{[S_-]}^{(\pi N)}=(0.120\pm 0.003 )\,{\rm fm}\,, \qquad   \quad a_{[S_+]}^{(\pi N)}=(0.002\pm 0.003 )\, {\rm fm}\,,
\label{chPT-deuteron}
\end{eqnarray}
which are consistent with our results.

\begin{table}[t]
\rescale
\setlength{\tabcolsep}{1.4mm}
\setlength{\arraycolsep}{3.2mm}
\renewcommand{\arraystretch}{1.4}
\begin{center}
\begin{tabular}{|l|c|c|c|c|c|}
\hline
&$c_1$ [$\rm{GeV}^{-1}$]&$c_2$ [$\rm{GeV}^{-1}$]&$c_3$ [$\rm{GeV}^{-1}$]&$c_4$ [$\rm{GeV}^{-1}$]\\
\hline
current work&$-1.31$&$2.03$&$-5.11$&$3.59$\\ \hline
K-matrix &$-1.60$&$3.25$&$-6.34$&$3.93$\\ \hline
$Q^2$ phases \cite{Krebs:2007rh}&$-0.57$&$2.84$&$-3.87$&$2.89$\\ \hline
$Q^3$ phases \cite{Fettes:1998ud} fit 1&$-1.23$&$3.28$&$-5.94$& $3.47$\\ \hline
$Q^3$ phases \cite{Fettes:1998ud} fit 2&$-1.42$&$3.13$&$-5.85$& $3.50$\\ \hline
$NN$ phases \cite{Rentmeester:2003mf}&&&$-4.78$&$3.96$\\ \hline
&$\bar d_1+\bar d_2$ [$\rm{GeV}^{-2}$]&$\bar d_3$ [$\rm{GeV}^{-2}$]&$\bar d_5$ [$\rm{GeV}^{-2}$]&$(\bar d_{14}-\bar d_{15})$ [$\rm{GeV}^{-2}$]\\
\hline
current work&$0.13$&$0.79$&$+0.68$&$-0.49$\\
\hline
K-matrix&$3.94$&$-3.68$&$-0.17$&$-7.46$\\
\hline
$Q^3$ phases  \cite{Fettes:1998ud} fit 1&$3.06$&$-3.27$& $+0.45$&$-5.65$\\ \hline
$Q^3$ phases  \cite{Fettes:1998ud} fit 2 &$3.31$&$-2.75$& $-0.48$&$-5.69$\\ \hline
\end{tabular}
\end{center}
\caption{Comparison of low-energy constants in pion-nucleon scattering.}
\label{table:LEC}
\end{table}

We continue with a discussion of our parameter set as determined from a fit to the s- and p-wave pion-nucleon phase shifts
and collected in Tab.~\ref{table:parameters}. The parameters related to the two CDD poles
reflect the presence of the $\Delta (1232)$ in the $P_{33}$ and $N(1440)$ in the $P_{11}$ waves.
They allow for an interpretation in terms of effective resonance vertices of the form
\begin{eqnarray}
&& \mathcal{L}^{}_{\rm eff}\,=
\frac{g_{\Delta}}{f}\, \,\bar \Delta_\mu \,\vec{T}\, (\partial^{\mu} \vec{\pi})\, N \,
+\frac{g_{N^*}}{2\,f} \,\bar{N}^*\,\gamma_5\,\gamma^{\mu} \,\big( \vec{\tau}\cdot (\partial_{\mu}\vec{\pi} )\big) \,N
  + {\rm h.c.} \,,\qquad
\nonumber\\
&& g_{\Delta}=\frac{\sqrt{3}\,g^{\pi N}_{P_{33}}\,f}{m_{P_{33}}} \simeq 1.43\,, \qquad
g_{N^*}=\frac{2\,g^{\pi N}_{P_{11}}\,f}{\sqrt{3}(m_{P_{11}}+m_N)} \simeq 1.09 \,,
\label{}
\end{eqnarray}
with the isospin transition matrices normalized by $T_i^\dagger\,T_j = \delta_{ij}-\tau_i\,\tau_j/3$.

It is interesting to compare our parameter set with the ones from different analyses.
A collection of various sets is offered in Tab.~\ref{table:LEC}, where  we recall central values
only. We do not show statistical errors  arguing that a comparison of different parameter sets is more significant in our case.
At chiral order $Q^2$ a strict $\chi$PT fit to the $\pi N$ phase shift was performed in \cite{Krebs:2007rh}. The
sizes of the $c_{1-4}$ counter terms agrees roughly with our values, which we recall in
Tab.~\ref{table:LEC} for convenience. A more quantitative agreement is expected at
higher order in the chiral expansion. The parameter sets of \cite{Fettes:1998ud} are based on a
strict $\chi$PT fit at order $Q^3$.  The empirical $\pi N$ phase shifts at energies $\sqrt{s} \leq 1100$
MeV are described. Indeed, the $c_{1-4}$ parameters agree significantly better
with our set. In \cite{Fettes:1998ud} different parameter sets are obtained, the most significant two of which are
represented in Tab.~\ref{table:LEC}. A further source of information on some low-energy constants is
a chiral analysis of the nucleon-nucleon scattering process \cite{Rentmeester:2003mf}. Here only the parameters $c_3$ and
$c_4$ are relevant. The values reported in  \cite{Rentmeester:2003mf} differ somewhat from the results of \cite{Fettes:1998ud}.
While the value for  $c_4$ is quite compatible with the result of \cite{Rentmeester:2003mf}, the value for $c_3$ is somewhat larger
than our result.

The size of the parameter  $c_1$ can be related to the pion-nucleon sigma term \cite{Bernard:1996gq}. To the order $Q^3$  it holds
\begin{eqnarray}
\sigma_{\pi N}=-4 \,c_1\, m_\pi^2 -\frac{9 \,m_\pi^3}{64\,\pi\, m_N^2}\,g_{\pi NN}^2  + {\mathcal O} \left( Q^4 \right)\,,
\label{sigma-piN}
\end{eqnarray}
which  for our value of $c_1$ would imply $\sigma_{\pi N}=77$ MeV. This value is in contradiction to the recent determination
of the sigma term based on unquenched but two-flavour QCD lattice simulations \cite{Procura:2006bj}, which suggests a much smaller
sigma term $\sigma_{\pi N} = (44 - 54)$ MeV. Such values appear compatible with an analysis of the subthreshold $\pi N$ amplitudes
performed to chiral order $Q^3$ \cite{Buettiker:1999ap}, which suggests a significantly smaller value for $|c_1|$. The scattering
amplitude was reconstructed inside the Mandelstam triangle by means of dispersion relations using empirical phase shifts. This may
hint at a significant sensitivity on how to determine the parameter set and reflect the influence of higher order effects. Our work
does not shed additional light onto this puzzle.

\begin{figure}
\includegraphics*[width=13cm]{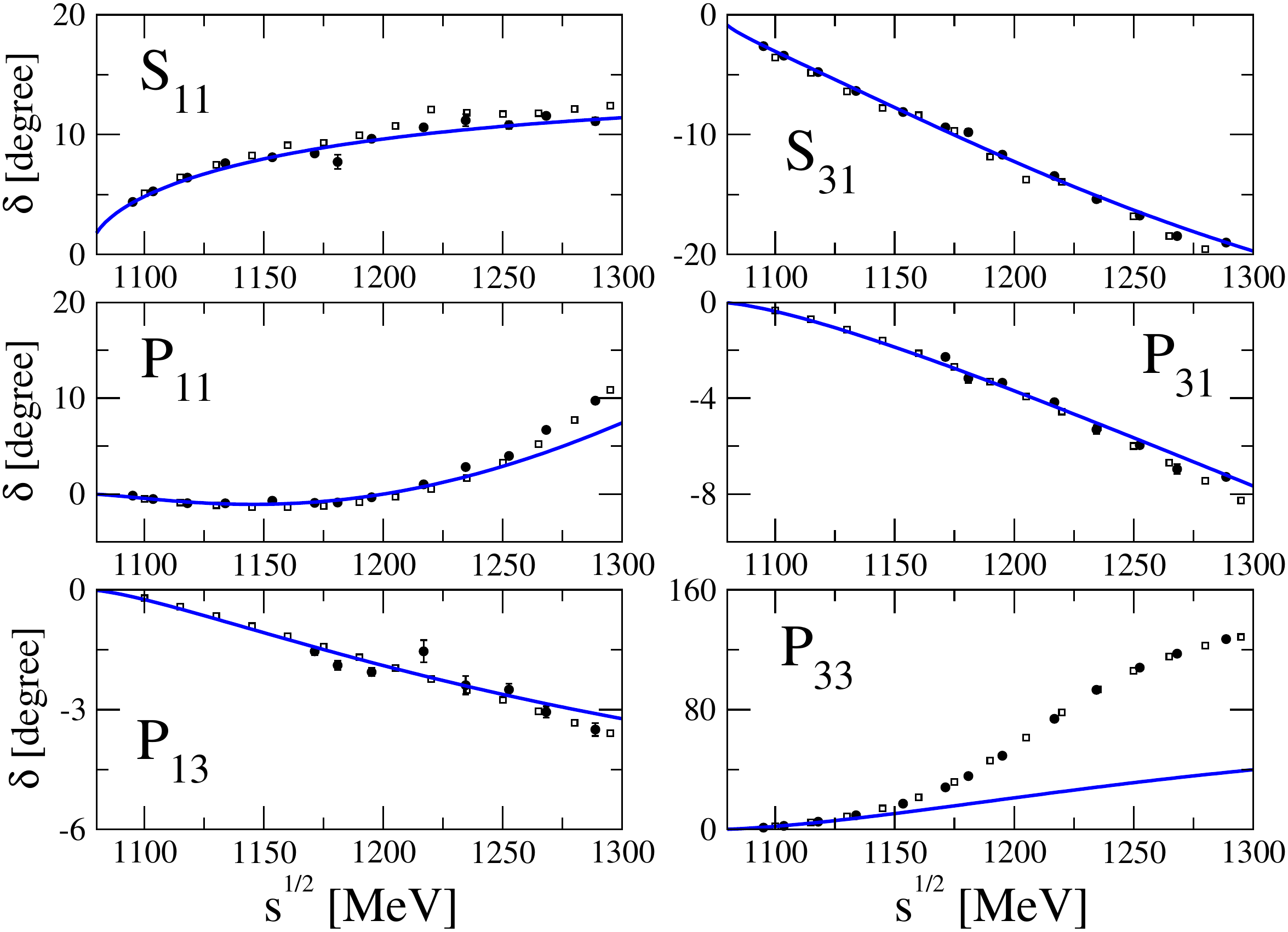}
\caption{Results of the fit for pion-nucleon  s- and p-wave phase shifts using a K-matrix ansatz.
The parameters are given in Tab.~\ref{table:LEC}.
The data are from \cite{Koch:1985bn}(circles) and \cite{Arndt:2006bf}(squares).}
\label{fig:piNKmatrix}
\end{figure}

A striking discrepancy we observe for the $Q^3$ counter terms. For instance our value for  $|\bar d_3|$ is about about four to five
times as small as the values of \cite{Fettes:1998ud,Fettes:2000xg}.
We traced the source of the discrepancy in the $Q^3$ counter terms as the consequence of rescattering effects. 
It is not related to the expansion of the potential using conformal mappings. If
instead of solving the non-linear equation \eqref{def-non-linear} we insist on a $K$-matrix ansatz as used
in \cite{Fettes:1998ud,Fettes:2000xg} the parameter set is altered significantly. More specifically, in this case we apply the
conformal mapping technique to the K-matrix. While the tree-level contributions to the K-matrix and the generalized potential
are identical, the one-loop contributions differ. The K-matrix does not receive a contribution of
the second order rescattering term (\ref{subtraction}) nor the CDD pole correction term (\ref{matching-CDD}).
Only the real part of the one-loop diagrams contribute to the K-matrix. It is  decomposed according to (\ref{expansion}), where 
we use the same truncation order as in our proper approach. The result of a low-energy  refit  of the counter terms in the 
K-matrix ansatz is included in Tab.~\ref{table:LEC} by the two rows labeled 'K-matrix'. The values are much closer to the $\chi$PT 
fits of \cite{Fettes:1998ud,Fettes:2000xg}. Note that the $c_2$ parameter is also affected significantly 
by rescattering effects bringing its value close to the one obtained in \cite{Fettes:1998ud}. 
The resulting $\pi$N phase shifts are shown in Fig.~\ref{fig:piNKmatrix}. Such an ansatz, which is at odds with the causality 
constraint of local quantum field theory, is nevertheless able to recover the empirical phase shifts amazingly well, with the 
exception of the $P_{33}$ phase. The latter is described only much below the isobar resonance.

\newpage

\subsection{Pion photoproduction}

The generalized potential for the $\gamma N \to \pi N$ reaction is calculated
to chiral order $Q^3$, where a multipole expansion is performed with the $\pi N$ channel in
an s- or p-wave state. At leading order the potential is determined by the mass parameters and
the pion-nucleon coupling constant (\ref{def-gpinn}). At subleading orders the
anomalous magnetic moments of the proton and neutron, $\kappa_p$ and $\kappa_n$, are probed. They determine
the isoscalar and isovector moments, $\kappa_s$ and $\kappa_v$, with
\begin{eqnarray}
&& \kappa_s=\kappa_p+\kappa_n\,, \qquad  \kappa_v=\kappa_p-\kappa_n\,, \qquad
\kappa_p\simeq 1.793\,,\qquad \kappa_n\simeq -1.913\,,
\label{magnetic-moments}
\end{eqnarray}
where we recall also their empirical values.
At order $Q^3$ the four counter term combinations $d_8, d_9, d_{20}, 2 \,d_{21}+d_{22}$ are relevant
in addition. Since we allowed for a CDD pole in the $P_{11}$ and $P_{33}$ elastic $\pi N$ amplitudes, there
will be additional parameters, $g^{\gamma N}_{P_{11}}$ and $g^{\gamma N}_{1, P_{33}}, g^{\gamma N}_{2, P_{33}}$,
characterizing the coupling of the CDD pole to the $\gamma N$ states. An interpretation in terms of effective resonance
vertices is provided by
\begin{eqnarray}
&& {\mathcal L}_{\rm eff}=
\frac{i\,e\,f^{(1)}_\Delta}{2\,m_N}\,\bar{\Delta}^{\mu}\,
\gamma_\nu \,\gamma_5\, T_3\,N\, F^{\mu \nu}
+\frac{e\,f^{(2)}_\Delta}{4\,m_N^2}\,\bar{\Delta}^{\mu}\,
\gamma_5\, T_3\,(\partial_\nu N) \,F^{\mu \nu}
+ \rm{h.c.}
\nonumber\\
&& \qquad \!-\frac{e}{4\,m_{N}}\,\bar{N}^*\,\sigma_{\mu\nu}\,\frac{\kappa^{N^*}_s+\kappa^{N^*}_v\,\tau_3}{2}\,N\,F^{\mu\nu}+ \rm{h.c.}  \,,
\nonumber\\
&& f^{(1)}_\Delta=\frac{m_N\,\sqrt{3}\,g^{\gamma N}_{1,P_{33}}(m_N-m_\Delta)
-2\,g^{\gamma N}_{2,P_{33}}m_\Delta}{e\,m_\Delta^2}\simeq 5.14 \,,
\nonumber\\
&& f^{(2)}_\Delta =-\frac{4\,\sqrt{3}\,g^{\gamma N}_{1,P_{33}}\,m_N^2}{e\,m_\Delta^2}\simeq 5.89\,,\nonumber\\
&&\kappa^{N^*}_s =-\frac{\sqrt{3}\,m_N}{e \,m_{P_{11}}}\left(g^{\gamma p}_{P_{11}}+
g^{\gamma n}_{P_{11}}\right)\simeq 1.28\,,\nonumber\\
&&\kappa^{N^*}_v =-\frac{\sqrt{3}\,m_N}{e \,m_{P_{11}}}\left(g^{\gamma p}_{P_{11}}-
g^{\gamma n}_{P_{11}}\right)\simeq 1.02 .
\label{}
\end{eqnarray}

\begin{figure}[b]
 \includegraphics*[width=13cm]{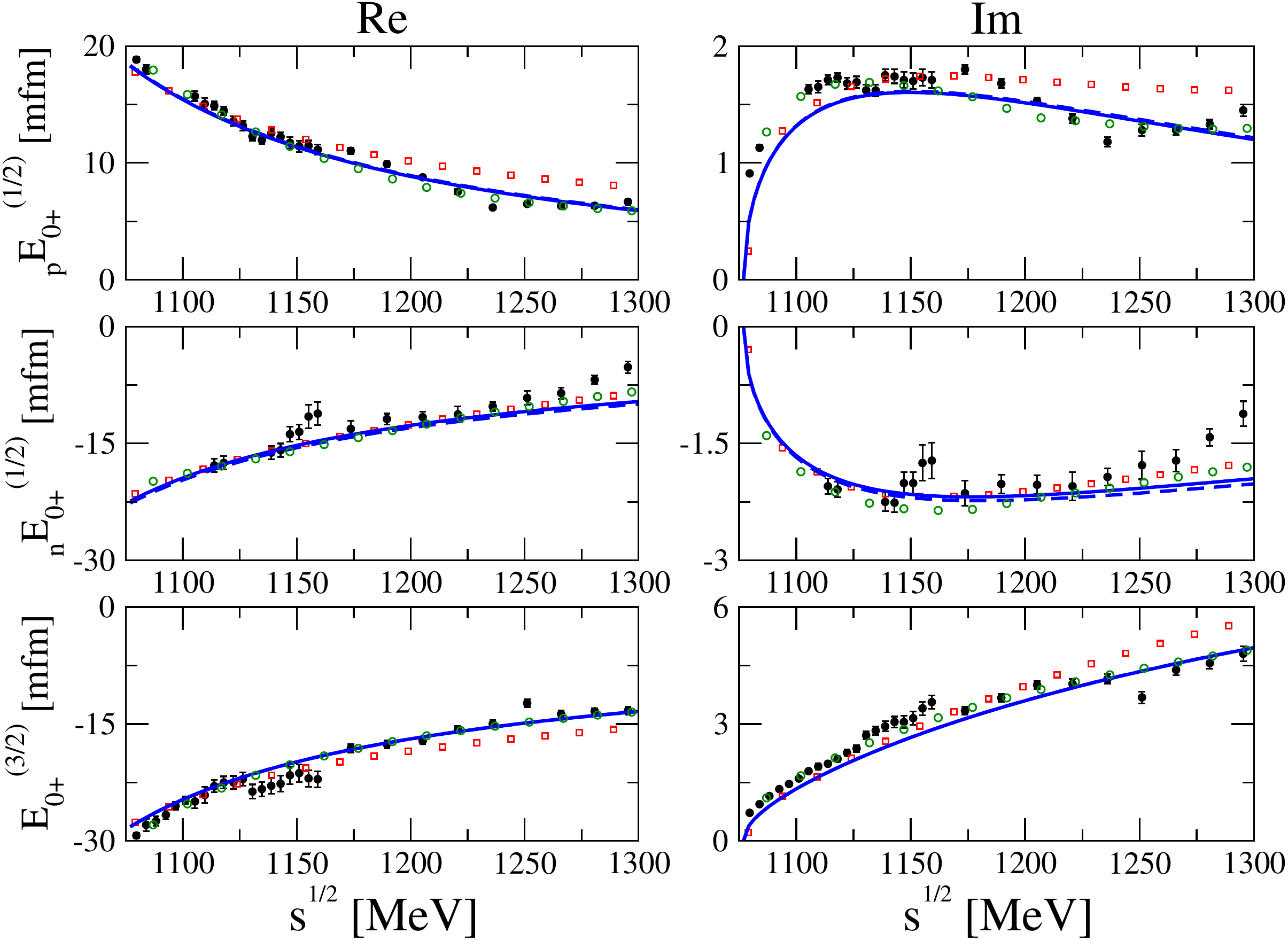}
\caption{Electric photoproduction multipoles $E_{0+}$. The data are from
\cite{Arndt:2002xv}(filled circles -- energy independent analyses, open circles
-- energy dependent analysis) and \cite{Drechsel:2007if}(squares). The solid line denotes the $Q^3$
calculation, the dashed line -- $Q^2$ calculation.}
\label{fig:gammaN1}
\end{figure}
\begin{figure}[t]
 \includegraphics*[width=13cm]{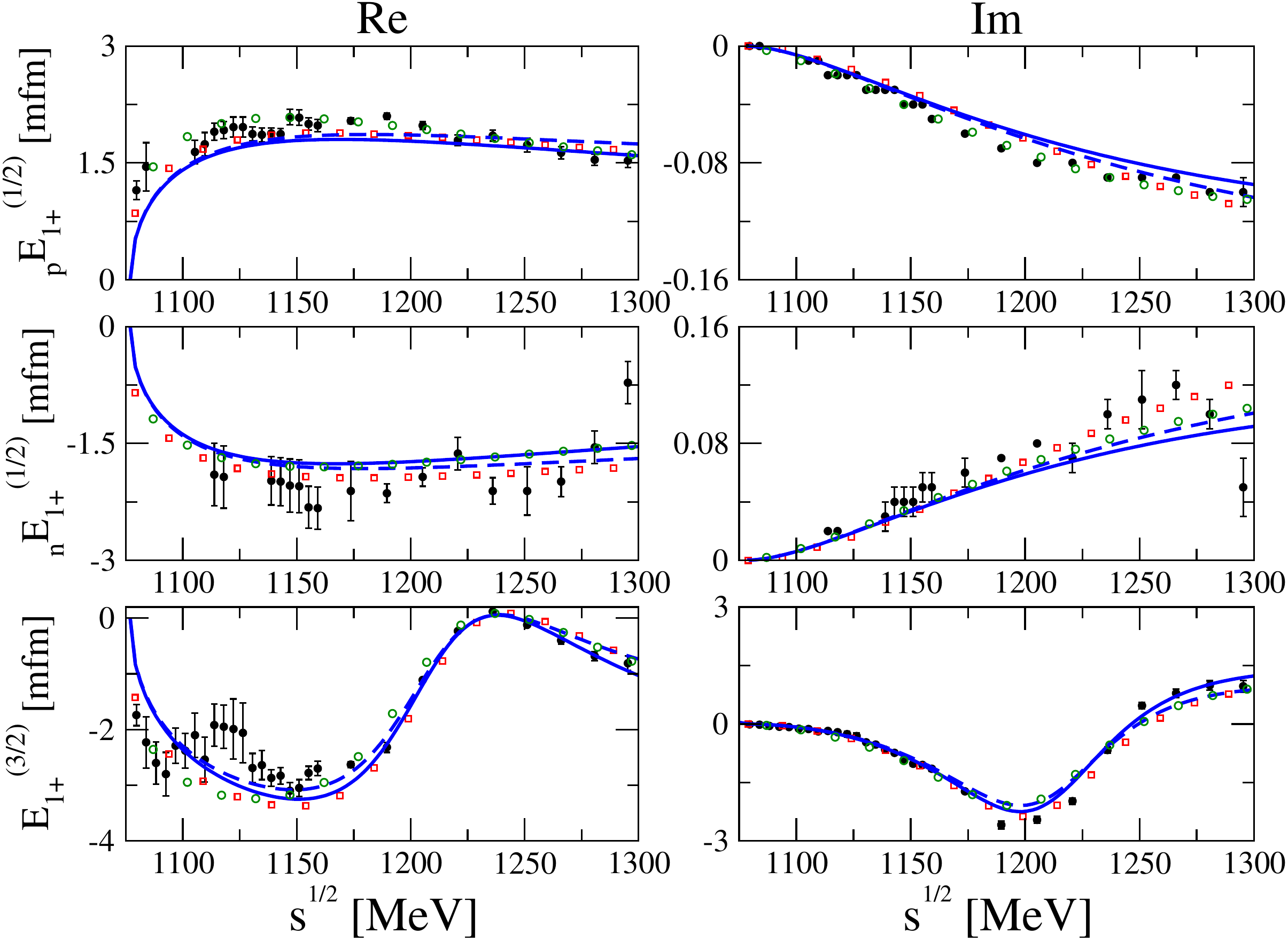}
\caption{Electric photoproduction multipoles $E_{1+}$. The data and line conventions are as in
Fig.~\ref{fig:gammaN1}.}
\label{fig:gammaN2}
\end{figure}
\begin{figure}[t]
 \includegraphics*[width=13cm]{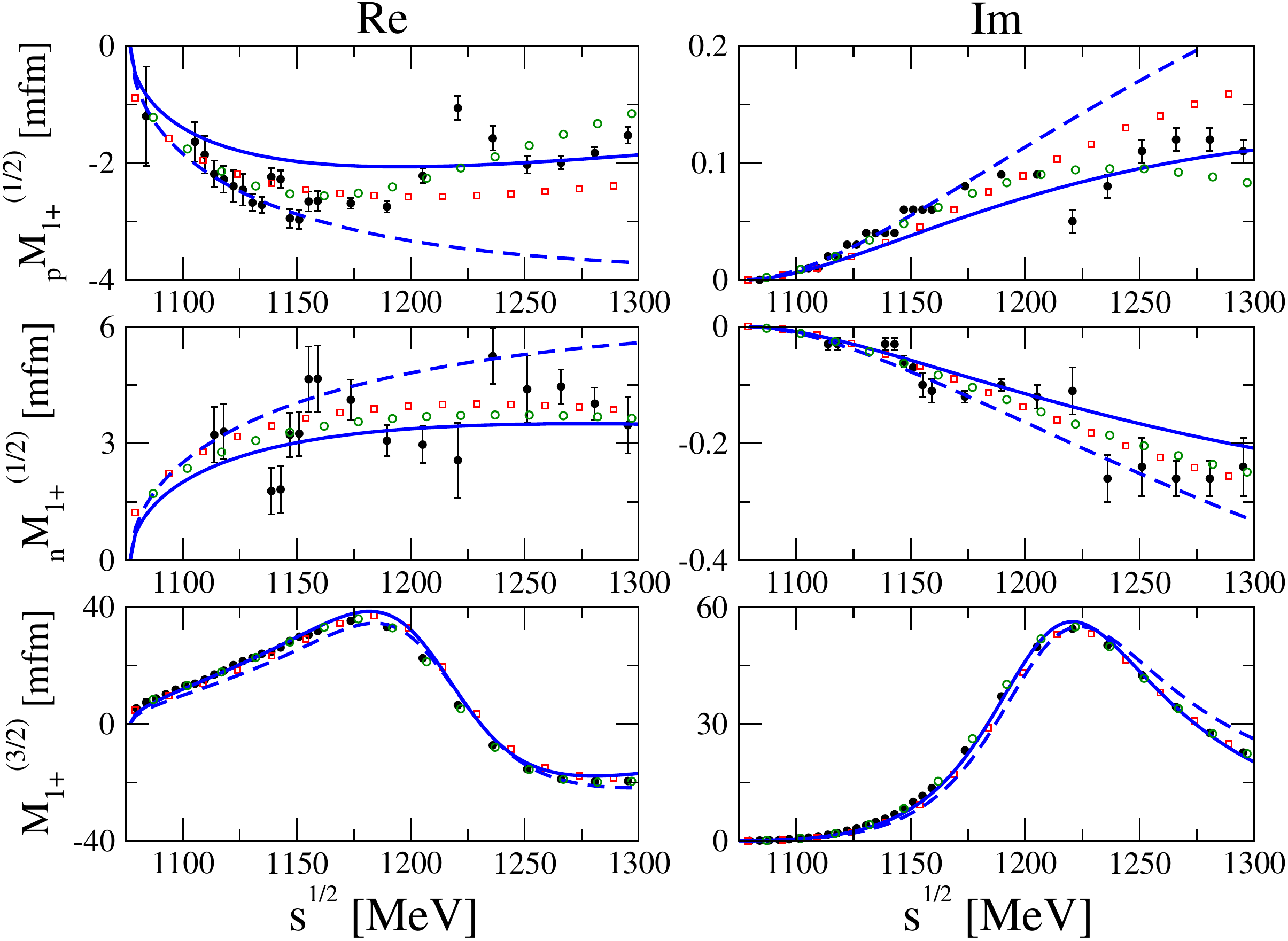}
\caption{Magnetic photoproduction multipoles $M_{1+}$. The data and line conventions are as in
Fig.~\ref{fig:gammaN1}.}
\label{fig:gammaN3}
\end{figure}
\begin{figure}[t]
 \includegraphics*[width=13cm]{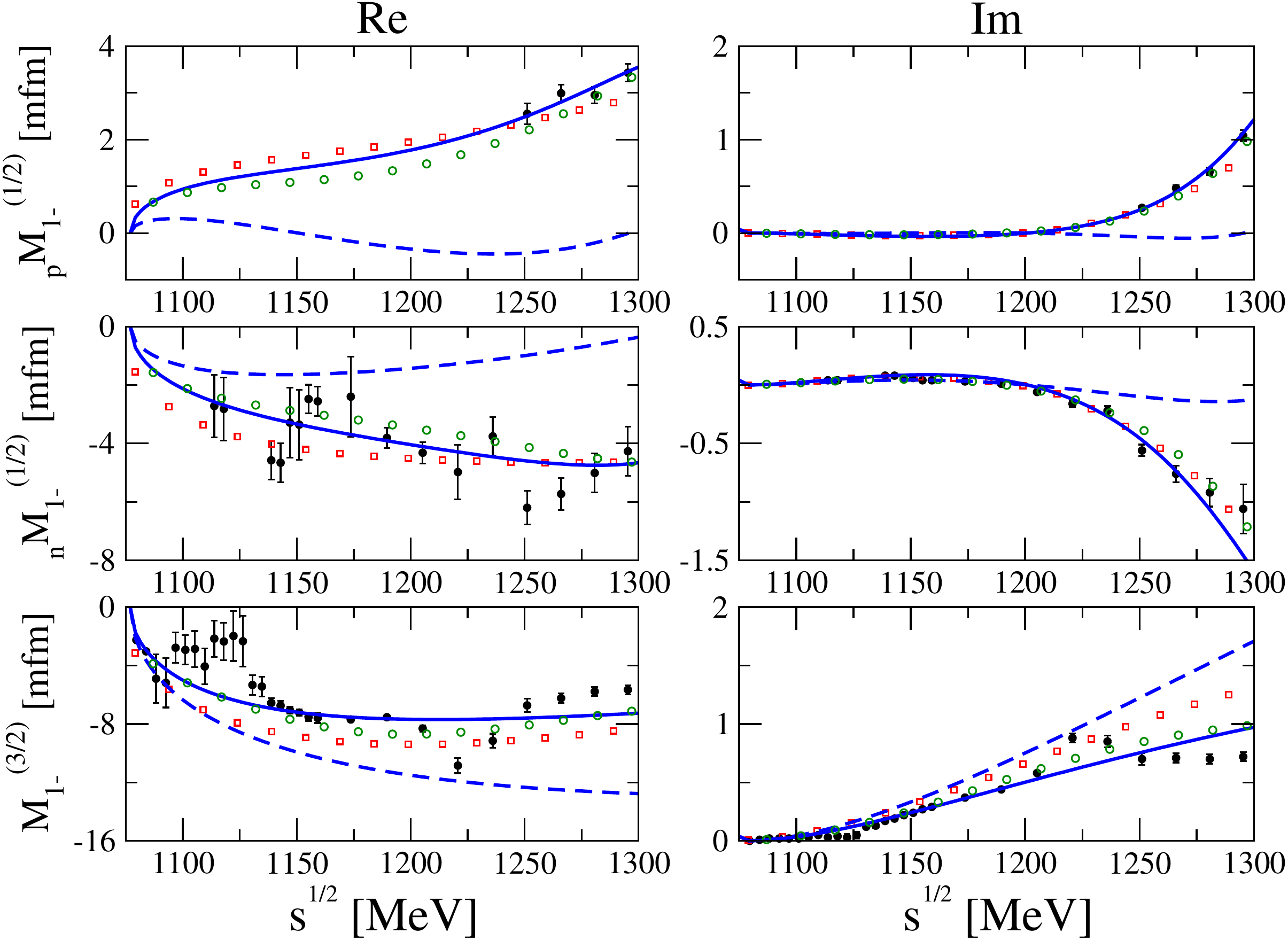}
\caption{Magnetic photoproduction multipoles $M_{1-}$. The data and line conventions are as in
Fig.~\ref{fig:gammaN1}.}
\label{fig:gammaN4}
\end{figure}

The generalized potential is decomposed into an inside and outside part according to the decomposition (\ref{expansion}).
The inside  part is determined unambiguously by $g_{\pi NN}$. It consists of
multiple pole terms at $\sqrt{s} = m_N$. The outside part contains branch cuts located outside the contour $C_3$ of
Fig.~\ref{fig:gammaN} only. It is expanded systematically using the conformal mapping (\ref{mapping2}) with
$\Lambda_0 = 991$ MeV and $\mu_E = m_N+ m_\pi/2$.  Like for the elastic $\pi N$ potential we find it advantageous to
perform a particular summation in the outside part of the photoproduction potentials. The one-pion exchange contribution
provides a left-hand cut at $\sqrt{s} \leq m_N-m_\pi$ that is characterized by multiple poles at the branch point
$\sqrt{s} = m_N-m_\pi$. Though such structures may be expanded systematically using (\ref{mapping2}), the convergence would be
unnaturally slow in particular at large $J$. In order to treat the pole structures accurately
we consider the region $m_N-2\,m_\pi\leq \sqrt{s} \leq  m_N-m_\pi$ explicitly in the outside  part of the potential.
For more technical details we refer to Appendix B. In the residual outside part of the potential we truncate the
expansion (\ref{expansion}) at $k=0$ for the s-waves and p-waves.

\begin{table}[t]
\begin{center}
\begin{tabular}{|c|c|c|c|c|c|c|c|}
\hline
$\bar d_8$ GeV$^2$ &$\bar d_9$ GeV$^2$&$\bar d_{20}$ GeV$^2$&
$(2\,\bar d_{21}-\bar d_{22}$) GeV$^2$&$g^{\gamma p}_{P_{11}}$ &$g^{\gamma n}_{P_{11}}$
&$g^{\gamma N}_{1,P_{33}}$ &$g^{\gamma N}_{2,P_{33}}$\\
\hline
$3.35$&$-0.06$&$0.61$&$0.05$&$-0.33$&$0.04$&$-0.44$&$-0.93$\\
\hline
\end{tabular}
\end{center}
\caption{Parameters obtained from photoproduction data. }
\label{table:photoproduction}
\end{table}

The parameter set is obtained from the empirical photoproduction s- and p-wave multipoles. There are in part
large discrepancies among different energy-dependent analyses \cite{Arndt:2002xv,Drechsel:2007if}. Therefore we try to
adjust the parameters to the energy independent partial-wave analysis from \cite{Arndt:2002xv}, which is less biased than the
energy dependent multipole analyses. We fit only the real part of the multipoles, since the imaginary parts are given by Watson's
theorem \cite{Watson:1954uc}. Our preferred parameters are given in Tab.~\ref{table:photoproduction}.

We discuss some details of the fit procedure. We find the $\pi N$ rescattering effects
important for the real parts of the s-wave multipoles. For the
$p$-wave multipoles, which do not have a CDD pole contribution, rescattering effects are mostly
responsible for generating the correct imaginary part, but do not modify the real parts much.
The s-wave multipoles shown by solid lines in Fig.~\ref{fig:gammaN1}
depend on the particular counter term combination
\begin{eqnarray}
\bar d_{20}+\frac{2\, \bar d_{21}-\bar d_{22}}{2}\,.
\label{s-wave-multipole}
\end{eqnarray}
Since none of the other multipoles depend on $\bar d_{20}$, the parameter combination (\ref{s-wave-multipole}) is determined
by the empirical s-wave multipoles. Given the spread in the different energy dependent  analyses a satisfactory description of the
s-wave multipoles is obtained.

The magnetic multipole $M_{1+}^{(3/2)}$ multipole provides a large and dominant contribution to the cross sections
in the $\Delta (1232)$ resonance region \cite{Arndt:2002xv,Drechsel:2007if}. As a  consequence the empirical
error bars for this multipole are very small. The multipole depends on the particular parameter combination
\begin{eqnarray}
\bar d_8+\frac{2\, \bar d_{21}-\bar d_{22}}{2} \,.
\label{Delta-multipole}
\end{eqnarray}
Since the same parameter combination enters the electric multipole $E_{1+}^{(3/2)}$ the parameter
combination (\ref{Delta-multipole}) together with the CDD pole parameters $g^{\gamma N}_{1, P_{33} }, g^{\gamma N}_{2, P_{33} }$
is determined by a fit of the $E_{1+}^{(3/2)}$ and $M_{1+}^{(3/2)}$ multipoles.
The solid lines in Fig.~\ref{fig:gammaN2} and Fig.~\ref{fig:gammaN3} show that a satisfactory description is obtained.
The electric $p$-wave multipoles $_pE_{1+}^{(1/2)}$, $_nE_{1+}^{(1/2)}$, also shown in Fig.~\ref{fig:gammaN2}, do not depend
on any of the parameters collected in Tab.~\ref{table:photoproduction}. Nevertheless, the empirical multipoles are recovered
reasonably by the solid lines.

There remain two parameter combinations, $\bar d_9$ and $2\, \bar d_{21}-\bar d_{22}$,
that cannot be determined unambiguously from the energy independent multipole analysis \cite{Arndt:2002xv}.
Incidentally, the energy dependent analyses \cite{Arndt:2002xv,Drechsel:2007if} differ most significantly in those
five magnetic multipoles, $_pM_{1\pm}^{(1/2)}$, $_nM_{1\pm}^{(1/2)}$ and $M_{1-}^{(3/2)}$, that are left to determine
$\bar d_9$, $2\, \bar d_{21}-\bar d_{22}$ and the CDD pole parameters
$g^{\gamma p}_{P_{11}}$ and $g^{\gamma n}_{P_{11}}$. In Tab.~\ref{table:photoproduction} we provide our preferred
parameter set that took into account additional constraints from photoproduction cross sections and Compton scattering data.
The five magnetic multipoles are shown in Fig.~\ref{fig:gammaN3} and Fig.~\ref{fig:gammaN4}.

Before scrutinizing in more depth the quality of the multipole description we illustrate the convergence properties
of our approach. In all Figs.~\ref{fig:gammaN1}-\ref{fig:gammaN4} the dashed lines show the effect of switching off the
contribution from the one-loop diagrams and counter terms in the photoproduction potentials.
The CDD pole parameters are unchanged. For most of the multipoles this affects the results by less than $20\%$. For the
$M_{1-}^{(3/2)}$ and $M_{1+}^{(1/2)}$  multipoles the difference at the highest energy considered constitute a factor of two.
Also the $_pM_{1-}^{(1/2)}$ and $_nM_{1-}^{(1/2)}$ multipoles prove sensitive
to the one-loop and counter term effects. This is analogous to our findings for the $P_{11}$ phase shift, the potential of which is a result
of subtle cancelations. We conclude that all together there is a satisfactory convergence pattern.

\begin{figure}[t]
 \includegraphics*[width=13.5cm]{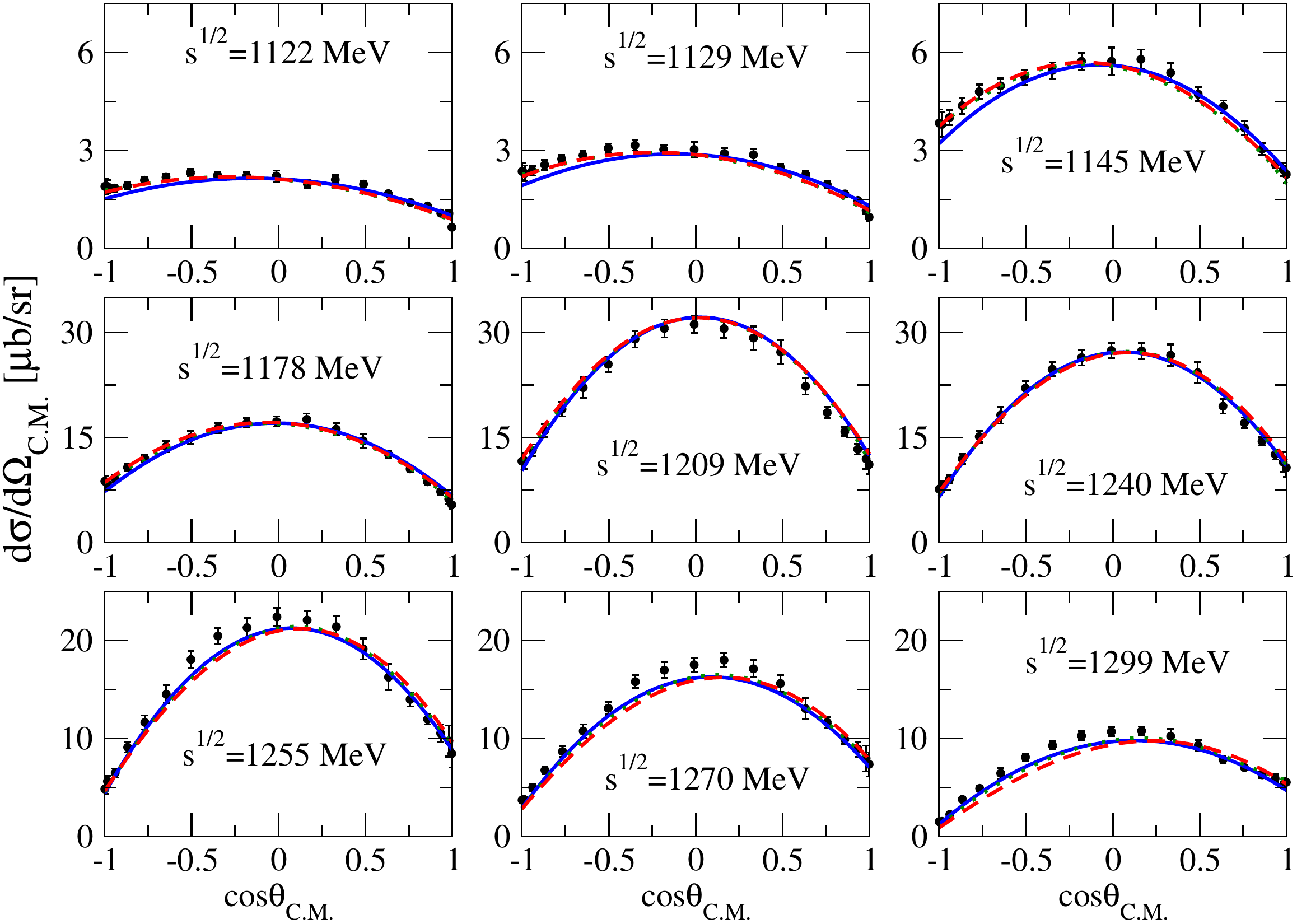}
\caption{Differential cross section for the reaction $\gamma p\to\pi^0 p$.
The solid line corresponds to our calculation with only $s$- and $p$-wave multipoles
included. The effect of higher partial waves is shown by the dashed and dotted lines as explained in the text.
The data are from \cite{Leukel:2001}.}
\label{fig:gammaN1dXS}
\end{figure}

\begin{figure}[t]
 \includegraphics*[width=13.5cm]{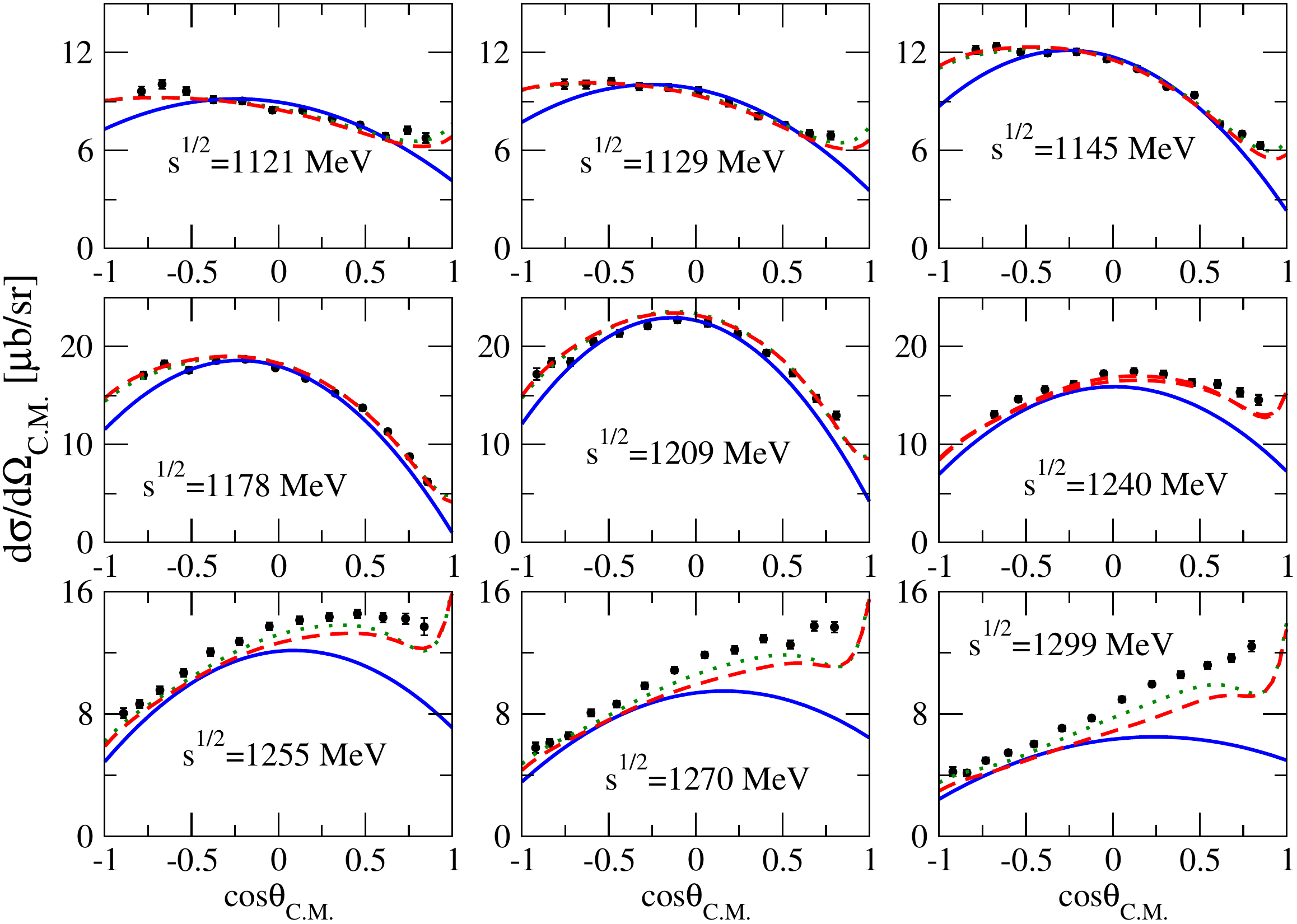}
\caption{Differential cross section for the reaction $\gamma p\to\pi^+ n$ with data taken from \cite{Preobrajenski:2001}.
The assembly  of the lines is as in Fig.~\ref{fig:gammaN1dXS}.}
\label{fig:gammaN2dXS}
\end{figure}

\begin{figure}[t]
 \includegraphics*[width=13.5cm]{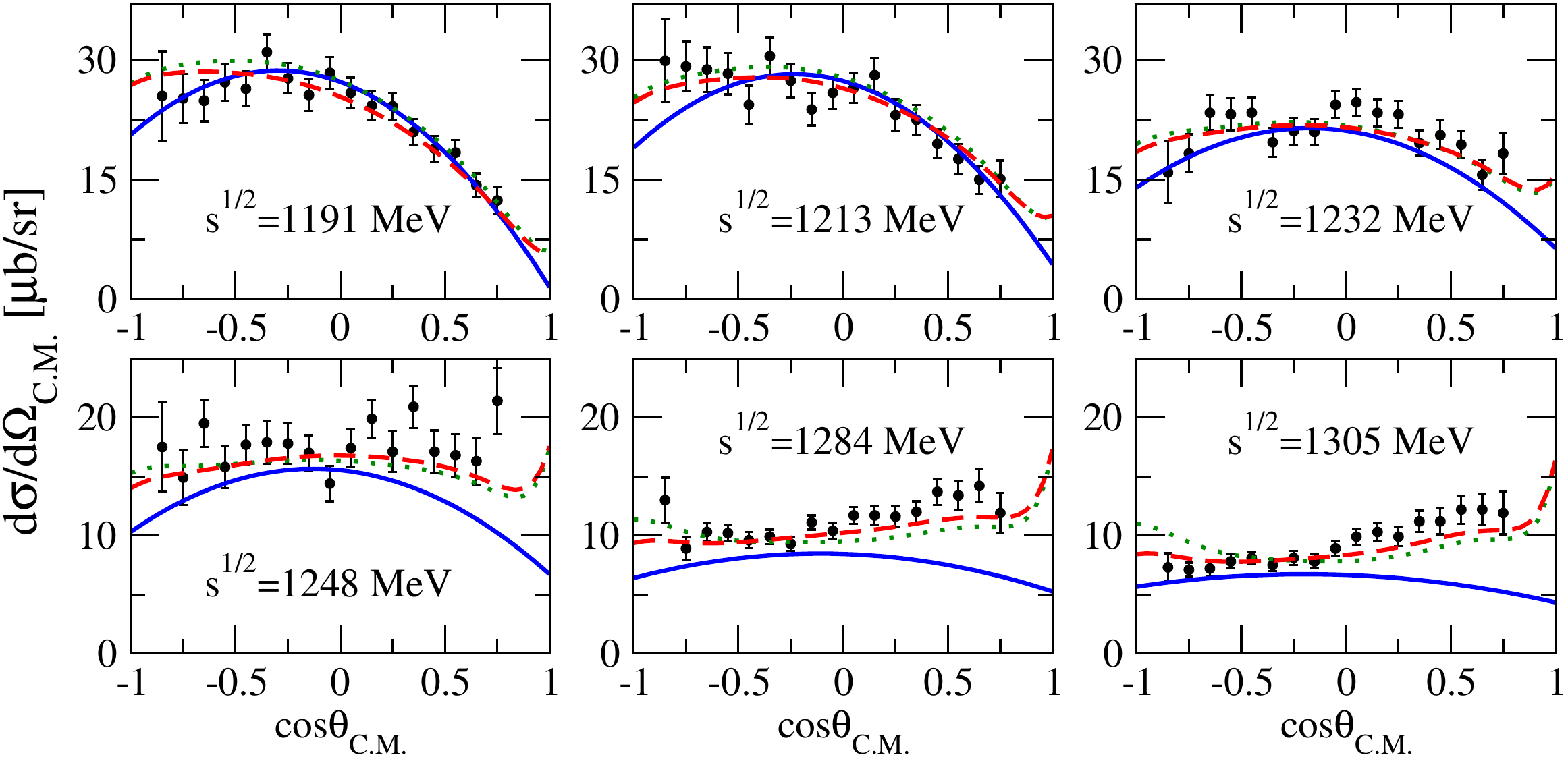}
\caption{Differential cross section for the reaction $\gamma n\to\pi^- p$ with data taken from \cite{Shafi:2004zn}.
The assembly of the lines is as in Fig.~\ref{fig:gammaN1dXS}.}
\label{fig:gammaN3dXS}
\end{figure}

\begin{figure}[t]
 \includegraphics*[width=13.5cm]{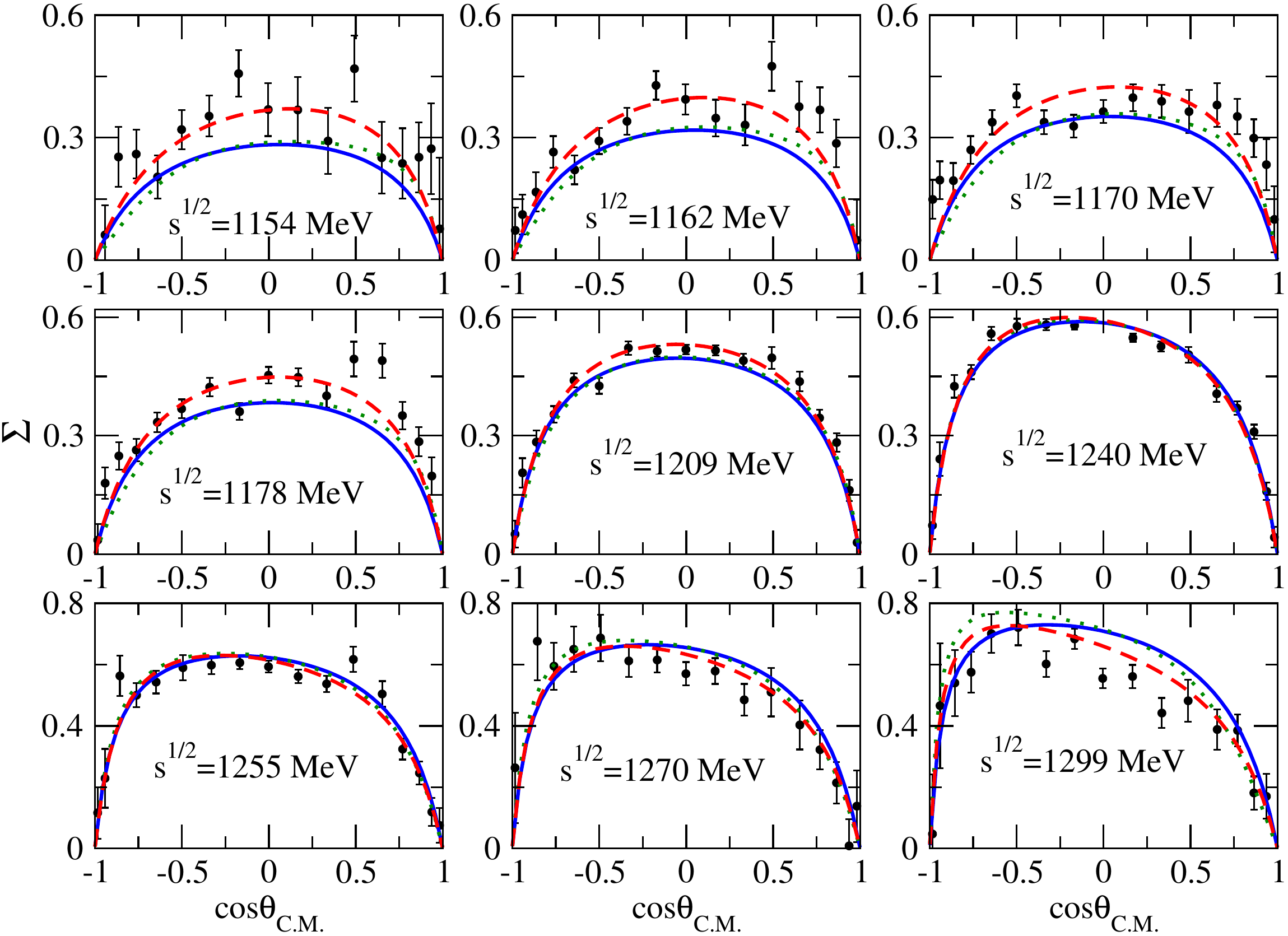}
\caption{Beam asymmetry for the reaction $\gamma p\to\pi^0 p$ with data taken from \cite{Leukel:2001}.
The assembly of the lines is as in Fig.~\ref{fig:gammaN1dXS}.}
\label{fig:gammaN1Sigma}
\end{figure}

\begin{figure}[b]
 \includegraphics*[width=13.5cm]{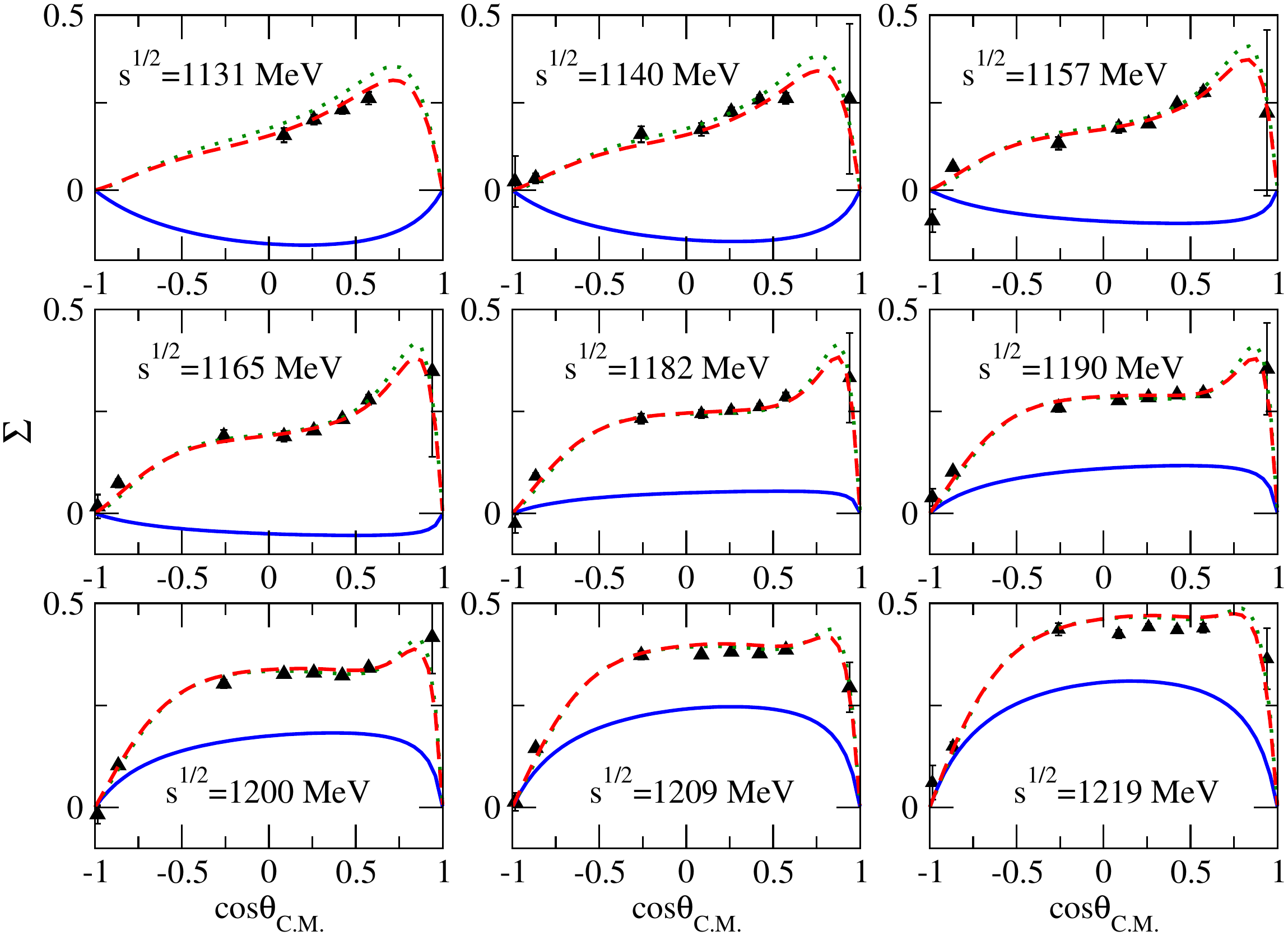}
\caption{Beam asymmetry for the reaction $\gamma p\to\pi^+ n$ with data taken from \cite{Blanpied:2001ae}.
The assembly of the lines is as in Fig.~\ref{fig:gammaN1dXS}.}
\label{fig:gammaN2Sigma}
\end{figure}

\begin{figure}[t]
 \includegraphics*[width=13.5cm]{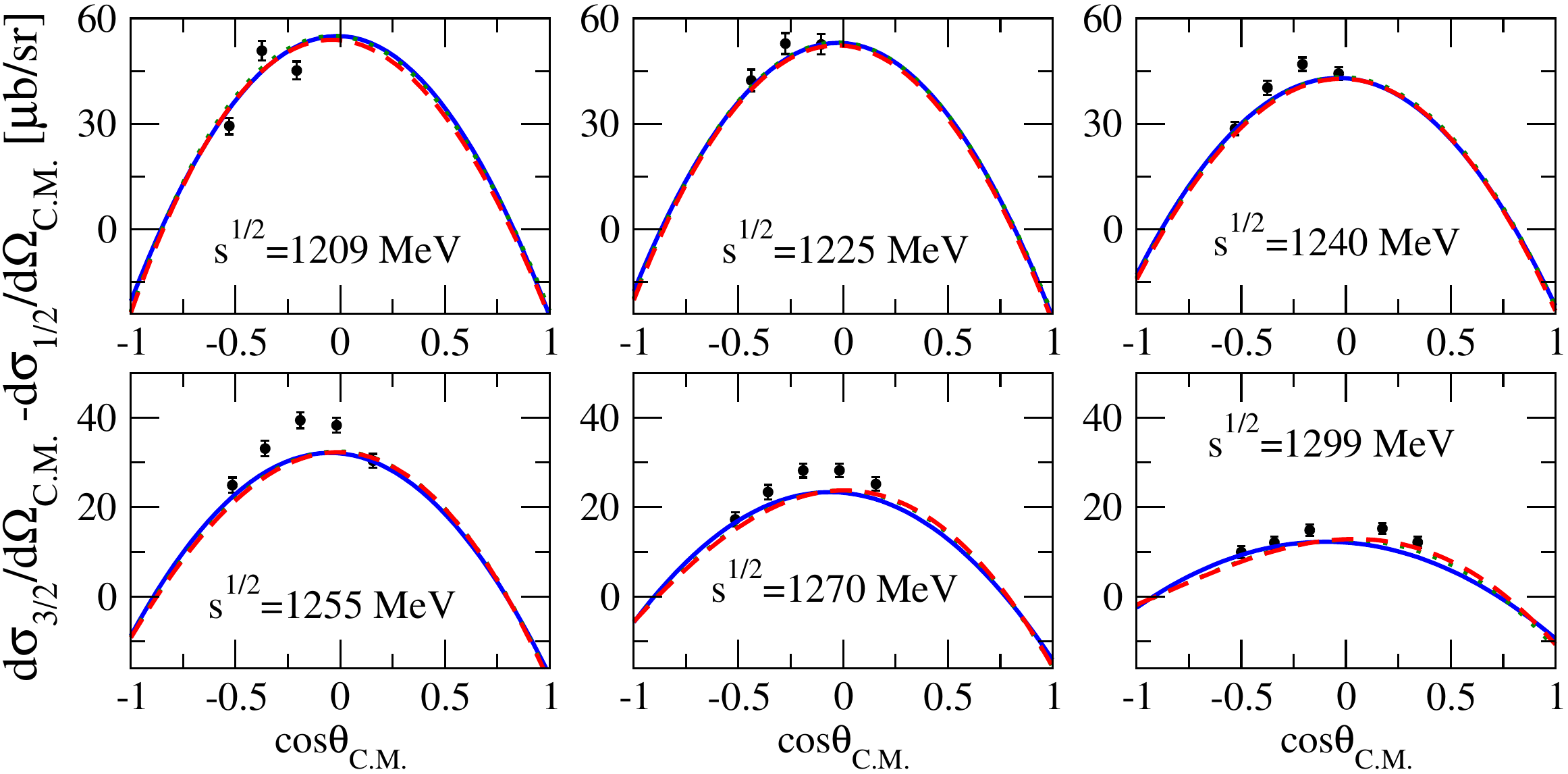}
\caption{Helicity asymmetry for the reaction $\gamma p\to\pi^0 p$ with data taken from \cite{Preobrajenski:2001}.
The assembly of the lines is as in Fig.~\ref{fig:gammaN1dXS}.}
\label{fig:gammaN1dsigma}
\end{figure}

\begin{figure}[b]
 \includegraphics*[width=13.5cm]{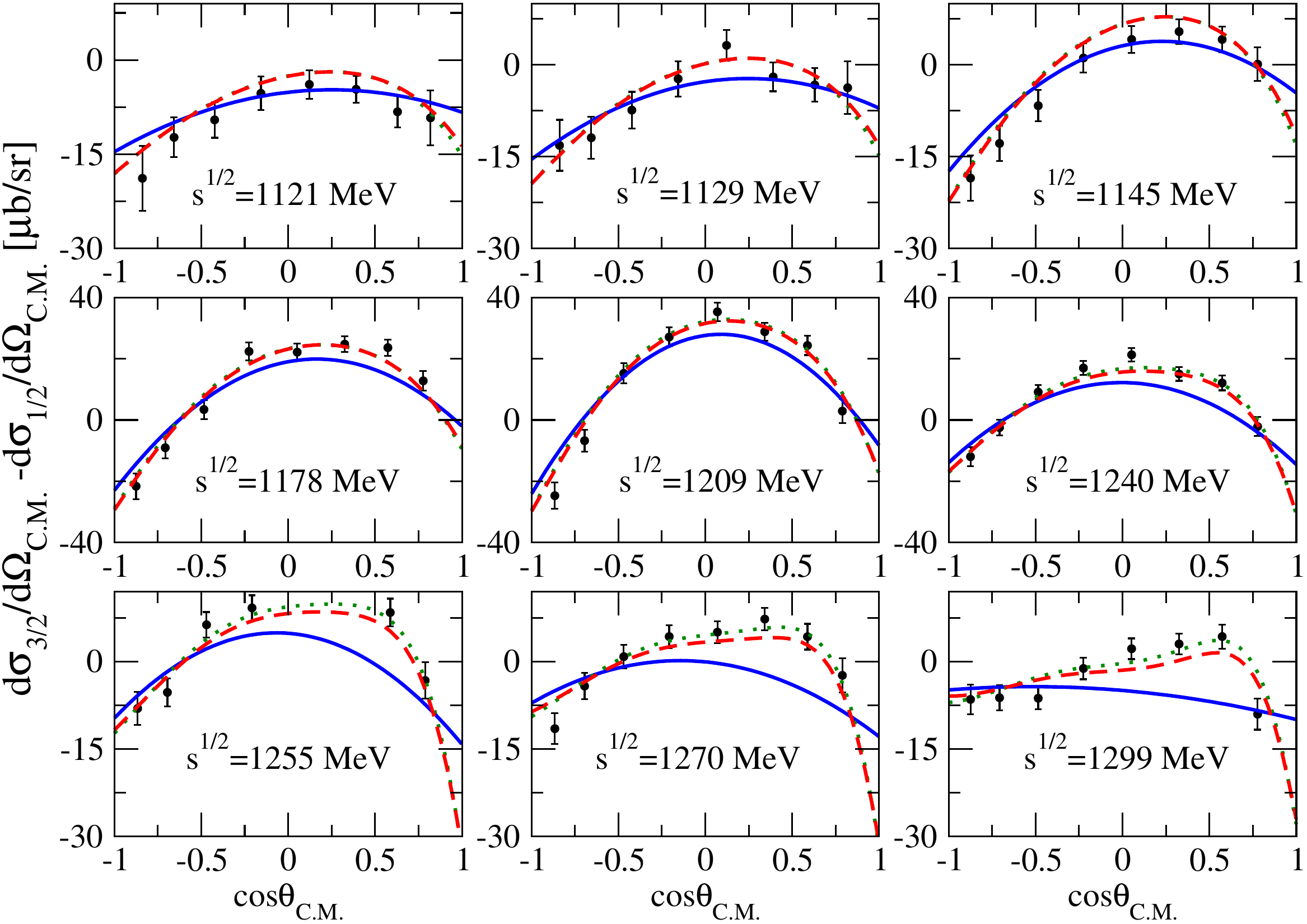}
\caption{Helicity asymmetry for the reaction $\gamma p\to\pi^+ n$ with data taken from \cite{Preobrajenski:2001}.
The assembly of the lines is as in Fig.~\ref{fig:gammaN1dXS}.}
\label{fig:gammaN2dsigma}
\end{figure}

We proceed with a more direct comparison of our results with empirical cross section data.
This is achieved by taking our s- and p-wave multipoles in the calculation of differential and polarization cross sections.
Higher partial waves are supplemented from two different sources. First we compute the higher multipoles within
our given scheme but neglect the final state interaction, i.e. we identify the generalized potential with the
partial-wave production amplitude. Note that the one-loop contribution of a strict $\chi$PT computation does not contribute
to any of the higher partial waves for which we neglect the final state interaction. Second we take the higher multipoles from
the energy dependent analysis \cite{Drechsel:2007if}.  In Figs. \ref{fig:gammaN1dXS}-\ref{fig:gammaN2dXS} we present differential
cross section data from several recent and precise measurements \cite{Leukel:2001,Shafi:2004zn,Preobrajenski:2001,Blanpied:2001ae}
for the $\gamma p\to\pi^0 p$  and $\gamma p\to\pi^+ n$ and $\gamma n\to\pi^- p$ reactions.
Fig.~\ref{fig:gammaN1dXS} illustrates that our multipoles describe the differential cross section for the $\pi^0$ production off
the proton accurately, the effect of higher multipoles being quite small. Recall the absence of the $t$-channel pion exchange
process in this reaction. After adding the effects of the higher multipoles the same is true for the $\pi^\pm$ production cross
sections of Fig.~\ref{fig:gammaN3dXS} and Fig.~\ref{fig:gammaN2dXS}. The solid lines show the contributions from the s- and
p-wave multipoles. The dashed and dotted lines result upon adding up the higher multipoles from our theory and \cite{Drechsel:2007if}
respectively. The effect of using different higher multipoles, dashed versus dotted line, is irrelevant for our conclusions.

We compute the beam and helicity asymmetries of the two
reactions $\gamma p \to \pi^0 p$ and $\gamma p \to \pi^+ \,n$, for which empirical data exist.
Their relation to the multipole amplitudes is given in Appendix~\ref{photoproduction_amplitude}.
The results are shown in Figs.~\ref{fig:gammaN1Sigma}-\ref{fig:gammaN2dsigma}.
After inclusion of higher partial wave contributions all considered observables are nicely reproduced
in the whole energy region considered. Higher partial wave contributions are significant in the production of
charged pions.

Like for $\pi N$ elastic scattering we scrutinize the near-threshold physics for pion photoproduction in some detail.
We compute the s- and p-wave threshold production parameters. Empirically it is well established that in neutral pion photoproduction
there is a strong cusp effect at the $\pi^+\,n$ threshold  \cite{Bergstrom:1996fq,Schmidt:2001vg}. To obtain accurate results we depart
from the isospin formulation and perform a coupled-channel computation in the particle basis using physical masses for the nucleons and
pions. Isospin breaking effects are not considered for the generalized potential being estimated to be of minor importance. No additional
parameters arise.

\clearpage

In Tab.~\ref{table:LOT} we present the s-wave threshold parameters for neutral and charged pion production.  Our results are compared
to the empirical values, which we reproduce quite accurately. Tab.~\ref{table:LOT} recalls  also the results of a $\chi$PT analysis
accurate to chiral order $Q^4$. While $\chi$PT  appears well converging for charged pion production a less convincing convergence
pattern is found for the neutral pion production \cite{Bernard:1994gm}. The order $Q^3$ value obtained in \cite{Bernard:1994gm} for the
$E_{0^+}$ threshold amplitude is $+0.90 \cdot 10^{-3}/m_{\pi^+}$, in striking conflict with experiment. Only the inclusion of the order $Q^4$ terms
lead to a value compatible with the empirical data. A precise prediction of the threshold amplitude is difficult in $\chi$PT  since it depends
sensitively on $Q^4$ counter terms that are not well known. The threshold value given in \cite{Bernard:2001gz} is based on an analysis of
the latest MAMI data set \cite{Schmidt:2001vg}. It is amusing to observe that we obtain accurate threshold parameters already
based on the chiral Lagrangian truncated at order $Q^3$. We reiterate that our threshold values are a consequence of a parameter
fit to the production data at energies excluding the  threshold region.

\begin{table}[t]
\begin{center}
\begin{tabular}{|cll|c|c|c|}
\hline
&&&Present work&$\chi$PT ($Q^4$)&Experiment\\
\hline
$E_{0+}$ &($\pi^+ n$) &[$10^{-3}/m_{\pi^+}$]&$\phantom{-}27.4$&$\phantom{-}28.2$ \cite{Bernard:1996ti}& $28.06\pm 0.27\pm 0.45$ \text{\cite{Korkmaz:1999sg}}\\
\hline
$E_{0+}$ &($\pi^- p$) &[$10^{-3}/m_{\pi^+}$]&$-31.5$&$-32.7$ \cite{Bernard:1996ti}&$-31.5\pm 0.8$ \text{\cite{Kovash:1997tj}}\\
\hline
$E_{0+}$ &($\pi^0\, p$) &[$10^{-3}/m_{\pi^+}$]&$-1.12$&$-1.16$ \cite{Bernard:1995cj}&
$\begin{array}{ll}-1.32\pm 0.05\pm 0.06\, \text{\cite{Bergstrom:1996fq}} \\
-1.23 \pm 0.08 \pm 0.03\, \text{\cite{Schmidt:2001vg}}
\end{array}
$\\
\hline
\end{tabular}
\end{center}
\caption{Threshold values for the $E_{0^+}$ multipole in the $\gamma \,p \to \pi^0 \,p$, $\gamma \,n \to \pi^- \,n$  and
 $\gamma \,p \to \pi^0 \,p$ reactions.}
\label{table:LOT}
\end{table}

\begin{figure}[b]
 \includegraphics*[width=13.5cm]{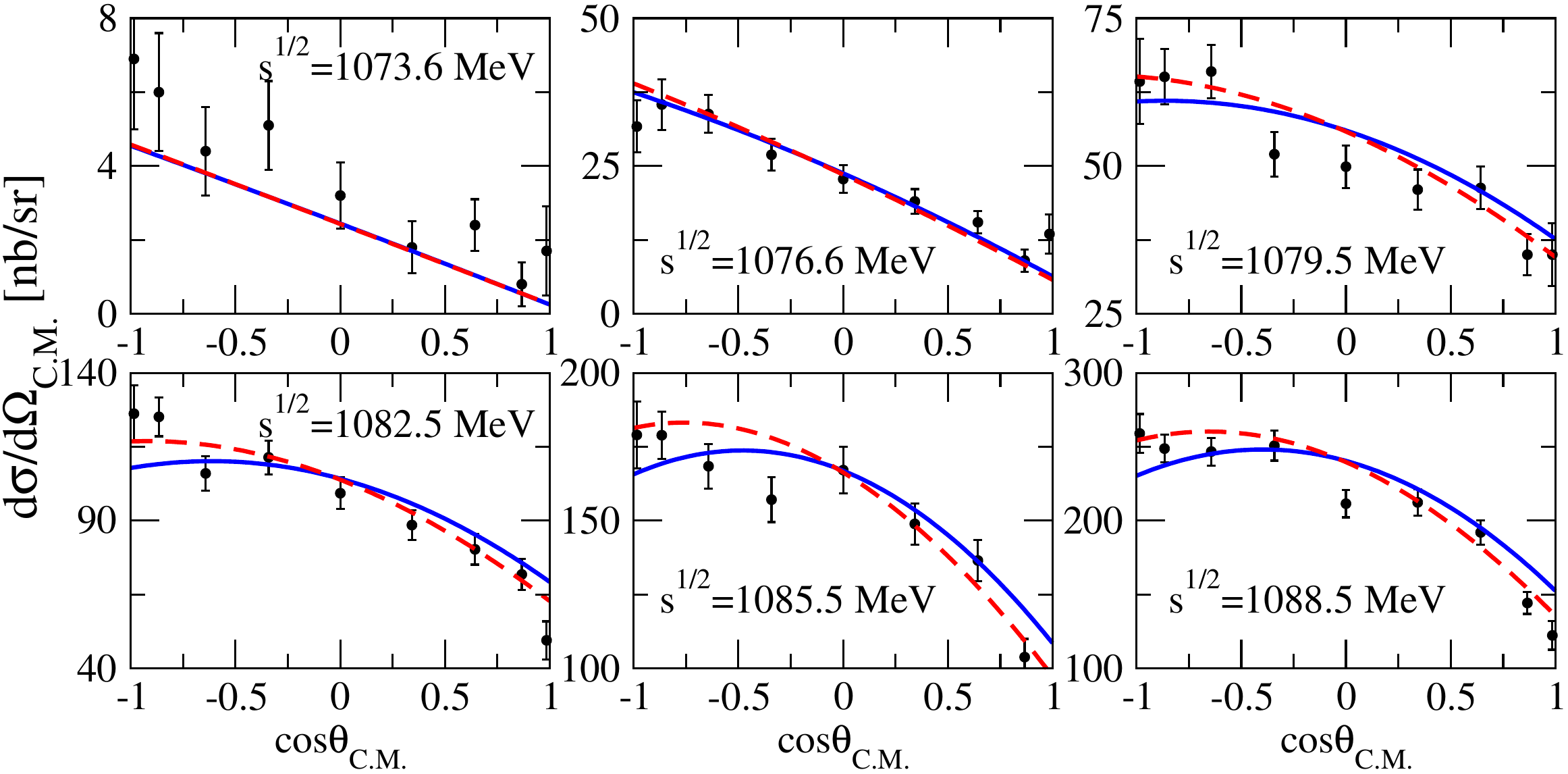}
\caption{Near threshold differential cross section for the 
reaction $\gamma p\to\pi^0 p$ with data taken from \cite{Schmidt:2001,Schmidt:2001vg}.
Shown are results from  our coupled-channel theory including isospin breaking effects as are implied by the use of empirical pion 
and nucleon masses. The solid lines correspond to our calculation with only $s$- and $p$-wave multipoles
included. The effect of higher partial waves is shown by the dashed lines. }
\label{fig:pi0threshold}
\end{figure}

Further significant information on the near-threshold region of neutral pion photoproduction is available.
The s-wave amplitude shows a strong energy dependence with a prominent cusp structure at the $\pi^+ \,n$ threshold
\cite{Schmidt:2001vg}. Since the threshold region shows an intriguing interplay of the s- and p-wave multipoles
we confront our theory in Fig. \ref{fig:pi0threshold} with empirical differential cross section of the Mainz group directly. Given the fact that
we did not fit the parameters to those data an excellent description is achieved.

\begin{table}[t]
\begin{center}
\begin{tabular}{|cll|c|c|c|}
\hline
&&&Present work&$\chi$PT ($Q^3$) &Experiment\\
\hline
$\bar P_1$ &($\pi^0\,  p$)&[$ 10^{-3}/m_{\pi^+}^2$]&$10.2$&$9.4$ \cite{Bernard:1995cj}&$9.46\pm 0.05\pm 0.28$ \cite{Schmidt:2001vg}\\
\hline
$\bar P_2$ &($\pi^0 \, p$) &[$10^{-3}/m_{\pi^+}^2$]&$-10.7$&$-10.0$ \cite{Bernard:1995cj}&$-9.5\pm 0.09\pm0.28$ \cite{Schmidt:2001vg}\\
\hline
$\bar P_3$ &($\pi^0\,  p$)&[$ 10^{-3}/m_{\pi^+}^2$]&$10.3 $&$10.6$ \cite{Bernard:1995cj}&$11.32\pm 0.11\pm 0.34$ \cite{Schmidt:2001vg}\\
\hline
\end{tabular}
\end{center}
\caption{Threshold values of p-wave multipoles in neutral pion photoproduction. }
\label{tab:P123}
\end{table}

\begin{figure}[b]
 \includegraphics*[width=12.5cm]{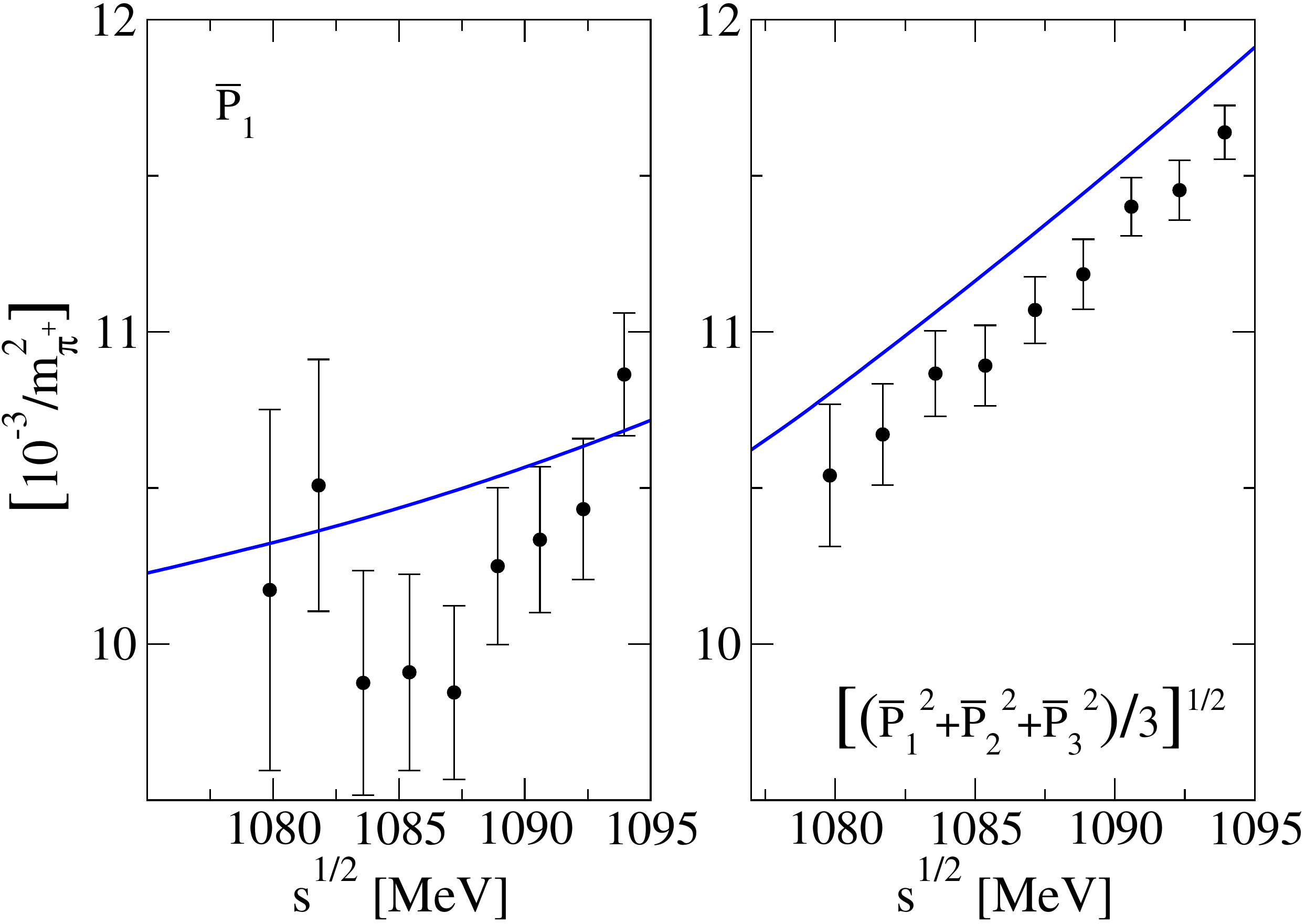}
\caption{Energy dependence of the p-wave amplitude $\bar P_1$ (l.h.p.) and $ (\bar P_1^2 +\bar P_2^2 +\bar P_3^2)^{1/2}/\sqrt{3}$ (r.h.p.) in
neutral pion photoproduction. The data are from \cite{Bergstrom:1997jc}. The solid lines follow from the coupled-channel theory. }
\label{fig:P1}
\end{figure}

From the near-threshold differential cross section two combinations
of p-wave threshold parameters may be extracted. It is customary to
introduce the following three combinations of p-wave multipole
amplitudes,
\begin{eqnarray}
&& \bar p_{\rm cm}\,\bar P_1=3\,E_{1+}+M_{1+}-M_{1-}\,,\qquad
\bar p_{\rm cm}\,\bar P_2=3\,E_{1+}-M_{1+}+M_{1-}\,,
\nonumber\\
&&\bar p_{\rm cm}\, \bar P_3 =2\,M_{1+}+M_{1-}\,,
\end{eqnarray}
which all vanish at the production threshold with $\bar p_{\rm cm}=0$. A complete determination of all three p-wave threshold amplitudes
requires additional information. For this purpose a measurement of the near-threshold photon asymmetry suffices \cite{Schmidt:2001vg}.

At leading orders in a chiral expansion $\bar P_1$ and $\bar P_2$  do not depend on any of the $Q^3$
counter terms. The threshold behavior of $\bar P_1$ and $\bar P_2$ is predicted
in terms of the electromagnetic charge, the pion-nucleon coupling constant and the masses of the pions and
nucleons only \cite{Bernard:1994gm,Bernard:1995cj,Bernard:2001gz}. In Tab. \ref{tab:P123} we recall their numerical values
from \cite{Bernard:1995cj}. The chiral corrections of order $Q^4$ were studied
in \cite{Bernard:2001gz} and shown to be subject to sizeable cancelation effects amongst chiral loops and further counter
terms dominated by isobar exchange processes. In contrast $\bar P_3$ receives a contribution from the $Q^3$ counter term
combination $\bar d_8 + \bar d_9$ and there is no parameter-free prediction accurate to order $Q^3$.

\begin{figure}[t]
 \includegraphics*[width=10.5cm]{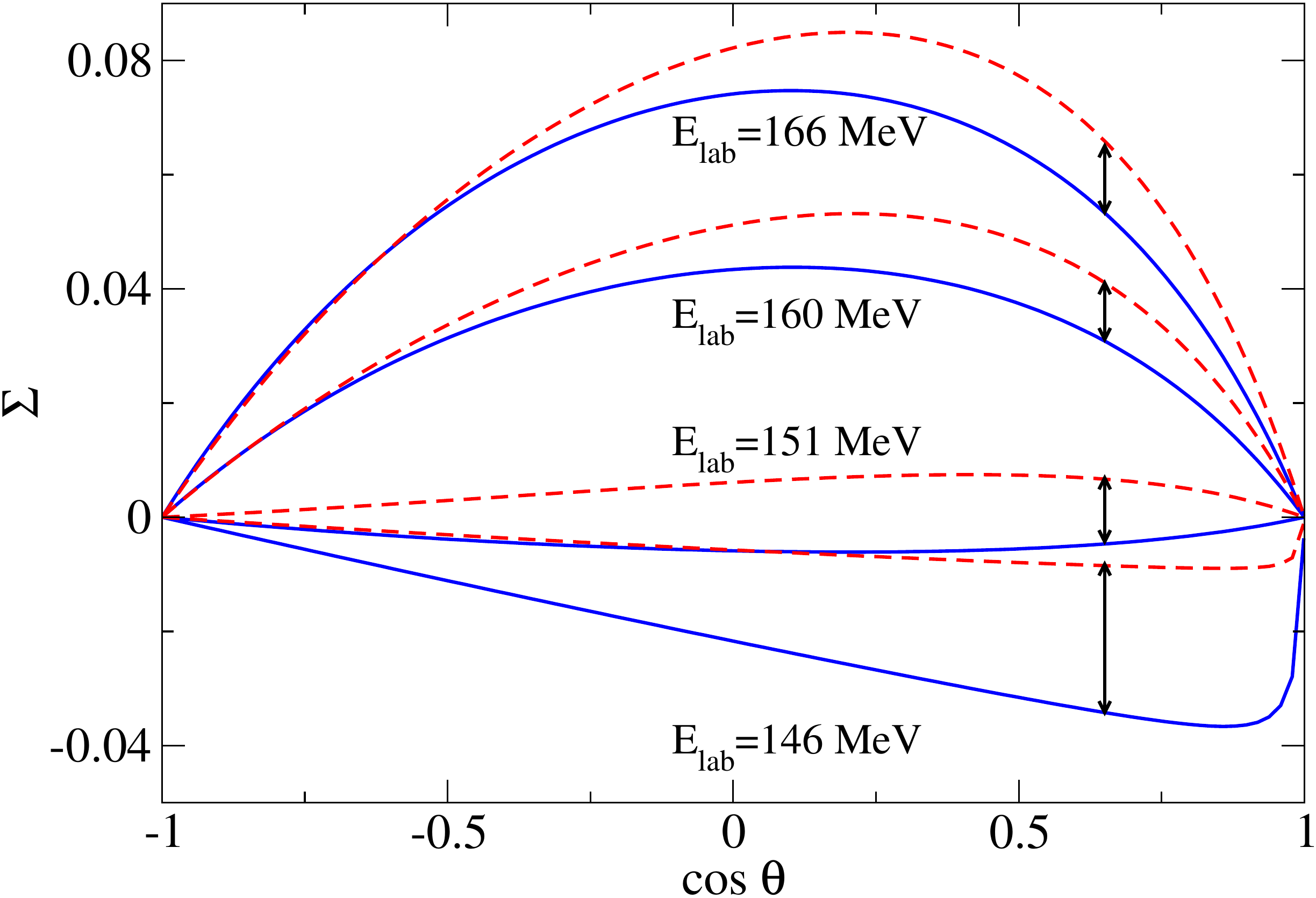}
\caption{Energy dependence of the photon asymmetry in
neutral pion photoproduction from the coupled-channel theory.
The solid lines correspond to our calculation with only $s$- and $p$-wave multipoles
included. The effect of higher partial waves is shown by the dashed lines.}
\label{fig:photon-asymmetry}
\end{figure}

In Tab. \ref{tab:P123} we confront our p-wave threshold parameters with those of \cite{Schmidt:2001vg,Bernard:1995cj}. Though
we are in qualitative agreement with the empirical values obtained by the Mainz group there is a significant discrepancy.
Contrasting conclusions were drawn from the $\chi$PT computation to order $Q^4$ in \cite{Bernard:2001gz}, which was able to
accommodate the threshold  values presented in
\cite{Schmidt:2001vg}. It is interesting to locate the source of the observed discrepancies. For this purpose we provide a
comparison with near-threshold data of the Saskatoon group \cite{Bergstrom:1997jc}. In Fig. \ref{fig:P1} the empirical energy dependence
of $\bar P_1$ is shown against our theoretical result. Given the energy dependence of our theory one would expect a threshold value of
around $10$, a value close to our result. This is striking disagreement with the value of $9.3 \pm 0.09$ extracted in \cite{Bergstrom:1997jc},
based on a different assumption on the energy dependence. In Fig.  \ref{fig:P1} we present also the empirical constraint on the mean p-wave
amplitudes. We find an excellent description of the energy dependence where the overall magnitude is overestimated by less than 2$\%$.

We finally turn to the near-threshold photon asymmetry measurement of the Mainz group \cite{Schmidt:2001vg}.
A direct comparison is not straightforward since an average  from threshold to 166 MeV laboratory photon energies
was performed. In Fig. \ref{fig:photon-asymmetry} we show the result of our computation at four different
energies demonstrating a sign change in the asymmetry at energies below the mean value of 159.5 MeV in the Mainz
experiment. At the mean energy our photon asymmetry is about a factor four to five smaller than the averaged value
presented in \cite{Schmidt:2001vg}. Within our scheme we have no freedom to significantly increase 
that value without destroying the successful description of the photoproduction multipoles at higher energies. It is 
interesting to recall that a negative photon asymmetry was obtained also in  \cite{Hanstein:1996bd,Kamalov:2001qg}
based on a dispersion-relation analysis of photo production data. Given such a sign change an average over the photon 
energy depends on the very details of the averaging procedure. 

A significant effect of higher partial wave contributions on the asymmetry is illustrated by the dashed lines in 
Fig. \ref{fig:photon-asymmetry}. The possible importance of d-wave amplitudes in the photon asymmetry was pointed 
out in \cite{FernandezRamirez:2009jb} recently. 

Since the near-threshold behavior of the photon asymmetry is subject to subtle cancelation effects it would be important 
to extend our analysis to order $Q^4$ and see whether the observed sign change persists.  Also further data taking on 
the photon asymmetry like planned and ongoing at Mainz \cite{Hornidge2003,Bernstein2009} is highly welcome.

\clearpage

\subsection{Proton Compton scattering}

\begin{figure}[b]
 \includegraphics*[width=13cm]{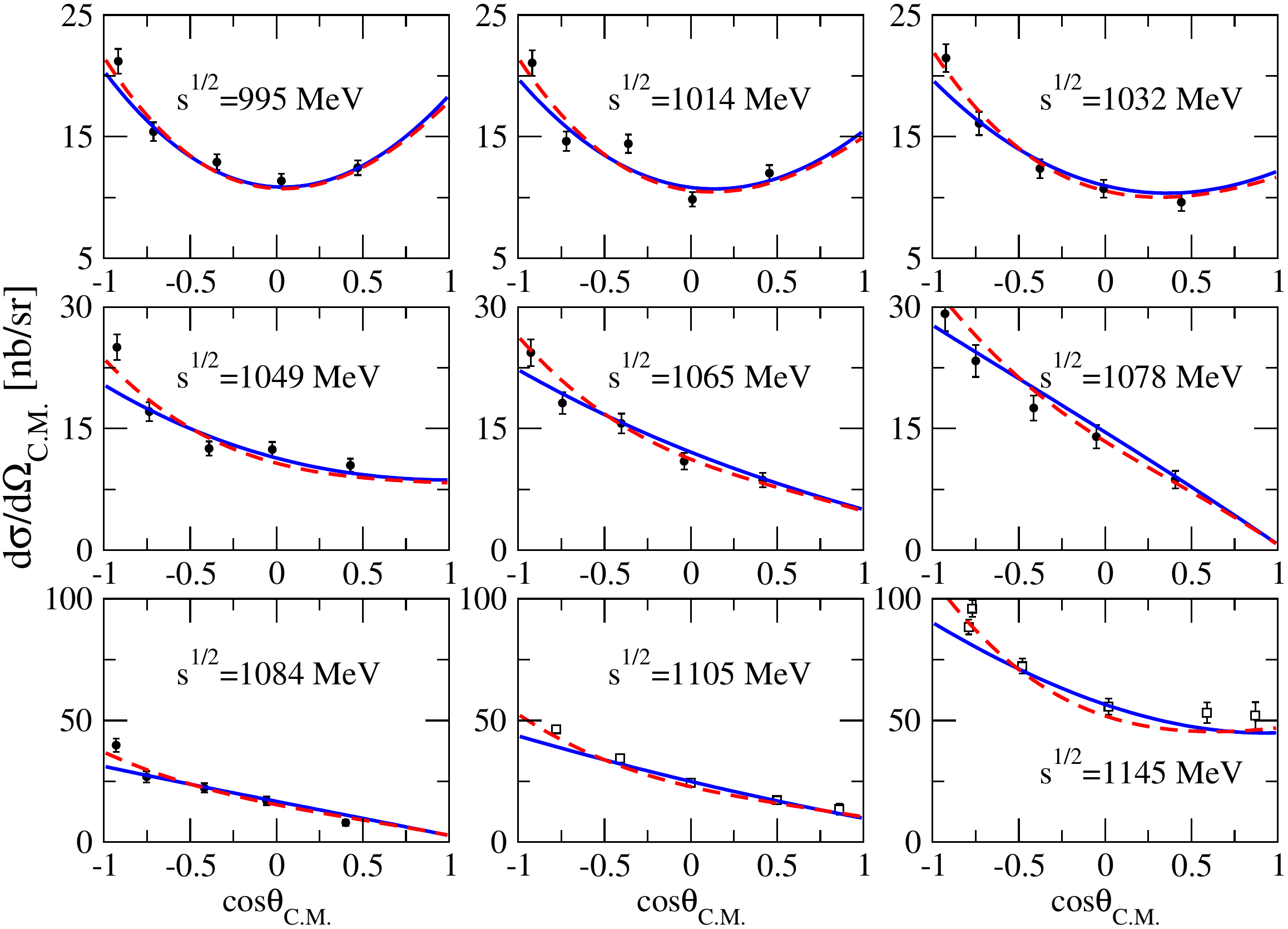}
\caption{Differential cross section for Compton scattering off the proton
as a function of the center-of-mass scattering angle. The data are from  \cite{OlmosdeLeon:2001zn}($\bullet$) and
 \cite{Hallin:1993ft} ($\square$). The solid lines follow from the our partial-wave amplitudes with $J \leq 3/2$.
 The dashed lines show the effect of partial-wave contributions with $J> 3/2$ as  explained in the text.}
\label{fig:ComptondXS}
\end{figure}

We turn to Compton scattering off the proton. Since counter terms start to contribute at the order $Q^4$ there
are no additional free parameters to be determined in our work. The CDD pole parameters are set already from
pion-nucleon scattering and pion photoproduction. In the following we present parameter free results based on a
generalized potential accurate to order $Q^3$.

Since we neglect the intermediate $\gamma N$ states the Compton amplitude reduces to the sum
of two contributions. There is a direct term and a rescattering term, which depends quadratically on the photoproduction
amplitude. The latter has already been calculated and presented in the previous section. In the rescattering part we
consider  $\pi N$ states in s- or  p-waves only. Higher partial-waves cannot be treated here because those were neither considered in
elastic pion-nucleon scattering nor in pion photoproduction. We argue that this suffices for calculating cross sections
in the energy region considered. Higher partial waves are largely suppressed by the $\pi N$ phase space proportional
to $p_{\rm cm}^{2\,L+1}$ entering the integral in (\ref{def-non-linear}). On the other hand for the direct contribution we
take into account all $J=\frac{1}{2}$ and  $J = \frac{3}{2}$ waves. The positive and negative parity partial-wave
amplitudes are of equal importance as a consequence of gauge invariance. We checked that these waves reproduce the Born
diagrams quite accurately up to energies of about 1300 MeV. According to our strategy we have to apply an analytic extrapolation
to the generalized Compton potential. Using the conformal mapping as detailed in section 2.2 we truncate this expansion
at zeroth order. Terms beyond the zeroth order cannot be justified since $Q^4$ counter terms may alter those significantly.

\begin{figure}[t]
 \includegraphics*[width=13cm]{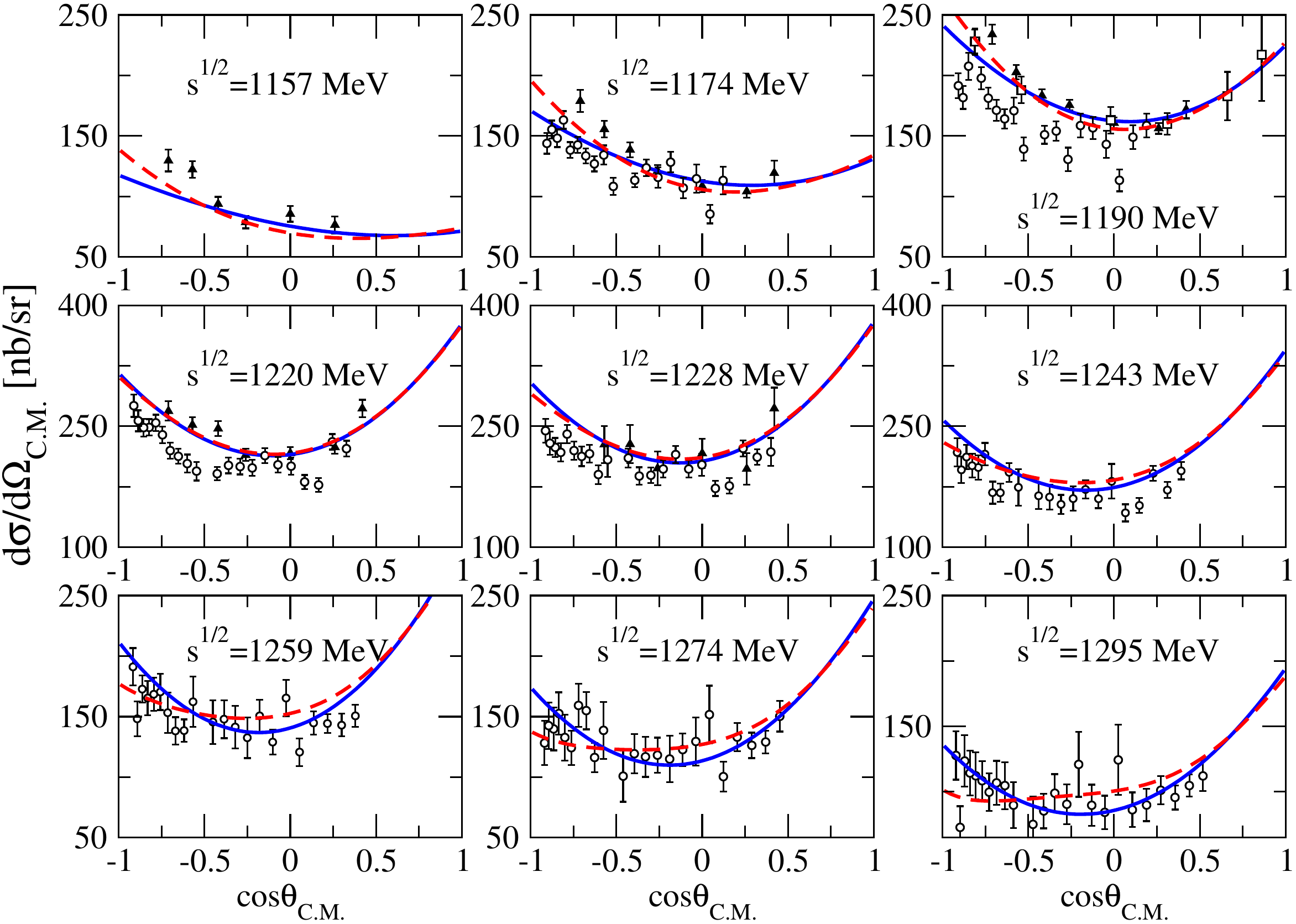}
\caption{Differential cross section for Compton scattering off the proton
as a function of the center-of-mass scattering angle.
The data are from   \cite{Hallin:1993ft} ($\square$),\cite{Blanpied:2001ae}($\blacktriangle$),
\cite{Wolf:2001ha}($\circ$) and the  lines are as in Fig. \ref{fig:ComptondXS}.}
\label{fig:ComptondXS2}
\end{figure}

\begin{figure}[t]
 \includegraphics*[width=13cm]{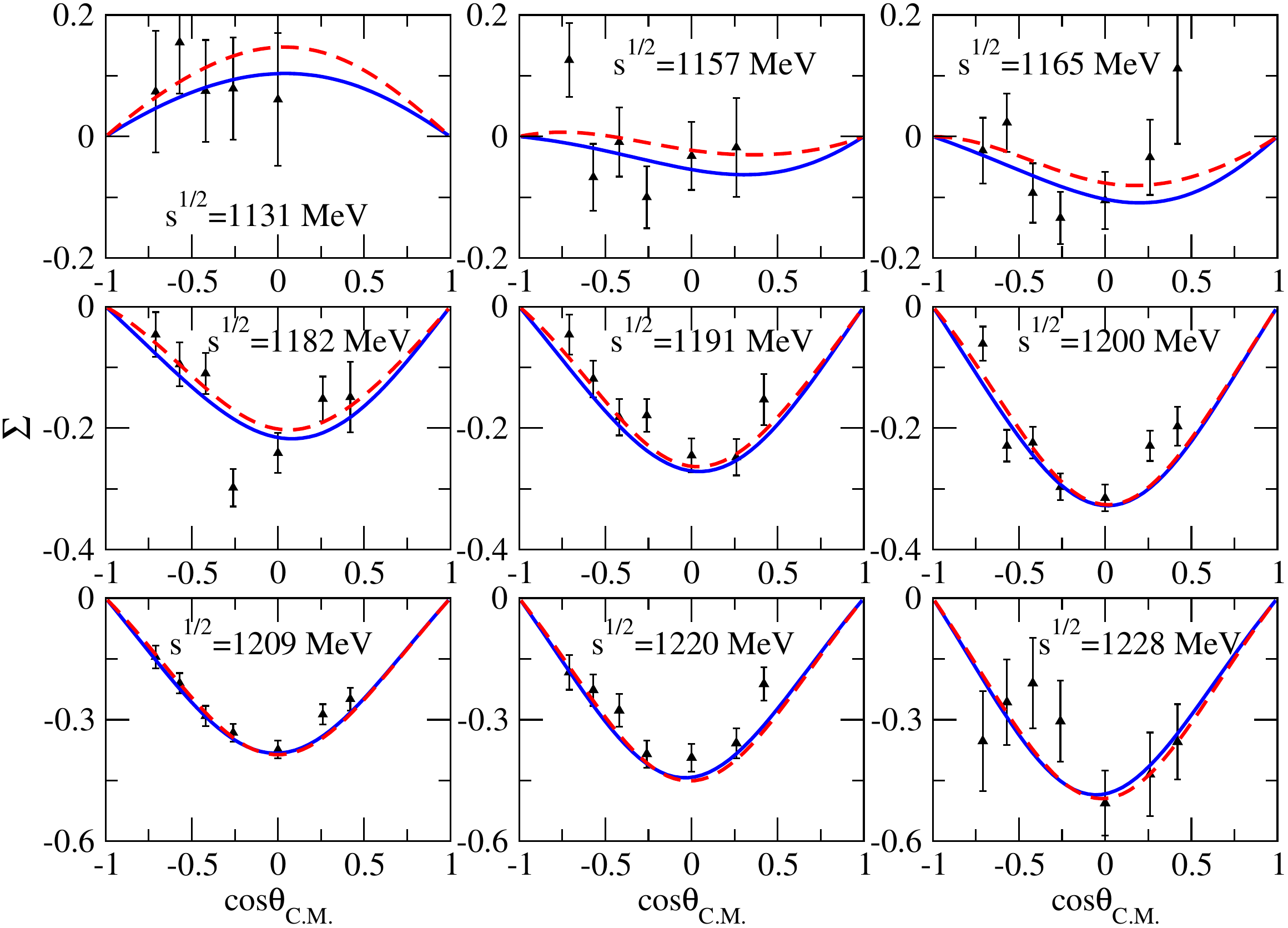}
\caption{Beam asymmetry for Compton scattering off the proton as a function of the center-of-mass scattering angle.
The data are from \cite{Blanpied:2001ae} and the lines  are as in Fig. \ref{fig:ComptondXS}.}
\label{fig:ComptonSigma}
\end{figure}

\begin{figure}[t]
 \includegraphics*[width=13cm]{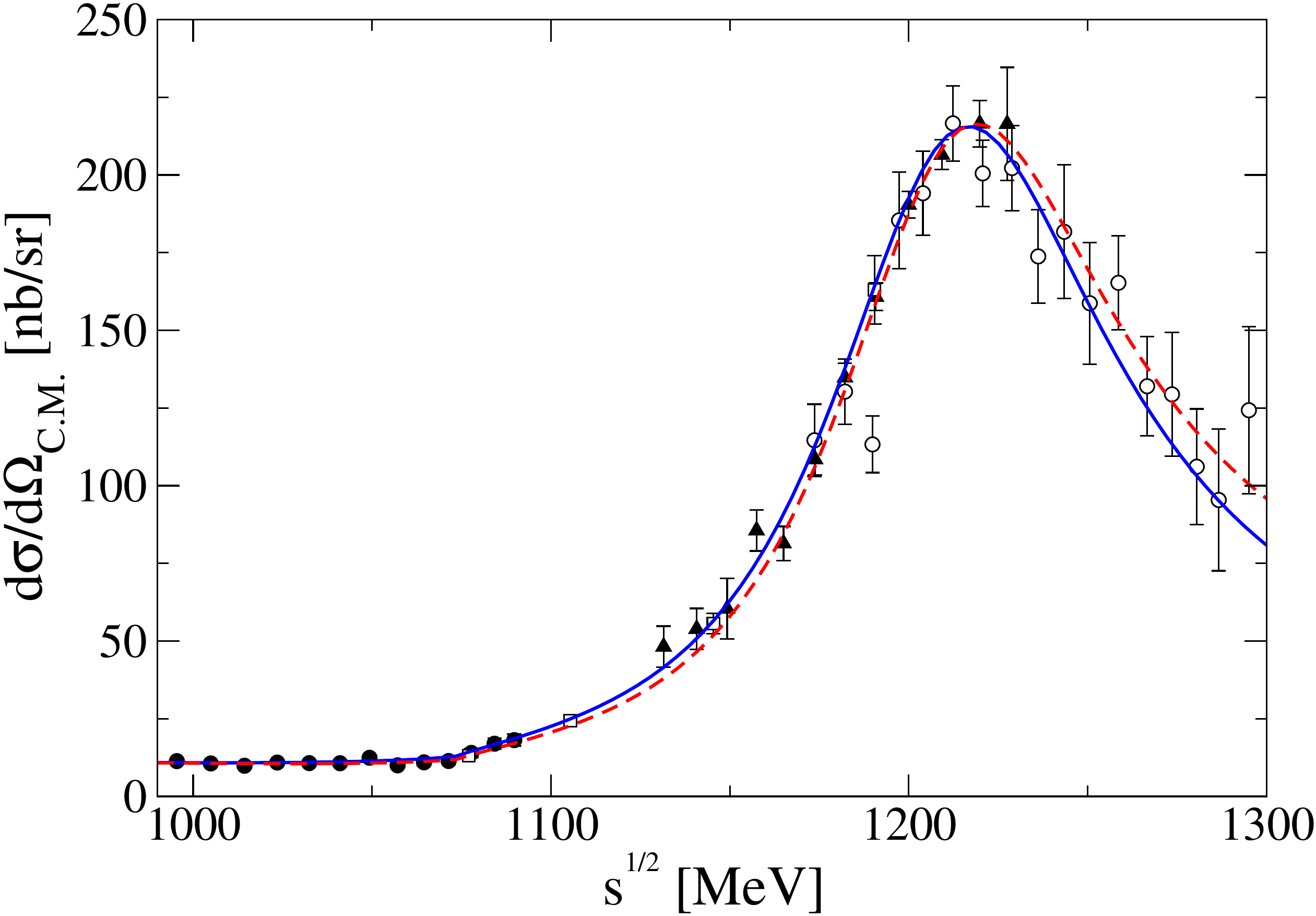}
\caption{Compton scattering off the proton as a function of energy at scattering angle $\theta=90^\circ$.
The data are from  \cite{OlmosdeLeon:2001zn}($\bullet$),
 \cite{Hallin:1993ft} ($\square$),\cite{Blanpied:2001ae}($\blacktriangle$),
\cite{Wolf:2001ha}($\circ$) and the  lines are as in Fig. \ref{fig:ComptondXS}.
The experimental values correspond to scattering angles closest to $\theta=90^\circ$ taken in the interval
$86^\circ<\theta<94^\circ$.}
\label{fig:Compton90}
\end{figure}

Our results for the differential cross section and beam asymmetry are presented in Figs.~\ref{fig:ComptondXS}-\ref{fig:ComptonSigma}
against empirical data. We find agreement with the data taking into account some discrepancy of the
different data sets.  The photon threshold region, the pionproduction threshold region, and the isobar region are equally well
reproduced. This is nicely illustrated by Fig. \ref{fig:Compton90}, which shows the energy dependence of the cross
section at fixed scattering angle  $\theta \simeq 90^\circ$. There seems to be a systematic undershooting of the backward
differential cross section for energies $\sqrt{s}<1150$ MeV. We believe, that a natural explanation for this effect are
missing higher order contributions. Indeed adding to the $J\leq 3/2$ result
contribution from additional partial-wave with $J> 3/2$ as implied by the direct term in our scheme the differential cross section
is increased towards the data in backward direction. This is shown by the dashed
lines in Figs.~\ref{fig:ComptondXS}-\ref{fig:ComptonSigma}.


The low-energy physics of Compton scattering is characterized
by various nucleon polarizabilities. For a definition of the polarizabilities see e.g. \cite{Babusci:1998ww}. It is instructive
to express the latter in terms of partial-wave amplitudes $T^{J}_{\pm,ab}(\sqrt{s}\,)$ as constructed in
Appendix A.4 and used in (\ref{def-non-linear}). In  our case the indices $a$ and $b$ run form one to two,
reflecting the two polarizations of the photon. The electric and magnetic dipole polarizabilities  $\alpha_p$ and $\beta_p$ probe partial-wave
amplitudes with $J=\frac{1}{2}$ and $J=\frac{3}{2}$ at threshold, where there are constraints set by crossing symmetry. To be specific the
following identity holds
\begin{eqnarray}
&& \Delta T^{\,\gamma p \to \gamma p,\frac{3}{2}}_{\pm ,22}(\sqrt{s}=m_p) =  \frac{3}{4}\,\Delta T^{\,\gamma p\to \gamma p,\frac{1}{2}}_{\mp,11}(\sqrt{s}=m_p) \,,
\label{res-crossing-Compton}
\end{eqnarray}
which can be derived from the expressions given in Appendix A.4 and \cite{L'vov:1996xd}. In (\ref{res-crossing-Compton}) the '$\Delta$' indicates the
need to subtract the contributions from the nucleon  exchange processes.
In our approach the crossing relation (\ref{res-crossing-Compton}) is obeyed strictly. This is a consequence of the matching scale
$\mu_M $ in (\ref{def-non-linear}) being identified with the nucleon mass. In turn we recover the one-loop $\chi$PT results
of \cite{Bernard:1995dp}:
\begin{eqnarray}
&& \alpha_p= \frac{3}{16\,\pi\,m_p^3}\,\Delta T^{\,\gamma p\to \gamma p,\frac{1}{2}}_{+ ,11}(m_p)= \frac{5\,e^2\,g_{\pi NN}^2}{384\,\pi^2 \,m_N^2 \,m_\pi}\simeq
13.0\cdot 10^{-4}\,\rm{fm}^3\,, \nonumber\\
&&\beta_p=\frac{3}{16\,\pi\,m_p^3}\,\Delta T^{\,\gamma p\to \gamma p,\frac{1}{2}}_{- ,11}(m_p) = \frac{e^2\,g_{\pi NN}^2}{768\,\pi^2 \,m_N^2 \,m_\pi} \simeq
\phantom{x}1.3\cdot 10^{-4}\,\rm{fm}^3\,,
\label{res-polarizabilities-chPT}
\end{eqnarray}
which are known to agree well with the experimental values
$\alpha_p=(12.0 \pm 0.6)\cdot 10^{-4}\,\rm{fm}^3$, $\beta_p=(1.9 \pm 0.6)\cdot 10^{-4}\,\rm{fm}^3$ \cite{Schumacher:2005an} (for a recent development see \cite{Lensky:2009uv}).
Further polarizabilities probe derivatives of partial-wave amplitudes at threshold. Since we expand the generalized potential
to the zeroth order in the conformal mapping only, we can extract reliably only electric and magnetic dipole polarizabilities.

\clearpage

\section{Summary}

We presented a uniform description of photon- and pion-nucleon scattering data up to and beyond the isobar
region. The results are based on partial-wave amplitudes derived from the chiral Lagrangian formulated
with photon, pion and nucleon fields. Electromagnetic gauge invariance is kept rigourously.

Our study is based on partial-wave amplitudes that have the MacDowell relations and that are free of kinematical constraints.
In the presence of spin the derivation of such amplitudes is tedious  and presented in this work for
pion photoproduction and Compton scattering  for the first time. The partial-wave scattering amplitudes were decomposed into parts
with left- and right-hand cuts only. Such a separation is electromagnetic-gauge invariant. The part with left-hand cuts only defines
a generalized potential, which is evaluated relying on the chiral Lagrangian truncated at order $Q^3$.  The part with right-hand cuts
only is derived as a solution of a non-linear integral equation. The latter combines the constraints set by causality and coupled-channel
unitary in a systematic fashion. The reparametrization invariance of local quantum field theory is obeyed strictly.

The non-linear integral equation does not permit solutions always. The unitarity constraint implies an asymptotic bound
on the generalized potential, which is at odds with any polynomial energy dependence as generated by $\chi$PT. This problem was resolved
by performing an analytic continuation constrained by the known asymptotic bound. The generalized potential
was evaluated in a strict chiral expansion to order $Q^3$. It followed an analytic extrapolation in terms of
suitably constructed conformal mappings. The non-linear integral equations were solved by N/D techniques,
where the presence of two CDD poles in the N/D ansatz reflect the physics of the isobar and Roper resonances. It was
demonstrated that the presence of a CDD pole corresponds to an infinite set of
local counter terms in the chiral Lagrangian.

Even though the parameter set was adjusted to the scattering data avoiding the near threshold region, we obtained results
that are compatible with the empirical threshold behavior nevertheless. The accurate pion-nucleon scattering lengths
as measured in pionic-hydrogen and deuterium systems are recovered. In order to compare with the near threshold data available
for neutral pion photoproduction we considered isospin breaking effects as implied by the empirical pion and nucleon masses.
Despite the fact that we reproduce the differential cross section accurately we predict s- and p-wave threshold parameters that differ somewhat from the
values obtained by the Mainz and Saskatoon groups. We traced this discrepancy to an additional energy dependence in the p-wave multipoles
that was not considered in the threshold analyses of the two groups. As a striking prediction we find a sign change in the photon asymmetry at photon
energies below 160 MeV. Further data on the photon asymmetry, that do not average over a large threshold region, would be highly
welcome. Our threshold analysis was completed by a determination of the electric and magnetic dipole  polarizabilties, for which
their empirical values are reproduced accurately.

A detailed comparison of threshold parameters  with results from $\chi $PT was given. In contrast to a strict $\chi$PT computation
that requires in many cases important contributions from the $Q^4$ terms, we obtained an accurate description of threshold
observable based on the chiral Lagrangian truncated at order $Q^3$. The apparent slow convergence of  $\chi$PT in particular for
the neutral pion photoproduction process is overcome largely by our resummation approach that protects causality and unitarity
rigourously. A more convincing convergence pattern is found in our scheme.

While in our work we focused on  s- and p-waves we provided a comprehensive comparison with scattering data up to energies of
$\sqrt{s} \sim 1300$ MeV. In case of pion-nucleon scattering this was achieved by a comparison with the phase shifts, which
are sufficiently well known from various partial-wave analyses and were reproduced quite accurately. In contrast the electric
and magnetic multipole amplitudes of the pion photoproduction
process suffer still from ambiguities. A comparison with recent analyses of the MAID and SAID groups was offered in this case.
Since we observe in part significant discrepancies amongst the two groups and our results we provided complementary results for
differential cross section, beam and helicity asymmetry. We focused on the processes which were measured most accurately.
For neutral pion photo production  the influence of additional partial waves was shown to be of minor importance.  For charged
pion photo production the role of higher partial wave contributions builds up significantly.
A comparison with a representative selection of high-quality data is offered by combining our results for the
s- and p-wave multipoles with the higher partial-waves from our theory
where we neglected the final state interaction. This way an accurate description of the data was achieved. The empirical data on
Compton scattering off the proton were reproduced with the $J=\frac{1}{2}$ and $J = \frac{3}{2}$ partial-waves. The influence of
higher partial wave contributions was shown to be of minor importance. Differential cross sections and beam asymmetries were
reproduced equally well from threshold to the isobar region.

\section*{Acknowledgments}

The authors acknowledge fruitful discussions with  E.E. Kolomeitsev, O. Scholten  and M. Soyeur.

After submission of our manuscript an improved $\chi$PT calculation of $a_{[S_-]}^{(\pi N)}$
and $a_{[S_+]}^{(\pi N)}$ was announced \cite{Baru:2010xn}. The new values
are $a_{[S_-]}^{(\pi N)}=(0.122\pm 0.001 )\,{\rm fm}$, $a_{[S_+]}^{(\pi N)}=(0.011\pm 0.005 )\, {\rm fm}$.

\clearpage

\begin{appendix}
\section{Partial-wave amplitudes and kinematic constraints}
\label{kinem_sing}

Throughout this appendix we adopt the notation $w_\mu=p_\mu+q_\mu
=\bar{p}_\mu+\bar{q}_\mu$,
where $p_\mu$ and $\bar{p}_\mu$ are the initial and final nucleon $4$-momenta
respectively, whereas $q_\mu$ and $\bar{q}_\mu$ are the  $4$-momenta of initial
pion (photon) and final pion (photon) respectively. The Madelstam variables
are $s= w^2$, $t = (\bar q-q)^2 = (\bar p-p)^2$ and $u= (\bar q-p)^2=(\bar p-q)^2$.
In the center of mass frame the energies
of the nucleons are denoted by $E$, $\bar E$, and the energies
of initial pion (photon) and final pion (photon) by $\omega$, $\bar \omega$.
For the initial state it holds
\begin{eqnarray}
&& E=\Bigg\{
\begin{array}{ll}
\frac{s+m_N^2-m_\pi^2}{2\,\sqrt{s}} \qquad & {\rm for} \; \; \pi N\\
\frac{s+m_N^2}{2\,\sqrt{s}} & {\rm for} \; \; \gamma N
\end{array}
\,,  \qquad \quad
\omega=\Bigg\{
\begin{array}{ll}
\frac{s-m_N^2+m_\pi^2}{2\,\sqrt{s}} \qquad & {\rm for} \; \; \pi N\\
\frac{s-m_N^2}{2\,\sqrt{s}}& {\rm for} \; \; \gamma N
\end{array}
\,,
\nonumber\\
&& E^2=m_N^2+p_{\rm cm}^2\,, \qquad \quad
\omega^2 = \Bigg\{
\begin{array}{ll}
m_\pi^2 + p_{\rm cm}^2\qquad & {\rm for} \; \; \pi N\\
p_{\rm cm}^2& {\rm for} \; \; \gamma N
\end{array} \,,
\label{def-kinematic-generic}
\end{eqnarray}
with the relative momentum $p_{\rm cm} $ of the center of mass frame. Analogous expressions hold for the final state.
The center-of-mass scattering angle $\theta$ is introduced  with
\begin{eqnarray}
p^\mu = (E,0,0, -p_{\rm cm })\,, \qquad \bar p^\mu =(\bar E, -\bar p_{\rm cm}\,\sin \theta ,0, -\bar p_{\rm cm}\,\cos \theta )\,.
\label{}
\end{eqnarray}

\subsection{Isospin decomposition}

\label{Isospin relations}
We recall the isospin decomposition for
the $\pi N$ elastic amplitude and the one for the photoproduction of the pion (see e.g. \cite{Chew:1957tf}).
The elastic amplitude $T^{ab}$ and the production amplitude $T^a$ may be decomposed as follows
\begin{eqnarray}
&& T_{\pi N \to \pi N}^{ab}=T^{(+)}_{\pi N}\,\delta^{ab}+T^{(-)}_{\pi N} \,i\,\varepsilon^{abc}\,\tau_c
= \Big(\frac{1}{3}\,\tau_a\,\tau_b\Big) \,T^{(\frac{1}{2})}_{\pi N}
+ \Big( \delta_{ab}- \frac{1}{3}\,\tau_a\,\tau_b \Big)T^{(\frac{3}{2})}_{\pi N}\,,
\nonumber\\
&& T_{\gamma N \to \pi N}^{a}=T^{(-)}_{\gamma N}\, i\,\varepsilon^{a3c}\,\tau_c + T^{(0)}_{\gamma N}\,\tau^a
+T^{(+)}_{\gamma N}\,\delta^{a3} =\sqrt{\frac{3}{2}}\,\Big(\delta_{a3}- \frac{1}{3}\,\tau_a\,\tau_3 \Big)\,T^{(\frac{3}{2})}_{\gamma N}
\nonumber\\
&& \qquad \quad \;\;\, + \,\sqrt{\frac{1}{3}}\,\tau_a \,\Big( \frac{1+\tau_3}{2}\,T^{(\frac{1}{2})}_{\gamma p} +
\frac{1-\tau_3}{2}\, T^{(\frac{1}{2})}_{\gamma n} \Big) \,,
\label{def-isospin-decomposition}
\end{eqnarray}
where $a$ and $b$ are the isospin indices of the final and initial pions respectively.
The elastic amplitudes $T^{(I)}_{\pi N}$ with isospin $I=\frac{1}{2}$ and $I=\frac{3}{2}$ are linear combinations of $T^{(\pm)}_{\pi N}$.
The production amplitudes $T^{(I)}_{\gamma N}$ with isospin $I=\frac{1}{2},\frac{3}{2}$ and $N=p,n$ are related to  $T^{(0,\pm)}_{\gamma N}$.
It holds
\begin{eqnarray}
&& T_{\gamma p}^{(\frac{1}{2})}= \sqrt{3}\,T^{(0)}_{\gamma N}+\frac{1}{\sqrt{3}}\,T^{(+)}_{\gamma N}+\frac{2}{\sqrt{3}}\,T^{(-)}_{\gamma N}\,,\qquad T^{(\frac{1}{2})}_{\pi N}=T^{(+)}_{\pi N}+2\,T^{(-)}_{\pi N}\,,
\nonumber\\
&& T_{\gamma n}^{(\frac{1}{2})}=\sqrt{3}\,T^{(0)}-\frac{1}{\sqrt{3}}\,T^{(+)}_{\gamma N}-\frac{2}{\sqrt{3}}\,T^{(-)}_{\gamma N}\,,
\nonumber\\
&&T^{(\frac{3}{2})}_{\gamma p}=T^{(\frac{3}{2})}_{\gamma n}=\sqrt{\frac{2}{3}}\,\Big( T^{(+)}_{\gamma N}-T^{(-)}_{\gamma N} \Big)
\,, \qquad \quad \;\;\, T^{(\frac{3}{2})}_{\pi N}=T^{(+)}_{\pi N}-T^{(-)}_{\pi N}\,,
\label{def-isospin-amplitudes}
\end{eqnarray}
where the amplitudes are introduced with respect to normalized states. This is convenient in a coupled-channel framework.
The isospin amplitudes of (\ref{def-isospin-amplitudes}) are related to the ones in the particle basis
as follows
\begin{eqnarray}
T_{\,\pi^0\, p\to\,\pi^0 \,p}&=&
\frac{1}{3}\,T^{(\frac{1}{2})}_{\pi N}+\frac{2}{3}\,T^{(\frac{3}{2})}_{\pi N}=T_{\pi^0 \,n\to\pi^0\, n}\,,
\nonumber\\
T_{\pi^+ n\to\pi^+ n}&=&
\frac{2}{3}\,T^{(\frac{1}{2})}_{\pi N}+\frac{1}{3}\,T^{(\frac{3}{2})}_{\pi N}= T_{\pi^- p\to\pi^- p}\,,\nonumber\\
T_{\pi^0\, p\to\pi^+ n}&=&
\frac{\sqrt{2}}{3}\left(T^{(\frac{1}{2})}_{\pi N} - T^{(\frac{3}{2})}_{\pi N}\right)=-T_{\pi^0 n\to\pi^- p}\,,
\label{}
\end{eqnarray}
and
\begin{eqnarray}
T_{\gamma p\to\pi^0\, p}&=&
\sqrt{\frac{1}{3}}\,T_{\gamma p}^{(\frac{1}{2})}+\sqrt{\frac{2}{3}}\,T_{\gamma p}^{(\frac{3}{2})}\,, \qquad
T_{\gamma n\to\pi^0 n}=-\sqrt{\frac{1}{3}}\,T_{\gamma n}^{(\frac{1}{2})}+\sqrt{\frac{2}{3}}\,T_{\gamma n}^{(\frac{3}{2})}\,,
\nonumber\\
T_{\gamma p\to\pi^+ n}&=&\sqrt{\frac{2}{3}}\,
T_{\gamma p}^{(\frac{1}{2})}-\sqrt{\frac{1}{3}}\,T_{\gamma p}^{(\frac{3}{2})}\,, \qquad
T_{\gamma n\to\pi^- p}=\sqrt{\frac{2}{3}}\,T_{\gamma n}^{(\frac{1}{2})}+\sqrt{\frac{1}{3}}\,T_{\gamma n}^{(\frac{3}{2})}\,.
\end{eqnarray}

\subsection{$\pi N$ elastic scattering}

The on-shell pion-nucleon scattering amplitude may be decomposed into
invariant amplitudes
\begin{eqnarray}
&&T(\bar q, q; w)=F_1(s,t)+F_2(s,t) \, \wslash
\nonumber\\
&& \qquad \quad  \;\;\;\;\, =F_+(\sqrt{s},t)\,\Bigg\{\frac{1}{2}\,+\frac{\wslash}{2\sqrt{w^2}} \Bigg\}+
F_-(\sqrt{s},t)\,\Bigg\{\frac{1}{2}\,-\frac{\wslash}{2\sqrt{w^2}}\Bigg\} \,,
\label{def-Fpm}
\end{eqnarray}
where we suppress the reference to the isospin channel and the on-shell nucleon Dirac-spinors.
The two sets of amplitudes introduced in (\ref{def-Fpm}) are related to each other
through
\begin{eqnarray}
F_\pm=F_1\pm F_2\, \sqrt{s} \,.
\label{piN-amplitude-relations}
\end{eqnarray}
Since the amplitudes $F_1$, $F_2$ are free of
kinematic singularities, so are the amplitudes $F_\pm$ except for the relation
\begin{eqnarray}
F_-(+\sqrt{s},t)=F_+(-\sqrt{s},t)\,.
\label{MacDowell-piN}
\end{eqnarray}
One may adopt the viewpoint that there is in fact only a single invariant amplitude $F_+(\sqrt{s},t)$ which
characterizes the scattering amplitude fully. The latter amplitude is free of any kinematic constraints.

The partial-wave amplitudes with definite parity, $P$, and total angular momentum, $J$, are
most economically given in terms of the $F_{\pm}$ amplitudes. We recall
from  \cite{Jacob:1959at,Lutz:2001yb}
\begin{eqnarray}
&&t_{\pm}^J (\sqrt{s}\,)
=\pm\,\frac{E\pm m_N}{2\,m_N}\left(p_{\rm cm}^2\right)^{J-\frac{1}{2}}
\int_{-1}^1\frac{d \cos \theta}{2}\,\Bigg\{ \frac{P_{J-\frac{1}{2}}(\cos \theta)}{\left(p_{\rm cm}^2\right)^{J-\frac{1}{2}}}\,
F_{\pm}(\sqrt{s},t)
\nonumber\\
&& \qquad\qquad  -\,\frac{P_{J+\frac{1}{2}}(\cos \theta)}{\left(p_{\rm cm}^2\right)^{J+\frac{1}{2}}}\,(E\mp m_N)^2 \,F_{\mp}(\sqrt{s},t) \Bigg\}\,,
\label{def-tpm}
\end{eqnarray}
where $\pm$ characterizes the helicity of the nucleon.  The parity of the amplitude $t_{\pm}^J(\sqrt{s}\,)$ is given by $P = \pm $ for
$J-\frac{1}{2}$ odd and $P= \mp$ for $J-\frac{1}{2}$ even.

The expressions (\ref{def-tpm}) imply the standard
normalization of the helicity amplitudes $t_{\pm}^J(\sqrt{s}\,)$.  We introduce partial-wave amplitudes
$T_\pm^J(\sqrt{s}\,)$,  that enjoy the MacDowell relation
\begin{eqnarray}
&& T_{-}^J(+\sqrt{s}\,)= T_{+}^J(-\sqrt{s}\,)\,,
\label{}
\end{eqnarray}
and that are free of kinematic singularities and zeros, where we
admit constraints at $s=0$ as the only exception. From (\ref{def-tpm}) it follows  that
the helicity amplitudes $t_{\pm}^J (\sqrt{s}\,)$ suffer from kinematic zeros at threshold and
pseudo-threshold defined by the conditions
\begin{eqnarray}
E\pm m_N=\frac{s+m_N^2-m_\pi^2}{2\,\sqrt{s}} \pm m_N=0\,.
\end{eqnarray}
This conclusion is possible only since the invariant functions $F_\pm (\sqrt{s},\,t)$ are kinematicly
unconstrained. With
\begin{eqnarray}
&& T_{\pm}^J(\sqrt{s}\,)=\frac{2\, m_N \sqrt{s}}{E\pm m_N}
\left(\frac{\sqrt{s}}{p_{\rm cm}}\right)^{2\,J-1}\,t_{\pm}^J(\sqrt{s}\,)\,,\quad
\label{def-Tamplitude}
\end{eqnarray}
we recover the covariant partial-wave amplitudes derived previously in \cite{Lutz:2001yb} up to
a factor $s^{J}$. That factor $s^{J}$ is introduced in the present work as to render the phase-space density
\begin{eqnarray}
\rho_{\pm}^J(\sqrt{s}\,)=
-\Im{\,\frac{1}{T_{\pm}^J(\sqrt{s}\,)} }= \frac{p_{\rm cm }\,(E\pm m_N)}{8\,\pi \,s}\left(\frac{p_{\rm cm}^2}{s}\right)^{J-1/2} \,,
\end{eqnarray}
constant at asymptotically large $\sqrt{s}$.

We turn to the technical details of the effective field theory.
The tree-level diagrams (\ref{piN-tree-level}) implied by the Lagrangian density (\ref{Lagrangian}) lead to the
invariant amplitudes
\begin{eqnarray}
&& F^{(I)}_{\pm}(\sqrt{s},t)=\Big\{ \frac{1}{2\, f^2}\,(\pm\sqrt{s} - m_N)
+ \frac{c_4}{f^2} \,( \bar{q}\cdot q - (\pm\sqrt{s}-m_N)^2)
\nonumber\\
&& \quad +\,
\frac{d_3 }{2 \,f^2 \,m_N^3}\,(s- m_N^2 -m_\pi^2)\, (s-m_N^2  - \bar{q}\cdot q)
\,(s-m_N^2 + m_\pi^2 - 2 \,\bar{q}\cdot q)
\nonumber\\
&& \quad +\,\Big[ 2\,\frac{ d_1+d_2}{f^2\,m_N}\,(\bar{q}\cdot q)+4\,\frac{d_5}{ f^2\,m_N}\,m_\pi^2\,
\Big]\,(s - m_N^2 - \bar q\cdot q)
\Big\}\,C_{-}^{(I)}
\nonumber\\
&& +\, \Big\{ -4\,\frac{c_1 }{f^2}  \,m_\pi^2
+\frac{c_2}{2\, f^2 \,m_N^2} (s-m_N^2 - m_\pi^2) \,(s-m_N^2 + m_\pi^2 - 2\, \bar{q}\cdot q )
\nonumber\\
&& \quad +\, 2\,\frac{c_3}{f^2} \,(\bar{q}\cdot q) + \frac{d_{14}-d_{15}}{f^2\, m_N}\,
(s-m_N^2 - \bar{q}\cdot q) \,( \bar{q}\cdot q-(\pm\sqrt{s}- m_N )^2)
\Big\}\,C_{+}^{(I)}
\nonumber\\
&& -\,\frac{(g_{A}-2\, m_\pi^2 \,d_{18})^2}{4 \,f^2} \frac{(\pm \sqrt{s}-m_N)^2}{m_N \pm \sqrt{s}}\,C_{s,N}^{(I)}
\nonumber\\
&& +\, \frac{(g_{A}-2\, m_\pi^2 \,d_{18})^2}{4\, f^2}
\Big(\frac{4\, m_N^2 (\pm\sqrt{s} - m_N)}{u - m_N^2}
 + m_N \pm \sqrt{s}\Big)\,C_{u,N}^{(I)}\,,
\label{res-Fpm-piN}
\end{eqnarray}
where the isospin coefficients $C^{(I)}_{...}$ are detailed in Tab. \ref{tab:isospin-piN}. The amplitudes (\ref{res-Fpm-piN})
receive contributions from different orders in a strict chiral expansion. While the counter terms $c_{...}$  start to contribute
at order $Q^2$, the counter terms $d_{...}$  contribute at $Q^3$.

At chiral order $Q^3$ one needs to consider a set of ultraviolet divergent one-loop diagrams.
The counter term combinations $d_1+d_2$, $d_3$, $d_5$ and $d_{14}-d_{15}$ are required for their renormalization.
In the heavy-baryon chiral perturbation theory the corresponding loop diagrams were computed in
\cite{Fettes:1998ud}. The results were expressed in terms of the frame-dependent amplitudes, $g$ and $h$, that arise
naturally in the center-of-mass frame with two-component nucleon spinors,
\begin{eqnarray}
T_{\rm cm} = \frac{E +m_N}{2\,m_N}\,\Big\{ g + i\,\vec \sigma \cdot ( \vec q_{\pi,\rm{out}} \times \vec q_{\pi,\rm{in}} )\,h\, \Big\}  \,.
\label{def-cm-amplitude-piN}
\end{eqnarray}
with the Pauli matrices $\sigma_i$. The invariant amplitudes $F_\pm$ of
(\ref{def-Fpm}) follow with
\begin{eqnarray}
F_+ = g - h \, p_{\rm cm}^2 \,\cos \theta \,, \qquad \quad F_- = -h\,(E + m_N)^2 \,.
\label{Fpmtogh}
\end{eqnarray}
We recall from \cite{Fettes:1998ud} the explicit result
\allowdisplaybreaks[1]
\begin{eqnarray}
g^{(I)}&=&\Big\{ i\,   \frac{\omega^2}{8\pi f^4}\,p_{\rm cm}
+\frac{g_{A}^2}{32\pi \,f^4} \,(m_\pi^2-2\, t) \,\Big[ m_\pi+
\frac{2 \,m_\pi^2-t}{2\,\sqrt{-t}} \arctan{\frac{\sqrt{-t}}{2\,m_\pi}}\Big]
\nonumber\\
&& \quad +\,\frac{g_{A}^4}{24\pi \,f^4 \,\omega^2}\,
\big(t-2\, m_\pi^2+2\, \omega^2\big)
 \big(i\,p_{\rm cm}^3-m_\pi^3\big) \Big\} \,C^{(I)}_+\, ,
\nonumber\\
&+&  \Big\{\frac{2\,\omega}{f^2}(2\,m_\pi^2-t)(\bar{d}_1+\bar{d}_2)
+\frac{4\,\omega^3}{f^2}\,\bar{d}_3+\frac{8\,\omega \,m_\pi^2}{f^2}\,\bar{d}_5
\nonumber\\
&& \quad +\,\frac{1}{32\pi^2 \,f^4}\,\Big[
\frac{\omega}{3}\,(4\,m_\pi^2-t)\sqrt{1-\frac{4\,m_\pi^2}{t}}
\log\frac{\sqrt{4\,m_\pi^2-t}+\sqrt{-t}}{2\,m_\pi}
\nonumber\\
&& \quad -\,4\,\omega^2\,p_{\rm cm}\Big(
\log{\frac{\omega+p_{\rm cm}}{m_\pi}}-i\,\frac{\pi}{2}\Big)
+\frac{\omega}{9}\left(18\,\omega^2-12\,m_\pi^2+\frac{5}{2}\,t\right)\Big]
\nonumber\\
&& \quad -\,\frac{g_{A}^2}{24 \pi^2 f^4}\,\Big[\frac{1}{4}\,\omega\,(5\,t-8\,m_\pi^2)\,
\sqrt{1-\frac{4\,m_\pi^2}{t}}\,
\log\frac{\sqrt{4\,m_\pi^2-t}+\sqrt{-t}}{2\,m_\pi}
\nonumber\\
&& \quad +\,2\,\omega\, m_\pi^2-\frac{13}{24}\,\omega\, t\Big]+
\frac{g_{A}^4}{96\pi^2 \,f^4 \,\omega^2}\,\Big[2\,p_{\rm cm}^3\Big(
\log{\frac{\omega+p_{\rm cm}}{m_\pi}}-i\,\frac{\pi}{2}\Big)
\nonumber\\
&&\quad+\,2\,m_\pi^2\,\omega
-\frac{5}{3}\,\omega^3 \Big]\,(2\,m_\pi^2-2\,\omega^2-t) \Big\} \,C^{(I)}_-\, ,
\nonumber\\
h^{(I)}&=& \Big\{\frac{2\,\omega}{f^2}\,(\bar{d}_{14}-\bar{d}_{15})
\nonumber\\
&& \quad -\,\frac{g_{A}^4}{24\pi^2 \,f^4\,\omega^2}\,
\Big[m_\pi^2\,\omega+\frac{\omega^3}{6}+p_{\rm cm}^3\,\Big(
\log{\frac{\omega+p_{\rm cm}}{m_\pi}}-i\frac{\pi}{2}\Big)     \Big] \Big\} \,C_+^{(I)}\, ,
\nonumber\\
&+& \Big\{ \frac{g_{A}^2}{32\pi \,f^4}\, \Big[
\frac{t-4\,m_\pi^2}{2\,\sqrt{-t}}\arctan{\frac{\sqrt{-t}}{2\,m_\pi}}-m_\pi
\Big]
\nonumber\\
&& \quad +\,\frac{g_{A}^4}{24\,\pi\,f^4\,\omega^2}\left(i\,p_{\rm cm}^3-m_\pi^3\right) \Big\}\,C_-^{(I)}\, ,
\label{quote:Fettes}
\end{eqnarray}
where  the $\bar{d}$'s are the scale independent
renormalized coupling constants according to \cite{Fettes:1998ud}.

At order $Q^3$ it is legitimate to add up the loop contribution (\ref{quote:Fettes}) to the tree-level contribution
(\ref{res-Fpm-piN}) via the relation (\ref{Fpmtogh}) provided the substitution rules
\begin{eqnarray}
&& \frac{g_A-2\, m_\pi^2 \,d_{18}}{f}  \to \frac{g_{\pi NN}}{m_N} \,, \qquad  d_{i}\to 0\,,
\label{def-substitution}
\end{eqnarray}
are used in (\ref{res-Fpm-piN}). In the one-loop result (\ref{quote:Fettes}) we replace $g_A \to f\,g_{\pi NN} /m_N$ for
convenience. The loop contributions are specified in a manner such that it is justified to use the empirical
pion-decay constant with $f = f_\pi$ in the tree-level expressions (\ref{res-Fpm-piN}).

\newpage

\subsection{Pion photoproduction amplitudes}
\label{photoproduction_amplitude}

We decompose the on-shell pion photoproduction amplitude into four invariant amplitudes, where we will suppress
reference to the channel specifics (see (\ref{def-isospin-decomposition})).
There are different choices possible. Chew et al. constructed in \cite{Chew:1957tf}
the CGLN amplitudes, $A(s,t), B(s,t), C(s,t)$ and $D(s,t)$ that are free of kinematic constraints \cite{Ball:1961zza,Bardeen:1969aw}.
Since we are interested in a partial-wave decomposition of the invariant
amplitudes it is advantageous to use a different set showing the  MacDowell symmetry \cite{MacDowell:1959zza} in addition.
Like for elastic pion-nucleon scattering it is possible to introduce invariant amplitudes $F_{1}^\pm(\sqrt{s},t)$  and
$F_{2}^\pm(\sqrt{s},t)$  with
\begin{eqnarray}
F^-_i(+\sqrt{s},t) = F^+_i(-\sqrt{s},t)\,,
\label{MacDowell-photoproduction}
\end{eqnarray}
that are free of kinematic constraints. We recall the definition of the CGLN amplitudes and relate them to the
$F_{1-2}^\pm(\sqrt{s},t)$ amplitudes
\begin{eqnarray}
&&i\,T_\mu(\bar q,q;w)=\gamma_5\,\gamma_\mu\, \qslash\,A(s,t) -2\,\gamma_5\,\big[\left(\bar{q}\cdot q\right) P_\mu - \left(P\cdot q\right) \bar{q}_\mu\big]\, B(s,t)
\nonumber\\
&&  \qquad  + \,\gamma_5\,\big[\left(\bar{q}\cdot q\right) \gamma_\mu - \qslash \,\bar{q}_\mu\big]\,C(s,t)
\nonumber\\
&&  \qquad +\,2\,\gamma_5\,\big[\left(P\cdot q\right) \gamma_\mu - \qslash \,P_\mu
-m_N \,\gamma_\mu \,\qslash\big]\, D(s,t)
\nonumber\\
&& \qquad = \sum_{i=1}^2\,\Big\{ F^+_{i}(\sqrt{s}\,,\,t)\, L^{(i,+)}_\mu(\bar q, q,w)
+ F^-_{i}(\sqrt{s}\,,\,t)\, L^{(i,-)}_\mu(\bar q, q,w)\Big\}\,,
\nonumber\\
&&L^{(1, \pm )}_\mu(\bar q,q,w)=\gamma_5\,\Big(  \frac{1}{2}\pm \frac{\wslash}{2\,\sqrt{s}} \Big)\,
\gamma_\mu\,\qslash \,, \qquad
\nonumber\\
&& L^{(2,\pm )}_\mu(\bar q,q,w)=
\gamma_5\,\Big(  \frac{1}{2}\pm \frac{\wslash}{2\,\sqrt{s}} \Big)\,
\Big( (\bar q \cdot  q)\,\gamma_\mu -\qslash\,\bar q_\mu \Big)\,,
\label{def-F12pm}
\end{eqnarray}
where  $P_\mu=(p_\mu+\bar{p}_\mu)/2$. The two representations in (\ref{def-F12pm}) are equivalent
in the presence of  an on-shell initial nucleon spinor, $p^2=m_N^2$ and
\begin{eqnarray}
F^\pm_{1}&=&A-B \,(\bar{q}\cdot q)- D \left(m_N\pm \sqrt{s}\,\right) ,
\nonumber\\
F^\pm_{2}&=&C-D-B\left(m_N\pm \sqrt{s}\,\right).
\label{F12pm-ABCD}
\end{eqnarray}
>From (\ref{F12pm-ABCD}) it follows our claim that the
amplitudes $F^\pm _{1}(\sqrt{s}\,,\,t)$, and $F^\pm_{2}(\sqrt{s}\,,\,t)$ are free of kinematic constraints
but (\ref{MacDowell-photoproduction}).

The partial-wave amplitudes with definite parity, $P$, and total angular momentum, $J$, corresponding to anti-aligned
and aligned helicities of the initial nucleon and photon are readily expressed in terms
of the invariant  amplitudes  $F^\pm_{1,2}$ introduced in (\ref{def-F12pm}). We derive
\begin{eqnarray}
&& t^J_{\pm,1} (\sqrt{s}\,)
=\frac{( \bar p_{\rm cm}\, p_{\rm cm})^{J-\frac{1}{2}}}{\sqrt{2\,m_N\,\sqrt{s}}}
\sqrt{\frac{\bar E\pm m_N}{2\, m_N}}\int_{-1}^1\,\frac{d \cos \theta}{2}\,
\Bigg\{ \Big[ \mp
2\,\omega\,\sqrt{s}\,F^\pm_{1 }(\sqrt{s},t)
\nonumber\\
&& \qquad + \big((\sqrt{s}\mp m_N)\,\bar \omega\,\omega
\pm m_N\,\bar  p_{\rm cm}\,p_{\rm cm} \,\cos \theta\big)\, F^\pm_{2 }(\sqrt{s},t)
\nonumber\\
&& \qquad +\, \omega\,\sqrt{s}\,(\bar{E}\pm m_N)\, F^\mp_{2}(\sqrt{s},t)  \Big]\,\omega\,(\bar{E}\mp m_N)\,
\frac{P_{J+\frac{1}{2}}(\cos \theta)}{(\bar p_{\rm cm}\, p_{\rm cm})^{J+\frac{1}{2}}}
\nonumber\\
&& \quad \pm\,
\Big[-\omega\,\sqrt{s}\,\big(2\,F_1^\mp(\sqrt{s},t) \pm(\bar E\mp m_N)\,F_2^\pm(\sqrt{s},t)\big)
\nonumber\\
&& \qquad +\,\Big(m_N\,\bar p_{\rm cm}\, p_{\rm cm}\,\cos \theta\mp(\sqrt{s}\pm m_N)\,\bar \omega\,\omega \Big)\,F^\mp_{2 }(\sqrt{s},t)\Big]\,
\frac{P_{J-\frac12}(\cos \theta)}{(\bar p_{\rm cm}\, p_{\rm cm})^{J-\frac{1}{2}}}\,\Bigg\}\,,
\nonumber\\
&&t^J_{\pm,2}(\sqrt{s}\,)
=( \bar  p_{\rm cm}\,p_{\rm cm})^{J-\frac{1}{2}}\sqrt{\frac{\sqrt{s}}{2\,m_N}}\sqrt{\frac{\bar E\pm\bar
    m_N}{2\, m_N}}\int_{-1}^1\frac{d\cos \theta}{2}\,
\Bigg\{
\nonumber\\
&& \quad \mp\,
\Big[{\textstyle{\sqrt{\frac{2\,J+3}{2\,J-1}}}}\,\omega \,(\bar{E}\pm m_N) \,F^\mp_{2 }(\sqrt{s},t)
\nonumber\\
&& \qquad +\,{\textstyle{\sqrt{\frac{2\,J-1}{2\,J+3}}}}\,\bar p_{\rm cm}\, p_{\rm cm}\, \cos \theta\, F^ \pm_{2}(\sqrt{s},t) \Big]
\,\omega\,(\bar{E}\mp m_N)\,\frac{P_{J+\frac12}(\cos \theta)}{(\bar p_{\rm cm}\, p_{\rm cm})^{J+\frac{1}{2}}}
\nonumber\\
&& \quad \pm\,
\Big[{\textstyle{\sqrt{\frac{2\,J+3}{2\,J-1}}}}\,\bar p_{\rm cm}\,p_{\rm cm} \,\cos \theta \,F^\mp_{2 }(\sqrt{s},t)
\nonumber\\
&&\qquad +\,{\textstyle{\sqrt{\frac{2\,J-1}{2\,J+3}}}}\,(\bar E\mp m_N)\,\omega \,F^\pm_{2 }(\sqrt{s},t) \Big]
\,\frac{P_{J-\frac12}(\cos \theta)}{(\bar p_{\rm cm}\, p_{\rm cm})^{J-\frac{1}{2}}} \Bigg\}\,,
\label{tviaF12pm}
\end{eqnarray}
except for  $J=\frac{1}{2}$ with   $t^{\frac{1}{2}}_{\pm,2}=0$. The parity of the amplitudes $t_{\pm,1}^J(\sqrt{s}\,)$ and
$t_{\pm,2}^J(\sqrt{s}\,)$ is given by $P = \pm $ for $J-\frac{1}{2}$ odd and $P= \mp$ for $J-\frac{1}{2}$ even.
The electric and magnetic multipoles \cite{Chew:1957tf,Donnachie:1966}
can be expressed in terms of helicity partial-wave amplitudes $t^{IJ}_{\pm , 1}(\sqrt{s}\,)$ and
$t^{IJ}_{\pm , 2}(\sqrt{s}\,)$ as follows
\begin{eqnarray}
&&E^{(I)}_{(J-\frac{1}{2})+}=\frac{m_N}{4\,\pi\,\sqrt{s}}\,\frac{-\sqrt{2}}{2\,J+1}\left(t^{IJ}_{+,1}-
\sqrt{\frac{2\,J-1}{2\,J+3}}\,t^{IJ}_{+,2}\right)
\left\{
\begin{array}{lll}
\sqrt{\frac{1}{3}} & \quad{\rm for} &\quad  I=\frac{1}{2} \\
\sqrt{\frac{3}{2}} &\quad  {\rm for} & \quad I=\frac{3}{2} \\
\end{array}
\right. , \nonumber\\
&&M^{(I)}_{(J-\frac{1}{2})+}=\frac{m_N}{4\,\pi\,\sqrt{s}}\,\frac{-\sqrt{2}}{2\,J+1}\left(t^{IJ}_{+,1}+
\sqrt{\frac{2\,J+3}{2\,J-1}}\,t^{IJ}_{+,2}\right)
\left\{
\begin{array}{lll}
\sqrt{\frac{1}{3}} & \quad{\rm for} &\quad  I=\frac{1}{2} \\
\sqrt{\frac{3}{2}} &\quad  {\rm for} & \quad I=\frac{3}{2} \\
\end{array}
\right. ,\nonumber\\ \\
&&E^{(I)}_{(J+\frac{1}{2})-}=\frac{m_N}{4\,\pi\,\sqrt{s}}\,\frac{+\sqrt{2}}{2\,J+1}\left(t^{IJ}_{-,1}+
\sqrt{\frac{2\,J+3}{2\,J-1}}\,t^{IJ}_{-,2}\right)\left\{
\begin{array}{lll}
\sqrt{\frac{1}{3}} & \quad{\rm for} &\quad  I=\frac{1}{2} \\
\sqrt{\frac{3}{2}} &\quad  {\rm for} & \quad I=\frac{3}{2} \\
\end{array}
\right. \,,
\nonumber\\
&&M^{(I)}_{(J+\frac{1}{2})-}=\frac{m_N}{4\,\pi\,\sqrt{s}}\,\frac{-\sqrt{2}}{2\,J+1}\left(t^{IJ}_{-,1}-
\sqrt{\frac{2\,J-1}{2\,J+3}}\,t^{IJ}_{-,2}\right)
\left\{
\begin{array}{lll}
\sqrt{\frac{1}{3}} & \quad{\rm for} &\quad  I=\frac{1}{2} \\
\sqrt{\frac{3}{2}} &\quad  {\rm for} & \quad I=\frac{3}{2} \\
\end{array}
\right. \,, \nonumber
\label{def-multipols}
\end{eqnarray}
where we made explicit the isospin structure. Note that the conventional electric and magnetic multipole amplitudes
are introduced with respect to isospin states that are not normalized. Since the partial-wave helicity amplitudes are
defined with respect to normalized states the factors $\sqrt{\frac{1}{3}}$ and $\sqrt{\frac{3}{2}}$ arise.

The differential cross section, $d\sigma /d \Omega $,  the beam asymmetry, $\Sigma$, and the
helicity asymmetry, $d\,(\sigma_{3/2}-\sigma_{1/2})/d \Omega $, are
expressed conveniently in terms of helicity matrix elements (see e.g. \cite{Arndt:1989ww}). It holds
\begin{eqnarray}
&& \frac{d\sigma}{d\Omega}=\frac{\bar{p}_{\rm cm}}{2\,p_{\rm cm}}
\left(|H_N|^2+|H_{SA}|^2+|H_{SP}|^2+|H_D|^2\right)\,,
\nonumber\\
&& \Sigma\,\frac{d\sigma}{d\Omega}=\frac{\bar{p}_{\rm cm}}{p_{\rm cm}}\,
\Re\left(H_{SP}H_{SA}^*-H_NH_D^*\right)\,,
\nonumber\\
&& \frac{d\sigma_{3/2}}{d\Omega}-\frac{d\sigma_{1/2}}{d\Omega}=
\frac{\bar{p}_{\rm cm}}{p_{\rm cm}}
\left(|H_{SP}|^2+|H_D|^2-|H_N|^2-|H_{SA}|^2\right)\,,
\end{eqnarray}
with $x= \cos \theta$ and\footnote{Eq. 3.1 of \cite{Arndt:1989ww}  misses a factor
$l$ in the expression for $H_{SA}$ in front of the $E_{(l+1)-}$ term.}
\begin{eqnarray}
&& H_{N\;\;}=\frac{m_N}{4\pi\,\sqrt{s}}\,\cos \frac{\theta}{2}\,
\sum_J\left(t^J_{+,1}-t^J_{-,1}\right)
\left(P'_{J+\frac{1}{2}}(x) -P'_{J-\frac{1}{2}}(x)\right)  \,,
\nonumber\\
&& H_{SA}=-\frac{m_N}{4\pi\,\sqrt{s}}\,\sin \frac{\theta}{2}\,
\sum_J \left(t^J_{+,1}+t^J_{-,1}\right)
\left(P'_{J+\frac{1}{2}}(x)+P'_{J-\frac{1}{2}}(x)\right)  \,,
\nonumber\\
&& H_{SP}=\frac{m_N}{4\pi\,\sqrt{s}}\,\frac{-\sin\theta \cos{\frac{\theta}2}}{\sqrt{(J-\frac{1}{2})(J+\frac{3}{2})}}\,
\sum_J \left(t^J_{+,2}-t^J_{-,2}\right)
\left(P''_{J+\frac{1}{2}}(x) -P''_{J-\frac{1}{2}}(x)\right) \,,
\nonumber\\
&& H_{D\;\;}=\frac{m_N}{4\pi\,\sqrt{s}}\,\frac{\sin\theta \sin {\frac{\theta}2}}{\sqrt{(J-\frac{1}{2})(J+\frac{3}{2})}}\,
\sum_J \left(t^J_{+,2}+t^J_{-,2}\right)
\left(P''_{J+\frac{1}{2}}(x)
+P''_{J-\frac{1}{2}}(x)\right)\,. \nonumber
\label{}
\end{eqnarray}

We construct partial-wave amplitudes, $T^J_{\pm ,1} (\sqrt{s}\,)$ and $T^J_{\pm ,2} (\sqrt{s}\,)$,
that are free of kinematic constraints. According to (\ref{tviaF12pm})
the helicity partial-wave amplitudes have zeros at thresholds and pseudothresholds
\begin{eqnarray}
\bar E\pm m_N=\frac{s+m_N^2-m_\pi^2}{2\,\sqrt{s}} \pm m_N=0\,.
\end{eqnarray}
Taking into account (\ref{def-kinematic-generic}) and
\begin{eqnarray}
&& \bar p_{\rm cm}\, p_{\rm cm} \,\cos \theta = \frac{t}{2}+\bar{E}\,E-m_N^2\,,
\label{def-kinematic-gammaN}
\end{eqnarray}
it is readily seen that the  two amplitudes
\begin{eqnarray}
&& \sqrt{2\,m_N\,\sqrt{s}}\,\sqrt{\frac{2\, m_N}{\bar E\pm m_N}}\,\frac{t^J_{\pm,1}(\sqrt{s}\,)}{(\bar p_{\rm cm}\, p_{\rm cm})^{J-\frac12}} \,,
\nonumber\\
&& \sqrt{\frac{m_N}{2\,\sqrt{s}}}\,
\sqrt{\frac{2\, m_N}{\bar E\pm m_N}}\,\frac{t^J_{\pm,2}(\sqrt{s}\,)}{(\bar  p_{\rm cm}\,p_{\rm cm})^{J-\frac12}}\,,
\end{eqnarray}
are free of kinematic constraints but one exception.  From (\ref{tviaF12pm}) it follows that
the amplitudes $t^J_{\pm ,1}(\sqrt{s}\,)$ and $t^J_{\pm , 2}(\sqrt{s}\,)$ are linear dependent at $\omega = p_{\rm cm}=0$
even after pulling out the common phase-space factor. In that limit the two partial-wave amplitudes are characterized
by the terms proportional to $ \bar p_{\rm cm}\, p_{\rm cm}\,\cos \theta\, F_2^\mp(\sqrt{s},t)$ in (\ref{tviaF12pm})\footnote{The amplitudes
$F^{\pm}_1(\sqrt{s}\,,\,t)$ and $F^{\pm}_2(\sqrt{s}\,,\,t)$
possess dynamical singularities at the $\gamma N$ threshold: while the s- and u-channel nucleon exchange processes are associated with
pole terms of the form $1/(s-m_N^2)$ and $1/(u-m_N^2)$, the pion-exchange process leads to a term $1/(t-m_\pi^2)$.
These dynamical singularities imply a singular behavior of the partial-wave amplitude at $\omega=p_{\rm cm}=0$.
They appear, however, only in the pole graphs which have to be treated explicitly.\label{footnote3}}. It follows
\begin{eqnarray}
\frac{t^J_{\pm,1}(\sqrt{s}\,)
-\frac{m_N}{\sqrt{s}}\,\sqrt{\frac{2\,J- 1}{2\,J+3}}\,
t^J_{\pm,2}(\sqrt{s}\,)}{(\bar p_{\rm cm}\,p_{\rm cm})^{J-\frac12}}\stackrel{p_{\rm cm}\to 0}{=}O(p_{\rm cm}) \,.
\label{eg-constraint}
\end{eqnarray}
Owing to (\ref{eg-constraint}) it is  straightforward to derive the following set of partial-wave amplitudes
\begin{eqnarray}
T^J_{\pm,a}(\sqrt{s}\,)&=&\sqrt{\frac{4\,m_N^2\,\sqrt{s}}{\bar E\pm m_N}}
\left(\frac{s}{\bar p_{\rm cm}\,p_{\rm cm}}\right)^{J-\frac12}\!t^J_{\pm,b}(\sqrt{s}\,)
\left(\begin{matrix}
\frac{\sqrt{s}}{p_{\rm cm}}&0\\
-\sqrt{\frac{2\,J-1}{2\,J+3}}\,\frac{m_N}{p_{\rm cm}}&1   \end{matrix}\right)_{ba} \!,
\label{t12toT12}
\end{eqnarray}
which are free of any constraints. As in the case of $\pi N$ scattering the factor of $s^J$
was chosen so that the $\gamma N$ phase-space density is asymptotically constant (see (\ref{phase-space-text})).
The MacDowell relations
\begin{eqnarray}
T_{-,1}^J(+\sqrt{s}\,)=- T_{+,1}^J(-\sqrt{s}\,)\,,\qquad T_{-,2}^J(+\sqrt{s}\,)= + T_{+,2}^J(-\sqrt{s}\,)
\label{MacDowell-photoproduction-partial-wave}
\end{eqnarray}
hold.

We turn to the details of the effective field theory. The Lagrangian density (\ref{Lagrangian}) leads with
(\ref{gN-piN-tree-level}, \ref{def-F12pm})
to the tree-level expressions
\begin{eqnarray}
F_1^{\pm}&=&-\frac{  e \,(g_{A} -2\,m_\pi^2\,d_{18}) \, (m_N \pm \sqrt{s}\,)
} {4 \,f\,m_N\, (m_N \mp \sqrt{s}\,)}\,\tilde C^{}_{s,N}
\nonumber\\
&-&\frac{  e\, (g_{A} -2\,m_\pi^2\,d_{18}) \, (m_N^2 \pm 2 \sqrt{s}\, m_N + u)
}{4\, f\,m_N \,(m_N^2 - u)}\,\tilde C^{}_{u,N}
\nonumber\\
&-&\frac{2 \, e\,
(2\, m_N^2 \pm 2 \sqrt{s}\, m_N - m_\pi^2 + \bar{q}\cdot q)}{f \,m_N}\,
(d_8 \, C^{}_{+}+d_9\,C^{}_{0})
\nonumber\\
 &-&\frac{e\, d_{20} \,(m_N \mp \sqrt{s}\,)\, (m_N^2 - s + \bar{q}\cdot q)}{4 \,f \,m_N^2}\,C^{}_{-}\,,
\nonumber\\
F_2^{\pm}&=&- \frac{ e \, (g_{A} -2\,m_\pi^2\,d_{18})\, (s - u) }{(m_N^2 - u) \,(m_N \mp \sqrt{s}\,)\,(t-m_\pi^2)}\,\frac{m_N}{f}\,C^{}_{-}
\nonumber\\
&-&\frac{e \,(g_{A} -2\,m_\pi^2\,d_{18})}{f\,(m_N^2 -  u)}\,\tilde C^{}_{u,N}
-\frac{  e \,(g_{A} -2\,m_\pi^2\,d_{18}) }{(m_N^2 - u)\, (m_N \mp \sqrt{s}\,)}\,\frac{m_N}{f}\,\big( C^{}_{+}+C^{}_{0}\big)
\nonumber\\
&-&\frac{2\,  e \,(m_N \mp \sqrt{s}\,)}{f\, m_N}\,(d_8\,C^{}_{+}+d_9\, C^{}_{0})
\nonumber\\
&+&\frac{e\, d_{20} \, ( m_N^2 + m_\pi^2 -s)}{4\, f\, m_N^2}\, C^{}_{-}
+\frac{ e\, (2\, d_{21} -d_{22})}{2\, f}\,C_- \,,
\label{F12pm-tree-level}
\end{eqnarray}
with the channel dependent coefficients $C$ and $\tilde C$ given in Tab. \ref{tab:isospin-gN}.
The tree-level expressions (\ref{F12pm-tree-level}) receive contributions at different orders in the chiral expansion.

At chiral order $Q^3$  a set of ultraviolet divergent one-loop diagrams contribute.
In the heavy-baryon chiral perturbation theory the corresponding loop diagrams were computed in
\cite{Bernard:1994gm,Fearing:2000uy}. Their renormalization  requires the presence of the $Q^3$ counter terms
$d_{8-9,20-22}$. The results of \cite{Bernard:1994gm,Fearing:2000uy} for the $\gamma\, p\to \pi^0 \,p$ production were expressed in terms
of the electric and magnetic multipoles (\ref{def-multipols}). The one-loop expressions for the $\gamma\, p \to \pi^+\,n$ reaction
were presented  in  \cite{Fearing:2000uy} in terms of amplitudes, that arise
naturally in the center-of-mass frame with two-component nucleon spinors \cite{Chew:1957tf}. We introduce such amplitudes in
a notation tailored for our purpose,
\begin{eqnarray}
&& i\,\vec T_{\rm cm} = \sqrt{\frac{\bar E +m_N}{2\,m_N}\,\frac{E +m_N}{2\,m_N}}\,
\Big\{ \vec \sigma\,f_1 -i\, (\vec \sigma \cdot \vec q_\pi)\, (  \vec \sigma \times \vec q_\gamma)\,f_2
\nonumber\\
&& \qquad \qquad \qquad \qquad \qquad \quad \;+\, (\vec \sigma \cdot \vec q_\gamma)\,\vec q_\pi\,f_3  + (\vec \sigma \cdot \vec q_\pi)\,\vec q_\pi\,f_4
 \Big\}  \,,
\label{def-cm-amplitude-photoproduction}
\end{eqnarray}
where we use a different overall normalization and pion and photon three momenta that are not normalized.
The Coulomb gauge with vanishing zero component of the photon wave function is assumed
in (\ref{def-cm-amplitude-photoproduction}). The original amplitudes of Chew, Goldberger and
Low \cite{Chew:1957tf} possess kinematical zeros at vanishing photon or pion momentum. In part such kinematical constraints are a consequence
of using normalized photon and pion momenta in the definition of the amplitudes. The amplitudes $f_2$ and $f_3$ will be shown to be free of
kinematical zeros at $\vec q_\pi =0$ or $\vec q_\gamma =0$. This is not the case for the remaining amplitudes $f_1$ and $f_4$  as can be seen by
expressing the invariant amplitudes of (\ref{def-F12pm}) in terms of the center-of-mass amplitudes
\begin{eqnarray}
 F_1^{+}&=&\frac{(\sqrt{s} + m_N)^2 - m_\pi^2}{2\,\sqrt{s}}
\left( - \,\frac{\sqrt{s} + m_N}{2\,\sqrt{s}}\,f_2 + \frac{\bar{q}\cdot q}{s- m^2_N}\,f_4\right)\, ,
\nonumber\\
 F_1^{-}&=& \frac{f_1}{\sqrt{s} -m_N} -  \frac{\sqrt{s} + m_N}{2\,\sqrt{s}}\,\frac{\bar{q}\cdot q}{\sqrt{s}-m_N}\,f_3 \, ,
\nonumber\\
 F_2^{+}&=&\frac{(\sqrt{s} + m_N)^2 - m_\pi^2}{2\,\sqrt{s}}\,\frac{f_4}{\sqrt{s} -m_N}\, , \qquad \quad \;\;
F_2^{-}= \frac{\sqrt{s} + m_N}{2\,\sqrt{s}}\,f_3\, .
\label{calFtoF12pm}
\end{eqnarray}
Since the invariant amplitudes $F_1^\pm$ and $F_2^\pm$ were shown to be free of kinematical constraints, it follows that the amplitudes
$f_1$ and $f_4$ must vanish at $\sqrt{s}=m_N$.

>From \cite{Bernard:1994gm,Fearing:2000uy} we reconstruct the four center-of-mass amplitudes
\begin{eqnarray}
&&\frac{f_1}{\omega} = 4\,e \left( \frac{\bar{d}_9}{f} \,C_0  +\frac{\bar{d}_8}{f} \,C_+ \right) \cos \theta\, \bar{p}_{\rm cm}
\nonumber\\
&& \quad \;+\,\Big\{ \frac{e \,g_A }{16\,\pi\, f^3 }\left(\frac{m_\pi^2}{\bar \omega}\,\arcsin{\frac{\bar{\omega}}{m_\pi} }
+i\,\bar{p}_{\rm cm}\right)
+ \frac{e\, g_A^3 \,\cos \theta \,\bar{p}_{\rm cm}}{48\,\pi\, f^3}
\Big(2\,\frac{ m_\pi^3}{\bar{\omega}^3}+ 3\, \frac{m_\pi}{\bar \omega }
\nonumber\\
&& \qquad \;-\,  \frac{3 \,m_\pi ^2}{\bar{\omega}^2} \,\arcsin{\frac{\bar{\omega}}{m_\pi} }
 + 2 \,i\,\bar{p}_{\rm cm}\,\frac{m_\pi ^2}{\bar{\omega}^3}   + i\, \frac{\bar{p}_{\rm cm}}{\bar \omega }  \Big) \Big\}\,C_+
\nonumber\\
&& \quad \;+ \,\Big\{ e \,\frac{\bar{d}_{20}}{f}\,\bar{\omega}
+ e \left( \frac{\bar{d}_{21}}{f}-\frac{\bar{d}_{22}}{2\,f} \right) (\bar{\omega}-\cos \theta\,\bar{p}_{\rm cm} )
\nonumber\\
&&\qquad \;+\,\frac{e\, g_A }{32\,\pi^2\, f^3}\,\Big(
\frac{m_\pi^2}{\bar \omega}\,\arcsin^2{\frac{\bar{\omega}}{m_\pi}}+ 2\,i\,\,\bar{p}_{\rm cm}\,\arcsin{\frac{\bar{\omega}}{m_\pi} }
\Big)
\nonumber\\
&& \qquad \;+\, \frac{e\, g_A^3\,\cos \theta \,\bar{p}_{\rm cm}}{32\,\pi^2 f^3 }
\Big(-\frac{m_\pi^2}{\bar{\omega}^2}\,\arcsin^2{\frac{\bar{\omega}}{m_\pi} }
+\Big(2\,i\,\frac{\bar{p}_{\rm cm}}{\bar \omega}+\frac{\pi \,m_\pi^2}{\bar{\omega}^2}\Big)\,\arcsin{\frac{\bar{\omega}}{m_\pi} }
\nonumber\\
&&\qquad \;-\,2\,\pi \,\frac{m_\pi}{\bar \omega }+2\, -i\,\pi\,\frac{\bar{p}_{\rm cm}}{\bar \omega } \Big) \Big\}\,C_-
\nonumber\\
&& f_2 =4\,e \left( \frac{\bar{d}_9}{f} \,C_0  +\frac{\bar{d}_8}{f} \,C_+ \right)
\nonumber\\
&& \quad \; +\,\frac{e \,g_A^3 }{32\,\pi\, f^3 }
\Big(\frac{m_\pi ^2}{\bar{\omega}^2}\,\arcsin{\frac{\bar{\omega}}{m_\pi} }-2\,\frac{m_\pi}{\bar \omega } -i\,\frac{\bar{p}_{\rm cm}}{\bar \omega }\Big)\, C_-\,,
\nonumber\\
&& f_3  =-4\,e \left( \frac{\bar{d}_9}{f} \,C_0  +\frac{\bar{d}_8}{f} \,C_+ \right)
+ e \,\left( \frac{\bar{d}_{21}}{f}-\frac{\bar{d}_{22}}{2\,f} \right) C_-
\nonumber\\
&& \quad  \;+\,\frac{e \,g_A^3}{48\,\pi^2\, f^3 }
\Big(2 \,\frac{m_\pi^3}{\bar{\omega}^3}-3\,\frac{m_\pi}{\bar \omega }+2\,i\,\frac{m_\pi^2}{\bar{\omega}^3}\,\bar{p}_{\rm cm}-2\,i\,\frac{\bar{p}_{\rm cm}}{\bar \omega } \Big)
\,C_+
\nonumber\\
&& \quad \;+\, \Big\{\frac{e \,g_A^3 }{32\, \pi\, f^3}
\Big(\frac{m_\pi ^2}{\bar{\omega}^2}\,\arcsin^2{\frac{\bar{\omega}}{m_\pi}} -\Big(\frac{\pi \,m_\pi^2}{\bar{\omega}^2}+ 2\,i \,\frac{\bar{p}_{\rm cm}}{\bar \omega }\Big) \,
\arcsin{\frac{\bar{\omega}}{m_\pi} }
\nonumber\\
&& \qquad \;+\, 2\,\pi\, \frac{m_\pi}{\bar \omega }  -2 +i\,\pi\, \frac{\bar{p}_{\rm cm}}{\bar \omega }\Big)
\Big\} \,C_-\,,
\nonumber\\
&&  f_4  = 0\,,
\label{loop-gNtopiN}
\end{eqnarray}
with renormalized and scale-independent counter terms $\bar d_{...}$ in the convention of \cite{Bernard:1994gm}.
The channel dependent coefficients $C_{...}$ are given in Tab. \ref{tab:isospin-gN}. The kinematical constraint
on the  amplitude $f_1$ at $\sqrt{s}=m_N$ is incorporated by pulling out the overall factor $\omega = (s-m_N^2)/(2\,\sqrt{s}\,)$.

At order $Q^3$ it is justified to add up the loop contribution (\ref{loop-gNtopiN}) to the tree-level contribution
(\ref{F12pm-tree-level}) via the relation (\ref{calFtoF12pm}) provided the substitution rules (\ref{def-substitution})
are used in (\ref{F12pm-tree-level}). In the one-loop result (\ref{loop-gNtopiN}) we replace
$g_A \to f\,g_{\pi NN} /m_N$.

\newpage

\subsection{Photon-nucleon scattering}
\label{compton_amplitude}

We decompose the Compton scattering amplitude into invariant amplitudes.
Bardeen and Tung \cite{Bardeen:1969aw} constructed six amplitudes $A^{BT}_{1-6}(s,t)$
that were proven to be free of kinematic constraints
\begin{eqnarray}
&&T^{\mu\nu}(\bar q, q;w)=\sum_{i=1}^6\,A^{BT}_i(s,t) \,L^{\mu \nu}_i \,, \qquad
Q_\mu= \frac{\bar q_\mu +q_\mu }{2}\,, \qquad P_\mu= \frac{\bar p_\mu +p_\mu }{2}\,,
\nonumber\\ \nonumber\\
&& L^{\mu \nu}_1=Q^2\,g^{\mu\nu}-2\,Q^\mu\, Q^\nu\,,
\nonumber\\
&& L^{\mu \nu}_2={\textstyle{\frac{1}{2}}}\,Q^2\,(\gamma^\mu\,\Qslash\,\gamma^\nu-\gamma^\nu\,\Qslash\,\gamma^\mu)
-(P\cdot Q)\,(Q^\mu\,\gamma^\nu+Q^\nu\,\gamma^\mu)
\nonumber\\
&&\qquad  +\;\Qslash\,(Q^\mu\, P^\nu +Q^\nu \,P^\mu)\,,
\nonumber\\
&& L^{\mu \nu}_3=(m_N\,\Qslash -P\cdot Q)\,g^{\mu\nu}
+i\,Q^2\,\sigma^{\mu\nu}-i\,\sigma^{\alpha\nu}\,Q_\alpha \,Q^\mu
-i\,\sigma^{\mu\alpha}\,Q_\alpha \,Q^\nu
\nonumber\\
&& \qquad -\,m_N \,(Q^\mu\,\gamma^\nu+Q^\nu\,\gamma^\mu)+(Q^\mu\,P^\nu+Q^\nu\,P^\mu)\,,
\nonumber\\
&& L^{\mu \nu}_4=Q^2\,(\gamma^\mu \,P^\nu+\gamma^\nu \,P^\mu)
-(P\cdot Q)\,(Q^\mu\,\gamma^\nu+Q^\nu\,\gamma^\mu)
\nonumber\\
&&\qquad  -\;\Qslash\,(Q^\mu\, P^\nu +Q^\nu\, P^\mu)+((P\cdot Q)\,\Qslash -m_N \,Q^2)\,g^{\mu\nu}+2\,m_N\,Q^\mu\, Q^\nu\,,
\nonumber\\
&& L^{\mu \nu}_5=Q^2\, P^\mu\, P^\nu-(P\cdot Q)\,(Q^\mu \,P^\nu +Q^\nu\, P^\mu)
\nonumber\\
&&\qquad -\,{\textstyle{\frac{1}{2}}}\,\big(P^2\,Q^2-(P\cdot Q)^2\big)\,g^{\mu\nu} +P^2\,Q^\mu \,Q^\nu \,,
\nonumber\\
&& L^{\mu \nu}_6=
-{\textstyle{\frac{1}{2}}}\,(P\cdot Q)\,(\gamma^\mu\, P^\nu+\gamma^\nu \,P^\mu)
+{\textstyle{\frac{1}{4}}}\,(P\cdot Q)\,(\gamma^\mu\,\Qslash\,\gamma^\nu-\gamma^\nu\,\Qslash\,\gamma^\mu)
\nonumber\\
&&\qquad -\,{\textstyle{\frac{i}{2}}}\,m_N\,Q^2\,\sigma^{\mu\nu}
+{\textstyle{\frac{1}{2}}}\,(m_N\,(P\cdot Q)-\Qslash\,P^2)\,g^{\mu\nu}+\Qslash \,(Q^\mu\, Q^\nu +  P^\mu \,P^\nu )
\nonumber\\
&&\qquad  +\,{\textstyle{\frac{1}{2}}}\,m^2_N\,(Q^\mu\,\gamma^\nu+Q^\nu\,\gamma^\mu)
-{\textstyle{\frac{1}{2}}}\,m_N\,(Q^\mu\, P^\nu +Q^\nu \,P^\mu)
\nonumber\\
&&\qquad +\,{\textstyle{\frac{i}{2}}}\,m_N\,\sigma^{\alpha\nu}\,Q_\alpha\, Q^\mu
+{\textstyle{\frac{i}{2}}}\,m_N\,\sigma^{\mu\alpha}\,Q_\alpha \,Q^\nu\,.
\label{def-Bardeen-Tang-amplitudes}
\end{eqnarray}
For our purpose it convenient to introduce an alternative set of invariant amplitudes, $F^\pm_{1-3}(\sqrt{s},t)$, for which
the MacDowell symmetry \cite{MacDowell:1959zza} is manifest and therefore more
transparent expressions for the partial-wave helicity amplitudes arise. We construct
amplitudes which enjoy the relations
\begin{eqnarray}
F^-_i(+\sqrt{s},t) = F^+_i(-\sqrt{s},t)\,,
\label{MacDowell-relation-Compton}
\end{eqnarray}
and are free of kinematic constraints. We find that the following decomposition
\begin{eqnarray}
&&T_{\mu\nu}(\bar q, q;w)= \sum_{i=1}^3\, \Big( F^+_{i}(\sqrt{s},t)\,L^{(i,+)}_{\mu \nu}(\bar q,q,w)
+F^-_{i}(\sqrt{s},t)\,L^{(i,-)}_{\mu \nu}(\bar q,q,w)\Big) \,,
\nonumber\\
&&  L^{(1,\pm )}_{\mu \nu} = \gamma_\nu \,\qslash \,
\Big( \frac{1}{2}\pm \frac{\wslash }{2\,\sqrt{s}} \Big)\,\barqslash\,\gamma_\mu \,, \qquad
L^{(2,\pm)}_{\mu \nu} =\gamma_\mu \,\barqslash \,\Big( \frac{1}{2}\pm \frac{\wslash }{2\,\sqrt{s}} \Big)\,\qslash\,\gamma_\nu  \,,
\nonumber\\
&& L^{(3,\pm)}_{\mu \nu} = \Big( \frac{1}{2}\pm \frac{\wslash }{2\,\sqrt{s}} \Big)\,
\Big((w\cdot \bar{q})\,g_{\mu\alpha}-w_\mu\,\bar{q}_\alpha\Big)\,
\Big(q_\alpha\gamma_\nu-\qslash \,g_{\alpha\nu}\Big)
\nonumber\\
&& \qquad \;\;\,+ \,\Big((w\cdot q)\,g_{\nu\alpha}-w_\nu\,q_\alpha\Big)\,
\Big(\bar q_\alpha\gamma_\mu-\barqslash \,g_{\alpha\mu}\Big)\,\Big( \frac{1}{2}\pm \frac{\wslash }{2\,\sqrt{s}} \Big) \,,
\label{def-Fpm123}
\end{eqnarray}
meets our requirements.
For on-shell conditions the amplitudes  $F^{\pm}_{1-3}(\sqrt{s},t)$ are linear combinations
of the amplitudes of Bardeen and Tung \cite{Bardeen:1969aw}. It holds
\begin{eqnarray}
&&F^{\pm}_1=\frac{m_N^2 \pm 2\,\sqrt{s}\,m_N - s}{8\,m_N^2}\,A^{BT}_1
-\frac{(m_N \mp \sqrt{s}\,)^2}{8\,m_N}\,A^{BT}_2+\frac{1}{4}\,A^{BT}_3
\mp\frac{\sqrt{s}}{4}\,A^{BT}_4
\nonumber\\
&& \quad \;\; \,\mp \,\sqrt{s}\,\frac{(2\,m_N \mp\sqrt{s}\, )\,s}{8\,m_N^2}\,A^{BT}_5
\mp\frac{\sqrt{s}}{8}\,A^{BT}_6\,,
\nonumber\\
&&F^{\pm}_2=\frac{m_N^2 \pm 2\,\sqrt{s}\,m_N + 2\,(\bar{q} \cdot q) - s}{8\,m_N^2}\,A^{BT}_1
\nonumber\\
&& \quad \;\;\, -\,\frac{m_N^2 \mp 2\sqrt{s}\,m_N - 2\,(\bar{q} \cdot q) + s}{8\,m_N}\,A^{BT}_2
-\frac{1}{4}\,A^{BT}_3\mp\frac{\sqrt{s}}{4}\,A^{BT}_4
\nonumber\\
&& \quad \;\; \,
+\,\frac{m_N^4 - s\,m_N^2 \mp 2\,\sqrt{s}\,s\,m_N + s^2 - 2\,(\bar{q} \cdot q)\,s}{8\,m_N^2}\,A^{BT}_5
\pm \frac{\sqrt{s}}{8}\,A^{BT}_6\,,
\nonumber\\
&&F^{\pm}_3=-\frac{1}{2\,m_N}\,A^{BT}_1-\frac{1}{2}\,A^{BT}_2
\mp \sqrt{s}\,\frac{m_N \mp 2\,\sqrt{s}}{4\,m_N}\,A^{BT}_5\,.
\label{def-trafo}
\end{eqnarray}
We observe that the transformation coefficients, expressing the  Bardeen and Tung  amplitude in terms of
the $F^\pm_{1-3}$ amplitudes, are finite everywhere but at $\sqrt{s}=0$. Furthermore
the determinant of the transformation (\ref{def-trafo}) equals $(\sqrt{s}\,s)/128$.  It follows our
claim that the amplitudes, $F^{\pm}_{1-3}(\sqrt{s},t)$, are free of kinematic constraints but (\ref{MacDowell-relation-Compton}).

The helicity partial-wave amplitudes with definite parity, $P$, and total angular momentum, $J$, are
conveniently derived from the invariant amplitudes introduced in (\ref{def-Fpm123}). We use a convention consistent
with the photoproduction amplitudes
\allowdisplaybreaks[1]
\begin{eqnarray}
&& t_{\pm,11}^J (\sqrt{s}\,) = \frac{\pm\,\omega^2}{2 \,m_N}\, \int_{-1}^{+1} \frac{dx}{2} \,\bigg\{\Big[
2 \,E\,F_1^{\mp}  + 2 \,\omega\,F_1^{\pm}    \pm (2 \,s -m_N^2 \pm m_N \,\sqrt{s}\, )\,F_3^{\pm}
\nonumber\\
&& \qquad - 4 \,\sqrt{s}\,F_2^{\mp}+ x\big( 2 \,E\,F_1^{\pm}   + 2 \, \omega\,F_1^{\mp}  \pm m_N \,(m_N \mp \sqrt{s}\,)\,F_3^{\mp}  \big)
\Big]\,P_{J-\frac{1}{2}}(x)
\nonumber\\
&& \quad +\,  \Big[
- 2 \,E\,F_1^{\pm}   - 2\, \omega\, F_1^{\mp}  + 4\,\sqrt{s}\, F_2^{\pm}
 \mp  (m_N^2 \pm m_N \,\sqrt{s} - 2\, s)\, F_3^{\mp}
\nonumber\\
&& \qquad +\,x\,\big(- 2 \,E\,F_1^{\mp}   - 2\,\omega\, F_1^{\pm}
\pm  m_N\,  (m_N \pm \sqrt{s}\,)\,F_3^{\pm}\big) \Big]
\,P_{J+\frac{1}{2}}(x)\bigg\}\,,
\nonumber\\
&& t_{\pm,12}^J (\sqrt{s}\,) =\frac{\pm\,\omega^2}{4 \,m_N}\,\int_{-1}^{+1} \frac{dx}{2} \,\bigg\{\Big[
{\textstyle{\sqrt{\frac{2 \,J-1}{2\, J+3}}}}\,\big(-4 \,m_N\,F_1^{\mp} +
  (m_N \pm \sqrt{s}\,)^2\,F_3^{\pm} \big)
\nonumber\\
&& \qquad +\, x\,{\textstyle{\sqrt{\frac{2\, J+3}{2 \,J-1}}}}\,
\big(4 \,  m_N\,F_1^{\pm} -(m_N \mp \sqrt{s}\,)^2\, F_3^{\mp} \big)
\Big]\,P_{J-\frac{1}{2}}(x)
\nonumber\\
&& \quad +\, \Big[ {\textstyle{\sqrt{\frac{2 \,J+3}{2 \,J-1}}}}\,\big(-4 \,m_N\,F_1^{\pm}  +
   (m_N \mp \sqrt{s}\,)^2\,F_3^{\mp}\big)
\nonumber\\
&& \qquad +\, x\,{\textstyle{\sqrt{\frac{2 J-1}{2 J+3}}}}\,\big(4 \,F_1^{\mp}  m_N - F_3^{\pm}  (m_N \pm \sqrt{s})^2\big)
\Big]\, P_{J+\frac{1}{2}}(x)\bigg\}\,,
\nonumber\\
&& \! \! t_{\pm,22}^J (\sqrt{s}\,) =\frac{\pm\,\omega^2}{2\, m_N}\int_{-1}^{+1}\! \frac{dx}{2} \,\bigg\{ \Big[
{\textstyle{\frac{2\,J+7}{2\, J+3}}}\,\big(  2 \, \omega\,F_1^{\pm} -2 \,E\,F_1^{\mp}
+   \sqrt{s} \,(m_N \pm \sqrt{s}\,)\,F_3^{\pm}\big)
\nonumber\\
&& \qquad +\, x\,{\textstyle{\frac{2\,J+3}{2 \,J-1}}}\,\big(2\, E\, F_1^{\pm}   - 2\,\omega\, F_1^{\mp}
-  \sqrt{s} \,(m_N \mp \sqrt{s}\,)\,F_3^{\mp}\big) \Big]\,P_{J-\frac{1}{2}}(x)
\nonumber\\
&& \quad +\, \Big[
{\textstyle{\frac{2\,J-5}{2\, J-1}}}\,\big(2\,E\, F_1^{\pm}  - 2 \,\omega\,F_1^{\mp}
-  \sqrt{s} \,(m_N \mp \sqrt{s}\,)\,F_3^{\mp} \big)
\nonumber\\
&& \qquad -\, x\,{\textstyle{\frac{2\,J-1}{2 \,J+3}}}\,\big(2\,E\, F_1^{\mp}    - 2 \,\omega\,F_1^{\pm}
-  \sqrt{s} \,(m_N \pm \sqrt{s}\,)\,F_3^{\pm}\big)
\Big]\,P_{J+\frac{1}{2}}(x)\bigg\}\,,
\label{helicity-PW-Compton}
\end{eqnarray}
with $x = \cos \theta$ and $P = \pm $ for $J-\frac{1}{2}$ odd and $P= \mp$ for $J-\frac{1}{2}$ even. Following  \cite{Pfeil:1974ib}
the differential cross section and photon asymmetry are expressed in terms of helicity matrix elements\footnote{We utilize a different
phase convention as compared to \cite{Pfeil:1974ib}. The expressions
for differential cross section and photon beam asymmetry are nevertheless the
same.}
\begin{eqnarray}
 \frac{d\sigma}{d\Omega}&=&\frac{1}{2}\left(
|\phi_1|^2+|\phi_2|^2+2\,|\phi_3|^2+2\,|\phi_4|^2+|\phi_5|^2+|\phi_6|^2\right)\nonumber\\
 \Sigma\,\frac{d\sigma}{d\Omega}&=&
\Re\,\big((\phi_1+\phi_5)\,\phi_3^*+(\phi_2-\phi_6)\,\phi_4^*\big)\,,
\end{eqnarray}
with
\begin{eqnarray}
\phi_1&=&\frac{m_N}{4\pi\,\sqrt{s}}\,\cos{\frac{\theta}2}\,\sum_J\left(t^J_{-,11}+t^J_{+,11}\right)
\big\{P'_{J+\frac{1}{2}}(\cos \theta) -P'_{J-\frac{1}{2}}(\cos \theta)\big\}\,,
\nonumber\\
\phi_2&=&-\frac{m_N}{4\pi\,\sqrt{s}}\,\sin{\frac{\theta}2}\,\sum_J \left(t^J_{-,11}-t^J_{+,11}\right)
\big\{P'_{J+\frac{1}{2}}(\cos \theta) +P'_{J-\frac{1}{2}}(\cos \theta)\big\}\,,
\nonumber\\
\phi_3&=&\frac{m_N}{4\pi\,\sqrt{s}}\,\frac{\sin\theta \,\sin{\frac{\theta}2}}{\sqrt{(J-\frac{1}{2})(J+\frac{3}{2})}}\,
\sum_J\left(t^J_{-,12}-t^J_{+,12}\right)
\big\{P''_{J+\frac{1}{2}}(\cos \theta) +P''_{J-\frac{1}{2}}(\cos \theta)\big\}\,,
\nonumber\\
\phi_4&=&\frac{m_N}{4\pi\,\sqrt{s}}\,\frac{\sin\theta \,\cos{\frac{\theta}2}}{\sqrt{(J-\frac{1}{2})(J+\frac{3}{2})}}\,
\sum_J\left(t^J_{-,12}+t^J_{+,12}\right)
\big\{P''_{J+\frac{1}{2}}(\cos \theta) -P''_{J-\frac{1}{2}}(\cos \theta)\big\}\,,
\nonumber\\
\phi_5&=&\frac{m_N}{4\pi\,\sqrt{s}}\,\frac{2\,\cos^3{\frac{\theta}2}}{(J-\frac{1}{2})(J+\frac{3}{2})}\,
\sum_J\left(t^J_{-,22}+t^J_{+,22}\right)
\big\{-3\,(J-\textstyle\frac{1}{2})\,P''_{J-\frac{1}{2}}(\cos \theta)
\nonumber\\
&&\quad +\,(J-\textstyle\frac{1}{2})\,P''_{J+\frac{1}{2}}(\cos \theta) +2\,\big(P'''_{J-\frac{1}{2}}(\cos \theta)
-P'''_{J-\frac{3}{2}}(\cos \theta)\big)\big\}\,,
\nonumber\\
\phi_6&=&-\frac{m_N}{4\pi\,\sqrt{s}}\,\frac{2\sin^3{\frac{\theta}2}}{(J-\frac{1}{2})(J+\frac{3}{2})}\,
\sum_J\left(t^J_{-,22}-t^J_{+,22}\right)
\big\{3\,(J-\textstyle\frac{1}{2})\,P''_{J-\frac{1}{2}}(\cos \theta)
\nonumber\\
&& \quad +\,(J-\textstyle\frac{1}{2})\,P''_{J+\frac{1}{2}}(\cos \theta)
+2\,\big(P'''_{J-\frac{1}{2}}(\cos \theta) +P'''_{J-\frac{3}{2}}(\cos \theta)\big)\big\}\,.
\end{eqnarray}

In order to obtain partial-wave amplitudes that are free of kinematic constraints we apply the transformations
derived in (\ref{def-Tamplitude}, \ref{t12toT12}) to the helicity partial-wave amplitudes of (\ref{helicity-PW-Compton}).
We introduce
\begin{eqnarray}
T^J_{\pm,ab}(\sqrt{s}\,)&=& 2\,m_N\,
\left(\frac{s}{p^2_{\rm cm}}\right)^{J-\frac12}
\label{gn-trafo}\\
&  \times & \sum_{c,d}
\left(\begin{matrix}
\frac{\sqrt{s}}{p_{\rm cm}}&-\sqrt{\frac{2\,J-1}{2\,J+3}}\,\frac{m_N}{p_{\rm cm}}\\
0&1   \end{matrix}\right)_{ac}
t^J_{\pm,cd}(\sqrt{s}\,)
\left(\begin{matrix}
\frac{\sqrt{s}}{p_{\rm cm}}&0\\
-\sqrt{\frac{2\,J-1}{2\,J+3}}\,\frac{m_N}{p_{\rm cm}}&1   \end{matrix}\right)_{db}\,,
\nonumber
\end{eqnarray}
which implies the photon-nucleon phase-space distribution
\begin{eqnarray}
\rho^{\gamma N,J}_{\pm}(\sqrt{s}\,)&=& -\Im \,\big[T^{J}_{\pm}(\sqrt{s}\,)\,\big]^{-1}
\nonumber\\
&=&\,\frac{p_{\rm cm}}{8\pi\,\sqrt{s}}
\left(\frac{p_{\rm cm}^2}{s}\right)^{J-\frac{1}{2}}
\left(\begin{matrix}
\frac{p_{\rm cm}^2}{s}&\sqrt{\frac{J-\frac{1}{2}}{J+\frac{3}{2}}}
\frac{m_N p_{\rm cm}}{s}\\
\sqrt{\frac{J-\frac{1}{2}}{J+\frac{3}{2}}}
\frac{m_N p_{\rm cm}}{s}&
1+\frac{J-\frac{1}{2}}{J+\frac{3}{2}}\frac{m_N^2}{s}        \end{matrix}\right)\,.
\label{gn-phase-space}
\end{eqnarray}
Inserting (\ref{helicity-PW-Compton}) into (\ref{gn-trafo}) we confirm by explicit calculations that
indeed the partial-wave amplitudes $T^J_{\pm,ab}(\sqrt{s}\,)$ are kinematically unconstrained. Moreover,
the MacDowell relations
\begin{eqnarray}
T_{-,ab}^J(+\sqrt{s}\,)= \left\{
\begin{array}{ll}
+ T_{+,ab}^J(-\sqrt{s}\,) \qquad {\rm for} & \quad a=b \\
- T_{+,ab}^J(-\sqrt{s}\,) \qquad {\rm for} & \quad a\neq b
\end{array}
\right.
\,,
\label{MacDowell-Compton}
\end{eqnarray}
are satisfied.

We turn to the specifics of the effective field theory. For the photon-proton amplitude the Lagrangian density
(\ref{Lagrangian}) and (\ref{gp-gp-tree-level}) imply the tree-level expressions
\begin{eqnarray}
F_1^{\pm} &=& \frac{e^2\, (g_{A}-2\,m_\pi^2\,d_{18})\, (m_N \mp \sqrt{s}\,)^2}
{(4 \pi\,f)^2\, m_N \, (m_\pi^2 - t)} \mp \frac{e^2\, \sqrt{s}}{ m_N \,( m_N^2 - u) \,( m_N \pm \sqrt{s}\,)}
\nonumber\\
&+&\frac{e^2 \kappa_p \,( m_N^2\,(m_N \pm \sqrt{s}\,) + s\, ( m_N \mp \sqrt{s}\,))}
  { m_N^3 \,( m_N^2 - u)\, ( m_N \pm \sqrt{s}\,)}
+\frac{e^2 \,\kappa_p^2\, ( m_N \pm \sqrt{s}\,)}{4\,  m_N^2\, ( m_N^2 - u)}\,,
\nonumber\\
F_2^{\pm} &=&\frac{e^2\, (g_{A}-2\,m_\pi^2\,d_{18})\,((m_N \mp \sqrt{s}\,)^2 +t )}{(4 \pi\,f)^2 \,m_N\, (m_\pi^2 - t)}
\nonumber\\
&+& \frac{e^2\, ( m_N\,(m_N^2-s) \pm \sqrt{s}\,(m_N^2-u))}
{ m_N \,( m_N^2 - u)\, ( m_N \mp \sqrt{s}\,) \,( m_N \pm \sqrt{s}\,)^2}
\nonumber\\
&+& \frac{e^2 \,( (\kappa_p +2) \,m_N \pm \kappa_p\,\sqrt{s}\,)^2 }
{4 \,m_N^2\, (m_N \mp \sqrt{s}\,) (m_N \pm \sqrt{s}\,)^2}
\nonumber\\
&+&\frac{e^2\, \kappa_p\, (2\,  m_N^4 + s \, m_N^2 \mp 2 \,\sqrt{s}\,s \, m_N - u\, s)}
  { m_N^3 \,( m_N^2 - u)\, ( m_N^2 - s)}\,,
\nonumber\\
F_3^{\pm} &=& \frac{e^2\, (g_{A}-2\,m_\pi^2\,d_{18})}{(2\pi \,f)^2 \, (m_\pi^2 - t)}
+\frac{4\, e^2}{( m_N^2 - u)\, ( m_N^2 - s)}
\nonumber\\
&+& \frac{2\, e^2\, \kappa_p \,( m_N^2 \mp \sqrt{s} \, m_N + 2 \,s)}
{ m_N^2 \,( m_N^2 - u) \,( m_N^2 - s)} \,,
\label{F123pm-tree-level}
\end{eqnarray}
which contribute at different orders in a chiral expansion.

A systematic computation at order $Q^3$ requires the evaluation of
one-loop diagrams. The relevant loops were established by Bernard, Mei\ss ner and Kaiser \cite{Bernard:1995dp} in terms
of six center-of-mass amplitudes
\begin{eqnarray}
&&T_{\rm cm}^{\,ij} = \frac{E + m_N}{2\,m_N}\,\Big\{\delta^{ij}\,A_1 + \bar q\,^i \,q^j \,A_2 + i\,\epsilon_{ijk}\,\sigma_k \,A_3
+ i\,\epsilon_{klm}\,\sigma_k \,\bar q\,^l\,q^m\,\delta^{ij}\,A_4
\nonumber\\
&& \qquad \!+ \, i\,\sigma_k \,[\epsilon_{kil}\,q^l \,\bar q\,^j - \epsilon_{kjl}\,\bar q\,^l \,q^i ]\,A_5
+ \, i\,\sigma_k \,[\epsilon_{kil}\,\bar q\,^l \,\bar q\,^j - \epsilon_{kjl}\,q^l \,q^i ]\,A_6 \Big\}\,,
\label{def-cm-amplitudes-Compton}
\end{eqnarray}
within a two-component nucleon spinor formulation. The Coulomb gauge is assumed in (\ref{def-cm-amplitudes-Compton}) with vanishing
zero components of the photon wave functions. The invariant amplitudes of (\ref{def-Fpm123}) are unambiguously reconstructed from their
center-of-mass amplitudes, $A_{1-6}(\omega, t)$, with
\begin{eqnarray}
&& F_i^\pm = \frac{1}{\sqrt{s}}\left(\frac{m_N+ \sqrt{s}}{2\,m_N}\right)^2 \,\sum_{j=1}^6 \,a^{\pm }_{ij}\,A_j\,,
\nonumber\\
&& a_{13}^+ = - a_{11}^+= \frac{\sqrt{s}}{(m_N + \sqrt{s}\,)^2}\,,
\qquad
a_{15}^+ = -\frac{m_N - \sqrt{s}}{m_N + \sqrt{s}} \,\big( m_N+\omega\,(x-1)\big)\,,
\nonumber\\
&& a_{12}^+ =-\frac{m_N^2\,(m_N-\sqrt{s}\,) \,(x-1)-m_N \,s \,(x-7) + \sqrt{s}\,s\, (x-3)}{4 \,\sqrt{s} \,(m_N + \sqrt{s}\,)}\,,
\nonumber\\
&& a_{14}^+ =\frac{m_N^2\, (x-1)+\sqrt{s}\,(\sqrt{s}-2\, m_N)\,(x +1)}{4 \,\sqrt{s}}\,,
\qquad a_{16}^+ =-(m_N - \sqrt{s}\,) \,,
\nonumber\\ \nonumber\\
&& a_{13}^- = - a_{11}^-= \frac{m_N + \omega}{2\, \omega^2}\,,
\qquad
a_{15}^- = \omega \,(x-1)+ m_N \,x \,,
\nonumber\\
&& a_{12}^- =\frac{ m_N\,(m_N-2 \,\sqrt{s}\,)\, (x-1) - s\, (x-3) }{4 \,\sqrt{s}}\,,
\qquad a_{16}^- =- \frac{2 \, (m_N + \omega)\, \sqrt{s}}{m_N - \sqrt{s}} \,,
\nonumber\\
&& a_{14}^- =-\frac{m_N^2\, (x-1) -\sqrt{s}\,(\sqrt{s}+2 \,m_N)\,(1 + x)}{4 \,\sqrt{s}}\,,
\nonumber\\ \nonumber\\
&& a_{23}^+ = -a_{21}^+ =  \frac{m_N\, s\, (x-3) + m_N^3 \,(x-1)
+ s^{\frac32} \,(1 + x) - m_N^2\,\sqrt{s} \,(1 + 3\, x)}{4 \,\omega\, s\, (m_N + \sqrt{s}\,)} \,,
\nonumber\\
&& a_{22}^+ =\frac{m_N^5\, (x-1)^2  + m_N^4 \,\sqrt{s} \,(1 + 2 \,x - 3 \,x^2)
+4 \,m_N^2\, \sqrt{s}\,s\, (x-1) \,x }{8\, \sqrt{s}\,s\, (m_N + \sqrt{s}\,)}
\nonumber\\
&& \quad\;\; +\,\frac{ \sqrt{s}\,s^2 \,(3 + 2\, x - x^2)
 - m_N\, s^2 \,(13 - 2\, x + x^2)}{8\, \sqrt{s}\,s\, (m_N + \sqrt{s}\,)}\,,
\nonumber\\
&&  a_{24}^+ =-\frac{m_N^4\, (x-1)^2 + 4\, m_N\, \sqrt{s}\,s \,(1 + x)
- s^2 \,(1 + x)^2}{8 \,\sqrt{s}\,s}
\nonumber\\
&& \quad\;\; -\,\frac{-4 \,m_N^3 \,\sqrt{s} \,(x^2-1) +  4\, m_N^2\, s \,(1 + x^2)}{8 \,\sqrt{s}\,s}\,,
\nonumber\\
&&  a_{25}^+ =-\frac{8 \,m_N^3 \,s + m_N^5\, (x-1)^2 - 3 \,m_N^4\, \sqrt{s} \,(x^2-1)}{4\, \sqrt{s}\,s\, (m_N + \sqrt{s}\,)}
\nonumber\\
&& \quad\;\; -\,\frac{-\sqrt{s}\,s^2 \,(x^2-1)+ 4\, m_N^2 \,\sqrt{s}\,s \,(2 + x^2)
 - m_N\, s^2 \,(5 - 2 \,x + x^2)}{4\, \sqrt{s}\,s\, (m_N + \sqrt{s}\,)}\,,
 \nonumber\\
&&  a_{26}^+ =\frac{m_N \,s \,(x-3) + m_N^3 \,(x-1) + \sqrt{s}\,s\, (1 + x)
 - m_N^2\,\sqrt{s} \,(1 + 3\, x)}{2\, s}\,,
\nonumber\\ \nonumber\\
&& a_{21}^- = - \frac{8\, m_N \,s \,E + m_N^4 \,(x-1)
 + s^2\, (1 + x) - 2\, m_N^2\, s \,(2 + x)}{8 \,\omega^2 \,\sqrt{s}\,s }\,,
\nonumber\\
&& a_{22}^- = -\frac{4\,m_N\,\sqrt{s}\,(m_N^2+2\,s)\,(x-1)
 + m_N^4 \,(x-1)^2}{8\, \sqrt{s}\,s}
\nonumber\\
&& \quad \;\,\,-\,\frac{s^2\, (-3 - 2\, x + x^2) -
    2 \,m_N^2\, s \,(-3 + 2\, x + x^2)}{8\, \sqrt{s}\,s}\,,
\nonumber\\
&& a_{23}^- = \frac{8\, m_N \,s \,E + m_N^4 \,(x-1)
 + s^2\, (1 + x) - 2\, m_N^2\, s \,(6 + x)}{8 \,\omega^2 \,\sqrt{s}\,s }\,,
\nonumber\\
&& a_{24}^- =\frac{m_N^4\, (x-1)^2 + 4\, m_N \,\sqrt{s}\,s \,(1 + x) + s^2 \,(1 + x)^2}{8\, \sqrt{s}\,s}
\nonumber\\
&& \quad \;\,\,+\,\frac{-2\, m_N^2\, s \,(-1 + 4\, x + x^2)}{8\, \sqrt{s}\,s}\,,
\nonumber\\
&& a_{25}^- =\frac{2 \,m_N^3 \,\sqrt{s}\, (x-1) + m_N^4\, (x-1)^2
 + 2\, m_N\, \sqrt{s}\,s \,(3 \,x-1)}{4\, \sqrt{s}\,s}
\nonumber\\
&& \quad \;\,\,+\,\frac{s^2 \,(x^2-1) -   2\, m_N^2 \,s (x^2+5\,x-2)}{4\, \sqrt{s}\,s} \,,
\nonumber\\
&& a_{26}^- = -\frac{8 \,m_N \,s\, E + m_N^4\, (x-1) + s^2\, (1 + x)
 - 2\, m_N^2\, s \,(2 + x)}{2\, s\, (m_N - \sqrt{s}\,)}\,,
\nonumber\\ \nonumber\\
&& a_{33}^+= -a_{31}^+ = \frac{2 \,m_N}{\omega\, (m_N + \sqrt{s}\,)^2}\,,\qquad
a_{32}^+ = \frac{m^2_N\, (x-1)-\sqrt{s}\,m_N\, (x-3) }{\sqrt{s}\, (m_N + \sqrt{s}\,)}\,,
\nonumber\\
&& a_{36}^+ = \frac{2\,m_N}{\omega}\,, \qquad
 a_{34}^+ =-\frac{m^2_N\,(x-1) - \sqrt{s}\,m_N\,(x+1)}{\sqrt{s} \,(m_N + \sqrt{s}\,)}\,, \qquad
\nonumber\\
&& a_{35}^+ =-\frac{2 \,m_N\,(m_N - \sqrt{s}\,)\, (x-1)}{\sqrt{s}\, (m_N + \sqrt{s}\,)}\,,
\nonumber\\ \nonumber\\
&& a_{33}^-= -a_{31}^- = \frac{m_N}{\omega^2 \,\sqrt{s}} \,,\qquad
a_{32}^- = - \frac{m^2_N \,(x-1)+\sqrt{s}\, m_N\,(x-3)}{\sqrt{s} \,(m_N + \sqrt{s}\,)}\,,
\nonumber\\
&& a_{36}^- = \frac{4 \,m_N}{\sqrt{s}-m_N}\,, \qquad
 a_{34}^- =\frac{m_N^2\, (x-1) - \sqrt{s}\, m_N\,(x +1 )}{\sqrt{s}\,(m_N - \sqrt{s}\,)}\,, \qquad
\nonumber\\
&& a_{35}^- =\frac{2 \,m^2_N\, \sqrt{s} - m_N^3 \,(x-1) + s \,m_N\,(x-1)}{\omega \,s}\,,
\label{F13toA16}
\end{eqnarray}
with $x = \cos \theta $ and $\omega, E$ as in (\ref{def-kinematic-generic}).
We specify the one-loop contribution to the invariant amplitudes $F_{1-3}^\pm$ in terms of (\ref{F13toA16}) and
the one-loop expressions for the center-of-mass $A_{1-6}(\omega, t)$ amplitudes. From \cite{Bernard:1995dp} we recall
\begin{eqnarray}
W&=&\sqrt{m_\pi^2-\omega^2\,z^2+t\,(1-z)^2\,x\,(x-1)}\,,\quad
R=\sqrt{m_\pi^2+t\,(1-z)^2\,x\,(x-1)}\,,
\nonumber\\ \nonumber\\
A_1&=&\frac{e^2\,g_{A}^2}{8\pi \,f^2}\,\Big\{m_\pi -\sqrt{m_\pi^2-\omega^2}
+\frac{2\,m_\pi^2-t}{\sqrt{-t}}\,\Big[\frac{1}{2}\,\arctan{\frac{\sqrt{-t}}{2\,m_\pi}}
\nonumber\\
&&\quad -\,\int_0^1 dz \arctan{\frac{(1-z)\,\sqrt{-t}}{2\,\sqrt{m_\pi^2-\omega^2\, z^2}}}
\Big]\Big\}\,,
\nonumber\\
A_2&=&\frac{e^2\,g_{A}^2}{8\pi \,f^2}\frac{t-2\,m_\pi^2}{(-t)^\frac{3}{2}}
\int_0^1 dz \bigg[\arctan{\frac{(1-z)\,\sqrt{-t}}{2\,\sqrt{m_\pi^2-\omega^2\, z^2}}}
\nonumber\\
&&\quad -\,\frac{2\,(1-z)\sqrt{t\,(\omega^2\, z^2-m_\pi^2)}}{4\,m_\pi^2-4\,\omega^2\,z^2-t\,(1-z)^2}\bigg]\,,
\nonumber\\
A_3&=&\frac{e^2\,g_{A}^2}{8\pi^2 \,f^2}\,\Big[\frac{m_\pi^2}{\omega}\,
\arcsin^2{\frac{\omega}{m_\pi}}-\omega\Big]
\nonumber\\
&+&\frac{e^2\,g_{A}^2}{4\pi^2 \,f^2}\,\omega^4\sin^2{\theta}\int_0^1 dx
\int_0^1 dz \,\frac{x\,(1-x)\,z\,(1-z)^3}{W^3}\,\Big[\arcsin{\frac{\omega \,z}{R}}
+\frac{\omega\, z\, W}{R^2}\Big]\,,
\nonumber\\
A_4&=&\frac{e^2\,g_{A}^2}{4\pi^2\, f^2}\,\int_0^1 dx \int_0^1 dz\,
\frac{z\,(1-z)}{W}\arcsin{\frac{\omega \,z}{R}}\,,
\nonumber\\
A_5&=&\frac{e^2\,g_{A}^2}{8\pi^2 \,f^2}\int_0^1 dx \int_0^1 dz\,
\Big[-\frac{(1-z)^2}{W}\arcsin{\frac{\omega \,z}{R}}
\nonumber\\
&&\quad +\, 2\,\omega^2\,\cos{\theta}\,\frac{x\,(1-x)\,z\,(1-z)^3}{W^3}
\Big(\arcsin{\frac{\omega z}{R}}
+\frac{\omega\, z \,W}{R^2}\Big)\Big]\,,
\nonumber\\
A_6&=&\frac{e^2\,g_{A}^2}{8\pi^2 f^2}\int_0^1 dx \int_0^1 dz\,
\Big[\frac{(1-z)^2}{W}\arcsin{\frac{\omega \,z}{R}}
\nonumber\\
&&\quad -\,2\,\omega^2\,\frac{x\,(1-x)\,z\,(1-z)^3}{W^3}\,
\Big(\arcsin{\frac{\omega\, z}{R}}
+\frac{\omega \,z\, W}{R^2}\Big)\Big]\,.
\label{gntogn-1loop}
\end{eqnarray}
At order $Q^3$ it is justified to add up the loop contribution (\ref{gntogn-1loop}) to the tree-level contribution
(\ref{F123pm-tree-level}) via the relation (\ref{F13toA16}) provided the substitution rules (\ref{def-substitution})
are used in (\ref{F123pm-tree-level}). In the one-loop result (\ref{gntogn-1loop}) we replace
$g_A \to f\,g_{\pi NN} /m_N$.

\newpage

\section{Separating singularities of tree-level diagrams}

\label{singularities}
In this appendix we show how to derive the representation (\ref{expansion}) for the
nucleon $s$- and $u$-channel and the one-pion-exchange diagrams. The treatment of the nucleon
$s$-channel poles is straightforward. While the pole term at $\sqrt{s}= m_N$ is part of
$U_{\rm inside}(\sqrt{s}\,)$, any residual background contribution is part of $U_{\rm outside}(\sqrt{s}\,)$.

Consider the $u$-channel nucleon exchange in $\pi N$ elastic scattering. It takes the generic form
\begin{eqnarray}
\int\limits_{-1}^{1}\frac{d \cos \theta}{2}\,\frac{\lambda (\sqrt{s},\cos \theta )}{u-m_N^2} \,,
\label{generic-u-channel}
\end{eqnarray}
where the function $\lambda(\sqrt{s},\cos \theta )$ reflects the specifics of the partial-wave considered.
It is a polynomial in $\cos \theta$. The contribution to the generalized potential
is derived in application of the following identity
\begin{eqnarray}
&&\int\limits_{-1}^{1}\frac{d \cos \theta}{2}\,\frac{1}{u-m_N^2}=
\int\limits_{-\infty}^0\frac{-1}{4\,p^2_{\rm cm}(s')}\,\frac{ds'}{s'-s}+
\int\limits_{(\Lambda^-_N)^2}^{(\Lambda^+_N)^2}\,\frac{-1}{4\,p^2_{\rm cm}(s')}\frac{ds'}{s'-s}\,.
\nonumber\\
&& \Lambda^+_N=\sqrt{m_N^2+2\,m_\pi^2} \,, \qquad \qquad \Lambda^-_N=\frac{m_N^2 - m_\pi^2}{m_N}\,,
\nonumber\\
&& p^2_{\rm cm}(s) = \big(s-(m_N-m_\pi)^2 \big)\,\frac{s-(m_N+m_\pi)^2}{4\,s}\,,
\label{u-start-s}
\end{eqnarray}
where we consider an s-wave type angular average with $\lambda(\sqrt{s},\cos \theta ) =1$ for simplicity.
A translation of (\ref{u-start-s}) from the variable $s $ to $\sqrt{s}$ is performed by means of the identities
\begin{eqnarray}
\int\limits_{(\Lambda^-_N)^2}^{(\Lambda^+_N)^2}\frac{(\alpha(s')+\beta(s')\,\sqrt{s}\,)\,ds'}{s'-s}
&=&\int\limits_{+\Lambda^-_N}^{+\Lambda^+_N}\frac{(\alpha(w^2)+\beta(w^2)\,w)\,dw}{w-\sqrt{s}}
\nonumber\\
& -&\int\limits_{-\Lambda^+_N}^{-\Lambda^-_N}\frac{(\alpha(w^2)+\beta(w^2)\,w)\,dw}{w-\sqrt{s}}\,,
\label{s_to_sqrts}
\end{eqnarray}
where $\alpha(s)$ and $\beta(s)$ are analytic in the integration region.
Any contribution of the $u$-channel nucleon exchange to the partial-wave potential can therefore be represented as,
\begin{eqnarray}
&&\int\limits_{-1}^{1}\frac{d \cos \theta}{2}\,\frac{\lambda(\sqrt{s},\cos \theta )}{u-m_N^2}=
-\int\limits_{\Lambda^-_N}^{\Lambda^+_N}\,\frac{\gamma(w,x_\theta)}{4\,p^2_{\rm cm}(w^2)}\,\frac{dw}{w-\sqrt{s}}
+u_{\rm outside}(\sqrt{s}\,)\,,
\nonumber\\
&& x_\theta=\frac{2\,m_\pi^2\,w^2-w^4+(m_N^2-m_\pi^2)^2}{4\,w^2\,p^2_{\rm cm}(w^2)}\,,
\label{disp_nucl_exch}
\end{eqnarray}
where  the function $u_{\rm outside}(\sqrt{s}\,)$ has no
singularities inside the domain enclosed by the contour $C_2$ of
Fig.~\ref{fig:piN}. The value $x_\theta $ corresponds to the
specific scattering angle determined by the condition $u=m_N^2$. The
cut from $\Lambda^-_N<\sqrt{s}< \Lambda^+_N$  in
(\ref{disp_nucl_exch}) contributes to  the inside and outside part
of the generalized potential as seen in Fig. ~\ref{fig:piN}.
Nevertheless we include both parts explicitly as implied by
(\ref{disp_nucl_exch}). This amounts to a well defined summation of
an infinite set of terms in $U_{\rm outside}(\sqrt{s}\,)$. For
higher partial-waves there would be a large cancelation of the
inside and outside parts, which we avoid by the suggested summation.
Such a procedure is justified, since there are no analogous cuts
stemming from higher order contributions.

We turn to the $\gamma N\to\pi N$ reaction. In this case we need to consider the nucleon u-channel and t-channel
one-pion exchange. It holds
\begin{eqnarray}
&& \int\limits_{-1}^{1}\frac{d \cos \theta}{2}\frac{1}{u-m_N^2}=
\int\limits_{-\infty}^0\frac{-1}{4\,\bar p_{\rm cm}(s')\,p_{\rm cm}(s')}\,\frac{ds'}{s'-s}
\nonumber\\
&& \qquad\qquad \qquad \qquad  -\,\frac{\arctan{\left(2\,m_N\,\frac{|\bar p_{\rm cm}(m_N^2)|}{2\,m_N^2-m_\pi^2}\right)}}
{|\bar p_{\rm cm}(m_N^2)|}\, \frac{m_N}{s-m_N^2}\,,
\nonumber\\
&& \int\limits_{-1}^{1}\frac{d \cos \theta}{2}\,\frac{1}{t-m_{\pi}^2}=
\int\limits_{-\infty}^0\frac{-1}{4\,\bar p_{\rm cm}(s')\, p_{\rm cm}(s')}\,\frac{ds'}{s'-s}
\nonumber\\
&& \qquad\qquad \qquad \qquad +\,\int\limits_{0}^{(m_N-m_\pi)^2} \,\frac{2}{4\,\bar p_{\rm cm}(s')\, p_{\rm cm}(s')}\,\frac{ds'}{s'-s}
\nonumber\\
&& \qquad\qquad \qquad \qquad -\,\frac{\arctan{\left(2\,m_N\,\frac{|\bar p_{\rm cm}(m_N^2)|}{m_\pi^2}\right)}}
{|\bar p_{\rm cm}(m_N^2)|}\, \frac{m_N}{s-m_N^2}\,, \label{split-photoproduction}
\\ \nonumber\\
&& \bar p^2_{\rm cm}(s) = \big(s-(m_N-m_\pi)^2 \big)\,\frac{s-(m_N+m_\pi)^2}{4\,s}\,, \qquad
p^2_{\rm cm}(s) = \frac{(s-m_N^2)^2}{4\,s} \,. \nonumber
\end{eqnarray}
From the various contributions in (\ref{split-photoproduction}) only the pole terms, singular at $\sqrt{s}=m_N$, will be part of
$U_{\rm inside}(\sqrt{s}\,)$. Like for the case of elastic $\pi N$ scattering it is advantageous to perform a summation
in $U_{\rm outside}(\sqrt{s}\,)$, as implied by the cut contribution from $m_N-2\,m_\pi <\sqrt{s}< m_N-m_\pi$ in the one-pion exchange
contribution. This term leads to numerically large pole
structures at $\sqrt{s} = m_N-m_\pi$. The lower bound at $\sqrt{s}=m_N-2\,m_\pi $ is somewhat arbitrary, but suffices to identify
the important pole terms. We assure that our results do not depend on that specific choice. As in the case of the nucleon exchange
in $\pi N$ elastic scattering no such cut structures appear at higher orders.

The generalization of (\ref{split-photoproduction}) to the case of an arbitrary partial-wave is analogous to (\ref{disp_nucl_exch}), where a subtraction of the following form
 \begin{eqnarray}
&&\,\int\limits_{m_N-2 m_\pi}^{m_N-m_\pi} \,\frac{f(w)}{\bar p_{\rm cm}(w^2)\, (w-m_N+m_\pi)^n}\,\frac{dw}{w-\sqrt{s}}\longrightarrow
-\sum_{i=1}^n \frac{R_i}{(\sqrt{s}-m_N+m_\pi)^i}
\nonumber\\
&&+\,\int\limits_{m_N-2 m_\pi}^{m_N-m_\pi} \,\frac{f(w)}{\bar p_{\rm cm}(w^2)\, (\sqrt{s}-m_N+m_\pi)^n}\,\frac{dw}{w-\sqrt{s}}
\,,
\label{B6}
\end{eqnarray}
is useful. The function  $f(w)$ is regular at the pseudo threshold.
The residui $R_i$ in (\ref{B6}) are to be adjusted such that the r.h.s. of (\ref{B6}) does not have any
contributions from isolated poles at the pseudo threshold $\sqrt{s} = m_N-m_\pi$.

We turn to the u-channel nucleon and t-channel exchange processes in Compton scattering. For an s-wave angle average we derive the representations
\begin{eqnarray}
&& \int\limits_{-1}^{1}\frac{d\cos \theta }{2}\,\frac{1}{u-m_N^2}=
\int\limits_{-\infty}^0\frac{-1}{4\,p^2_{\rm cm}(s')}\,\frac{ds'}{s'-s}
-\frac{1}{s-m_N^2}\,,
\nonumber\\\\
&& \int\limits_{-1}^{1}\frac{d \cos \theta}{2}\,\frac{1}{t-m_{\pi}^2}=
\int\limits_{m_\pi^2}^{\infty}\frac{1}{4\,p^2_{\rm cm}\,C_+(m_t^2)}\,\frac{dm_t^2}{C_+(m^2_t)-s}\,\frac{d C_+(m_t^2)}{dm_t^2}\,
\nonumber\\
&& \qquad\qquad \qquad \qquad\! \! +\, \int\limits_{m_\pi^2}^{\infty}\frac{1}{4\,p^2_{\rm cm}\,C_-(m_t^2)}\,
\frac{d m_t^2}{C_-(m_t^2)-s}\,\frac{dC_-(m_t^2)}{d m_t^2}\,,
\nonumber\\
&& C_{\pm}(m_t^2)=m_N^2-\frac{m_t^2}{2} \pm \frac{1}{2}\,\sqrt{m_t^2\,(m_t^2-4\,m_N^2)}\,,\qquad
p^2_{\rm cm}(s) = \frac{(s-m_N^2)^2}{4\,s}\,. \nonumber
\label{split-Compton}
\end{eqnarray}
The pole term, singular at $\sqrt{s}=m_N$, from the nucleon u-channel exchange
contributes to $U_{\rm inside}(\sqrt{s}\,)$. The t-channel one-pion exchange contributes in the region with
$m_\pi < m_t < 2\,m_\pi$. Right at $m_t = 2\,m_\pi$ the functions $C_\pm(m^2_t)$ touch the contour
line $C_4$ of Fig.~\ref{fig:compton}. The generalization to the case of non-s-wave type contributions is
analogous to (\ref{disp_nucl_exch}).

\newpage

\end{appendix}
\bibliography{1}

\begin{thebibliography}{10}
\bibitem{Baru:2010xn}
V.~Baru, et~al., {Precision calculation of the $\pi^-$ deuteron scattering
  length and its impact on threshold pi-N scattering}.
\end{thebibliography}
\bibliographystyle{elsart-num}
\end{document}